\documentclass{article}
\usepackage{amssymb}
\usepackage{caption}
\usepackage{amsfonts}
\usepackage{amsmath}
\usepackage{amsthm}
\newtheorem{lemma}{Lemma}
\newtheorem{theorem}{Theorem}
\usepackage{empheq}
\usepackage{rotating}
\usepackage{graphicx}
\usepackage{subcaption}

\usepackage{fancyhdr}
\usepackage{lastpage}
\usepackage{bbm}
\usepackage{booktabs}
\usepackage{xurl}
\usepackage{graphicx}
\usepackage{hyperref}
\usepackage[paperheight=11in,paperwidth=8.5in,portrait,left=2.5cm,right=2.5cm,top=2.5cm,bottom=2cm]{geometry}
\usepackage{tabularx}
\usepackage[title]{appendix}\usepackage{rotating}
\usepackage{makecell}
\usepackage{diagbox}
\usepackage{longtable}
\usepackage{lscape}
\usepackage{footnote}

\usepackage{thmtools}
\declaretheoremstyle[
spaceabove=6pt, spacebelow=6pt,
headfont=\normalfont\bfseries,
notefont=\mdseries, notebraces={(}{)},
bodyfont=\normalfont,
postheadspace=0.6em,
headpunct=:
]{mystyle}
\declaretheorem[style=mystyle, name=Hypothesis, preheadhook={}]{hyp}
\usepackage{cleveref}
\crefname{hyp}{hypothesis}{hypotheses}
\Crefname{hyp}{Hypothesis}{Hypotheses}

\usepackage{mathtools}
\usepackage{siunitx}
\DeclarePairedDelimiterXPP\BigOSI[2]%
  {\mathcal{O}}{(}{)}{}%
  {\SI{#1}{#2}}

\makesavenoteenv{tabular}
\makesavenoteenv{table}

\usepackage{comment}
\providecommand{\keywords}[1]{\textbf{\textit{\hspace{0pt}Keywords}} #1}

\newcommand{\eqmath}[3][l]{\eqmakebox[#2][#1]{$\displaystyle\if#1l{}\fi#3$}}
	
\title{Predator Extinction arose from Chaos of the Prey: the Chaotic Behavior of a Homomorphic Two-Dimensional Logistic Map in the Form of Lotka-Volterra Equations}
\author{
		 Wei Shan Lee\footnote{email: \href{mailto:WSLEEemails}{wslee@g.puiching.edu.mo} }, Hou Fai Chan, Ka Ian Im, Kuan Ieong Chan, and U Hin Cheang
	}
\date{Pui Ching Middle School Macau\\ Macao Special Administrative Region, People's Republic of China.} 
\begin{document}
\maketitle

\begin{abstract}
		A two-dimensional homomorphic logistic map that preserves features of the Lotka-Volterra equations was proposed. To examine chaos, iteration plots of the population, Lyapunov exponents calculated from Jacobian eigenvalues of the $2$D logistic mapping, and from time series algorithms of Rosenstein and Eckmann et al. were calculated. Bifurcation diagrams may be divided into four categories depending on topological shapes. Our model not only recovered the $1$D logistic map, which exhibits flip bifurcation, for the prey when there is a nonzero initial predator population, but it can also simulate normal competition between two species with equal initial populations. Despite the possibility for two species to go into chaos simultaneously, where the Neimark-Sacker bifurcation was observed, it is also possible that with the same interspecies parameters as normal but with a predator population $10$ times more than that of the prey, the latter becomes chaotic, while the former dramatically reduces to zero with only a few iterations, indicating total annihilation of the predator species. Interpreting humans as predators and natural resources as preys in the ecological system, the above-mentioned conclusion may imply that not only excessive consumption of natural resources, but its chaotic state triggered by an overpopulation of humans may backfire in a manner of total extinction of the human species. Fortunately, there is little chance for the survival of the human race, as isolated fixed points in the bifurcation diagram of the predator reveal. Finally, two possible applications of the phenomenon of chaotic extinction are proposed: one is to inhibit viruses or pests by initiating the chaotic states of the prey on which the viruses or pests rely for existence, and the other is to achieve the superconducting state with the chaotic state of the applied magnetic field.
	\end{abstract}

\section{Introduction}
Understanding interactions between human beings and natural resources plays a vital role in the establishment of a sustainable economy and society. The prey and predator model may study the relationships of these two after we realize that humans may be regarded as predators. At the same time, natural resources may be considered preys$\cite{Roopnarine}$. Subsequently, research on prey-predator models can be implemented in this field instinctively$\cite{A Kumar et al}$-$\cite{Kin et al}$.	 

Generally speaking, there are two main approaches of studies in the literature to this prey and predator model. The first is to study differential equations, while the other is to check the iterations in difference equations, whose forms may be inspired by directly applying the forward Euler scheme to acquire a counterpart of the former$\cite{Hasan et al}$-$\cite{Hu Teng and Zhang}$. The discrete model could be more promising than the continuous one because it has more abundant dynamic characteristics in chaotic behaviors$\cite{Wang and Li}$. In contrast, it would be more difficult for solutions to continuous models to reach chaos in low-dimensional cases. Taking some examples of the first approach, studies$\cite{Cencini et al}$-$\cite{Vano et al}$ have performed on the chaos of Lotka-Volterra differential equations with dimensions higher than three, and researchers$\cite{Landau Paez and Bordeianu}$ claimed that it is impossible to reach chaos for two species in the form of differential Lotka-Volterra equations, whose general solutions were obtained in sinusoidal forms by Evans and Findley$\cite{Evans and Findley}$. Additionally, based on the Lotka-Volterra model, Dunbar$\cite{Dunbar}$ confirmed the existence of traveling wave solutions for two reaction-diffusion systems. In addition to that, Das and Gupta$\cite{Das and Gupta}$ proposed solutions to the fractional-order time derivative Lotka-Volterra equations using an analytical approach for nonlinear problems known as the homotopy perturbation method (HPM).

In addition, there are also several studies on the discrete difference equations. For example, Bessoir and Wolf $\cite{Bessoir and Wolf}$ made pioneering contributions to the application of $1$D logistic equation on biological and ecological studies. Many researchers also used the same equation to interpret, analyze and predict data according to COVID-19$\cite{Pelinovsky et al}$. Mareno and English$\cite{Mareno and English}$ implemented the $1$D logistic equation to the coupled $2$D logistic, and demonstrated that for a high growth rate the system underwent a Neimark-Sacker bifurcation. Li et al.$\cite{Li He Chen}$ imposed an equal intensity of individual effects, corresponding to the same growth rate in the $1$D logistic map, on the two oligopolists in the homomorphic Kopel model and observed three different kinds of bifurcation. Furthermore, Elhadi and Sprott$\cite{Elhadj and Sprott}$ proposed a two-dimensional mapping, one of which is the ordinary $1$D logistic map, while the other consists of a perturbation term of the former and is also modulated by the first. Shilnikov and Rulkov$\cite{Shilnikov and Rulkov}$ studied chaos behaviors in two-dimensional difference equations that reproduced spike bursting activities in biological neurons, further improving previous research based on the three-dimensional system of ODEs. Despite applying the forward Euler scheme to acquire the difference equation, researchers also used exponential forms corresponding to solutions in the differential equations. For example, Ishaque et al.$\cite{Ishaque et al}$ studied a three-dimensional predator-prey-parasite model with an exponential form that describes interactions between healthy or infected Tilapia fish as prey and Pelican birds as predator. Tassaddiq et al.$\cite{Tassaddiq et al}$ worked on the discrete-time exponential difference equation of the Leslie-Gower predator-prey model together with a Holling type III functional response and indicated the advantage in this type of discretization method. A previous study$\cite{Elsadany et al}$ suggested a heteromorphic term that describes the decreasing effects on the predator that was only linear to the population of that species, contrary to the corresponding quadratic term in the prey. Hassell et al.$\cite{Hassell et al}$ applied the predator-prey model to insect parasitoids and anthropods, and found that local movements of the two species may cause the extermination of the entire ecological system with chaos, and it is difficult to maintain population stability for a large growth rate of anthropods. Furthermore, researchers$\cite{Berryman and Millstein}$ also noted that human misbehavior may be the reason why an ecological system goes into chaotic states, and Danca et al.$\cite{Danca et al}$ studied predator extinction behavior after long iterations and provided solutions to prevent it from happening. Unfortunately, their system cannot see extinction directly in the bifurcation diagram due to the asymmetric form in their evolution equation.

However, there is no convincing reason for prey and predator to have different forms in the difference equations. Intuition in mathematical symmetry naturally came to our mind that a successful predator-prey difference model should resemble the symmetry structure as in the Lotka-Volterra differential equations. Moreover, solutions to Lotka-Volterra equations in sinusoidal forms cannot explain the extinction of species.

We proposed homomorphic two-dimensional logistic maps that preserve both forms of the Lotka-Volterra Equations and the $1$D logistic equation. In our model, we conjectured a quadratic form in both the corresponding terms of the prey and the predator, treating both species on the equal stance. The structures of the bifurcation diagrams showed that there could be four different categories in our dynamic system. For each category, we examined population iterations, phase portraits, phase space diagrams, and topological types of fixed points. Lyapunov exponents calculated from Jacobian eigenvalues of the $2$D mapping or from time-series algorithms, either Rosenstein$\cite{Rosenstein}$ or Eckmann et al.$\cite{Eckmann}$, were also calculated. Comparisons among those results were also discussed.  

The advantages of our model include the following. First, we may be able to establish a standard bifurcation diagram of $1$D logistic map about the prey with a non-zero initial predator population, growth rates in both species, and predation parameters. Second, our model may also describe the normal behavior of the rise and fall in the population of the two species when they interact with each other. Third, in addition to simultaneous chaos in both species, the main discovery in our research was that the predator may go extinct under the circumstance of chaos in the prey for which the predator overpopulation should be blamed.

\section{Model Formulation}
We first review the one-dimensional logistic equation and the two-dimensional Lotka-Volterra equations, comparing the similarities and differences between the two sets of equations. This inspires us to construct two hypotheses to establish our two-dimensional logistic equations that maintain the essential features of the original ones.

\subsection{Hypotheses inspired from $1$D logistic equation and Lotka-Volterra Equations}

To begin with, the one-dimensional logistic equation may be written as 
\begin{equation}\label{eq:1D logistic eq}
	x_{n+1} = \mu_{0}x_{n}(1-x_{n}),
\end{equation}
where $\mu_{0}$ denotes to the growth rate. 

On the other hand, the two-dimensional Lotka-Volterra equations$\cite{Lotka},\cite{Volterra}$ describe interactions between prey and predator in an environment where there is sufficient food supply for the prey, whose only natural enemy is the predator. The formula may be written as follows:
\begin{subequations}\label{eq:LV Equation}
	\begin{empheq}[left=\empheqlbrace]{align}
		\frac{dx}{dt}&=\mu_{0}x - \mu_{1}xy;\label{eq:diff for prey}\\	
		\frac{dy}{dt}&=-\nu_{0}y + \nu_{1}xy,\label{eq:diff for predator}
	\end{empheq}
\end{subequations}
where $\mu_{0}$, $\mu_{1}$, $\nu_{0}$, and $\nu_{1}$ are all positive parameters. $x$ denotes the prey population, while $y$ is the predator population, both belong to $[0,1]$. $\mu_{0}$ is the (per capita) growth rate of the prey, and $\mu_{1}$ refers to the inter-species parameter for the prey in the presence of the predator. $\nu_{0}$ refers to the (per capita) death rate, resulting from all effects that decrease the population of the predator, which may include disease, death, or emigration$\cite{wiki LV}$. At last, $\nu_{1}$ refers to the (per capita) growth rate of the predator at the presence of the prey.

With forward Euler's scheme, one may immediately write the difference version of Eq.($\ref{eq:LV Equation}$) as follows
\begin{subequations}
	\begin{empheq}[left=\empheqlbrace]{align}
		x_{n+1}-x_{n}&=\mu_{0}x_{n} - \mu_{1}x_{n}y_{n};\\
		y_{n+1}-y_{n}&=-\nu_{0}y_{n} + \nu_{1}x_{n}y_{n}.
	\end{empheq}
\end{subequations}

However, there is an obvious drawback about the above simultaneous equations: it does not preserve the feature of $1$D logistic equation because the first term on the right-hand side is linear to either $x_{n}$ or $y_{n}$, while the right-hand side in Eq.($\ref{eq:1D logistic eq}$) is quadratic to $x_{n}$. 

We now discuss establishing a two-dimensional map that resembles the interaction terms of the prey and the predator, as in Lotka-Volterra. First, we may rewrite the linear term of $x$ on the right-hand side of Eq.($\ref{eq:diff for prey}$) into quadratic, which looks like $\mu_{0}x(1-x)$, together with the homomorphic corresponding term in Eq.($\ref{eq:diff for predator}$) as $\nu_{0}y(1-y)$. Thus, the modified Lotka-Volterra Equations$\cite{Bomze}$ are 
\begin{subequations}\label{eq:modified LV Equation}
	\begin{empheq}[left=\empheqlbrace]{align}
		\frac{dx}{dt}&=\mu_{0}x(1-x) - \mu_{1}xy;\label{eq:modified LV for prey}\\	
		\frac{dy}{dt}&=-\nu_{0}y(1-y) + \nu_{1}xy,\label{eq:modified LV for predator}
	\end{empheq}
\end{subequations}
whose resemblance in the form of difference equations is, therefore,
\begin{subequations}\label{eq:version 6}
	\begin{empheq}[left=\empheqlbrace]{align}
		x_{n+1}-x_{n}&=\mu_{0}x_{n}(1-x_{n}) - \mu_{1}x_{n}y_{n};\\ 
		y_{n+1}-y_{n}&=-\nu_{0}y_{n}(1-y_{n}) + \nu_{1}x_{n}y_{n}.
	\end{empheq}
\end{subequations}
Although it seems more reasonable to the direct resemblance of Lotka-Volterra Equations, Eq.($\ref{eq:version 6}$) (Type 1) fails to restore Eq.($\ref{eq:1D logistic eq}$) when parameters other than $\mu_{0}$ are all set to zero. Fortunately, we may modify this by dropping $x_{n}$ and $y_{n}$ terms on the left-hand side of the above equations
\begin{subequations}\label{eq:2D logistic eq}
	\begin{empheq}[left=\empheqlbrace]{align}
		x_{n+1}&=\mu_{0} x_{n}(1-x_{n}) - \mu_{1}x_{n}y_{n};\label{version 5 for prey}\\	
		y_{n+1}&=-\nu_{0} y_{n}(1-y_{n}) + \nu_{1}x_{n}y_{n},\label{eq:version 5 for predator}
	\end{empheq}
\end{subequations}
which is the desired form. On the basis of the above discussions, we state the following two hypotheses:
\begin{hyp}\label{hyp:a}The mathematical formulation of the population is the \textit{same} for every species in the ecological system. This is also an intuition for mathematical symmetry.  \end{hyp}
\begin{hyp} \label{hyp:b}The predation of the prey in the presence of the predator (interspecies constant), death rate and growth rate of the predator are \textit{all} proportional to the growth rate of the prey with different proportionality. \end{hyp} Since  Eq.($\ref{eq:version 6}$) and Eq.($\ref{eq:2D logistic eq}$) both satisfiy \Cref{hyp:a,hyp:b}, we divide into two cases to further study the properties of Eq.($\ref{eq:version 6}$) (Type 1) and Eq.($\ref{eq:2D logistic eq}$)  (Type 2) in Sect. $\ref{subs:version 6}$ and Sect. $\ref{subs:version 5}$. But before that, in the next subsection, we first discuss a fundamental lemma that allows us to study the stability behaviors of fixed points.

\subsection{Stability of fixed points}\label{subs:stability}
Suppose that a mapping of two-dimensional iterations $x_{n+1}$ and $y_{n+1}$ is written as $x_{n+1} = f(x_{n},y_{n})$ and $y_{n+1} = g(x_{n},y_{n})$. The Jacobian is, therefore, 
\begin{subequations}\label{eq:Jacobian general case}
	\begin{empheq}[left=\empheqlbrace]{align}
		J &=	\frac{\partial (f,g)}{\partial (x,y)}\\				
		&=	\begin{pmatrix}
			\frac{\partial f}{\partial x} & \frac{\partial f}{\partial y} \\ 
			\frac{\partial g}{\partial x} & \frac{\partial g}{\partial y} 
		\end{pmatrix},
	\end{empheq}
\end{subequations}	
whose eigenvalus are $\omega_{0}$ and $\omega_{1}$. It is well known$\cite{Chen and Chen}$, $\cite{Wang and Li}$ that a fixed point can be divided into the following four topological types according to their stability behaviors. First, it could be a \textit{sink} and locally asymptotic stable if eigenvalues of Eq.($\ref{eq:Jacobian general case}$) satisfy $|\omega_{0}|<1$ and $|\omega_{1}|<1$. Second, it could be a \textit{source} and locally unstable if the eigenvalues satisfy $|\omega_{0}|>1$ and $|\omega_{1}|>1$. Third, a fixed point could be a \textit{saddle} if one of the absolute values of the eigenvalues is greater than $1$ while the other is smaller than $1$. Finally, a fixed point could be \textit{non-hyperbolic} if one of the absolute values of the eigenvalues is equal to $1$. The stability of a non-hyperbolic fixed point is fragile$\cite{Strogats}$, which means that the small nonlinear terms easily influence its stability. 

Instead of directly calculating the range of eigenvalues, most of the time it is more convenient to work with the quadratic formula consisting of eigenvalues $\omega_{0}$ and $\omega_{1}$, namely, \(\Omega(\omega)=\omega^2-\mathrm{Tr(J)}\omega+\mathrm{det(J)}, \) where $\mathrm{Tr(J)}$ and $\mathrm{det(J)}$ are trace and determinant of Jacobian in Eq.($\ref{eq:Jacobian general case}$), respectively, and there could be a correspondence on the stability behavior around a fixed point between the roots of the quadratic formula, $\omega_{0}$ and $\omega_{1}$, which are also Jacobian eigenvalues, through the following Lemma$\cite{Wang and Li}$
\begin{lemma}\label{lemma} Let \(\Omega(\omega)=\omega^2-\mathrm{Tr(J)}\omega+\mathrm{det(J)}, \) be a quadratic formula where $\mathrm{Tr(J)}$ and $\mathrm{det(J)}$ are trace and determinant of Jacobian in Eq.$\mathrm{(\ref{eq:Jacobian general case})}$, respectively. Then
	\begin{enumerate}
		\item If $\Omega(1)>0$, then
		\begin{enumerate}
			\item $|\omega_{0}|<1$ and $|\omega_{1}|<1$ and hence the fixed point is a sink \textit{if and only if} $\Omega(-1)>0$ and $\mathrm{det(J)}<1$;
			\item $|\omega_{0}|>1$ and $|\omega_{1}|>1$ and therefore the fixed point is a source \textit{if and only if} $\Omega(-1)>0$ and $\mathrm{det(J)}>1$;
			\item One of $|\omega_{0}|$ and $|\omega_{1}|$ is smaller than $1$ while the other greater than $1$ and hence the fixed point is a saddle \textit{if and only if} $\Omega(-1)<0$;	
			\item Either $|\omega_{0}|$ or $|\omega_{1}|$ is equal to $1$ and hence the fixed point is a non-hyperbolic whenever
			\begin{enumerate}
				\item $\omega_{0}=-1$ and $\omega_{1}\ne-1$ \textit{if and only if}  $\Omega(-1)=0$ and $\mathrm{Tr(J)}\ne2$.
				\item $\omega_{0}$ and $\omega_{1}$ are a pair of complex conjugates and $|\omega_{0}|=|\omega_{1}|=1$ \textit{if and only if} $|\mathrm{Tr(J)}|<2$ and $\mathrm{det(J)}=1$.
				\item $\omega_{0}=\omega_{1}=-1$ \textit{if and only if}  $\Omega(-1)=0$ and $\mathrm{Tr(J)}=2$.
			\end{enumerate} 
		\end{enumerate}
		\item If $\Omega(1)=0$, then either $|\omega_{0}|$ or $|\omega_{1}|$ has to be equal to $1$. Therefore, the fixed point is a non-hyperbolic. The absolute value of the other root is greater than, equal to, or smaller than $1$ \textit{if and only if}; correspondingly, the absolute value of $\mathrm{det(J)}$ is greater than, equal to, or smaller than $1$. 		
		\item If $\Omega(1)<0$, then either $|\omega_{0}|$ or $|\omega_{1}|$ must be greater than $1$. Therefore, the fixed point is a saddle. Also,
		\begin{enumerate}
			\item the other root is smaller or equal to $-1$ \textit{if and only if}, correspondingly, $\Omega(-1)<-1$ or $\Omega(-1)=-1$.
			\item absolute value of the other root is less than $1$ \textit{ if and only if}  $\Omega(-1)>0$.
		\end{enumerate}  
	\end{enumerate}
\end{lemma}
Lemma \textbf{\ref{lemma}} makes it easier for us to study the stability of fixed points analytically. In our analysis, we omitted all cases with negative parameters, as well as $\mu_{0}<1$.
\subsection{Properties of Eq.($\ref{eq:version 6}$) (Type 1)}\label{subs:version 6}
Setting up $x_{n+1} = f(x_{n+1},y_{n+1})$ and $y_{n+1} = g(x_{n+1},y_{n+1})$, the two-dimensional logistic equations in Eq.($\ref{eq:version 6}$) (Type 1) have the mappings
\begin{subequations}\label{eq:2D mapping version 6}
	\begin{empheq}[left=\empheqlbrace]{align}
		f(x,y)&=\mu_{0}x(1-x)-\mu_{1}xy+x;\\	
		g(x,y)&=-\nu_{0}y(1-y) + \nu_{1}xy+y.
	\end{empheq}
\end{subequations}
Eq.($\ref{eq:2D mapping version 6}$) has Jacobian, as indicated in Eq.($\ref{eq:Jacobian general case}$),
\begin{equation}\label{eq:Jacobian of version 6}
	J =	
	\begin{pmatrix}
		\mu_{0}(1-2x)-\mu_{1}y+1 & -\mu_{1}x \\ 
		\nu_{1}y & \nu_{0}(-1+2y) + \nu_{1}x+1
	\end{pmatrix},
\end{equation}			   		
with eigenvalues $\omega_{0}$ and $\omega_{1}$ being, respectively,
\begin{subequations}\label{eq: evals in version 6}
	\begin{empheq}[left=\empheqlbrace]{align}
		\omega_{0}&= -\mu_{0}x + \nu_{0}y + 1 + \frac{1}{2}(x\nu_{1} - y\mu_{1}) + \frac{1}{2}(\mu_{0} - \nu_{0}) + \frac{\omega}{2}\\ 
		\omega_{1}&= -\mu_{0}x + \nu_{0}y + 1 + \frac{1}{2}(x\nu_{1} - y\mu_{1}) + \frac{1}{2}(\mu_{0} - \nu_{0}) - \frac{\omega}{2} 
	\end{empheq}
\end{subequations}
where
\begin{center}
	\begin{align}\label{eq:omega}
		\omega&=\Bigg[4(\nu_{0}+\frac{\mu_{1}}{2})^2y^2+\Bigg(\bigg((4\mu_{0}-2\nu_{1})\mu_{1}+8\nu_{0}(\mu_{0}+\frac{\nu_{1}}{2})\bigg)x-4(\nu_{0}+\mu_{0})(\nu_{0}+\frac{\mu_{1}}{2})\Bigg)y\\\nonumber&+4\bigg((\mu_{0}+\frac{\nu_{1}}{2})x-\frac{\mu_{0}}{2}-\frac{\nu_{0}}{2}\bigg)^2\Bigg]^\frac{1}{2}
	\end{align}
\end{center}
Further, fixed points, at which pairs of $x$ and $y$ stay still irrespective of time-series iterations$\cite{Strogats}$, are those pairs of points $(x^*,y^*)$ such that
\begin{subequations}
	\begin{empheq}[left=\empheqlbrace]{align}
		x^*&=\mu_{0}x^*(1-x^*)-\mu_{1}x^*y^* + x^*\\
		y^*&=\nu_{0}y^*(1-y^*)-\nu_{1}x^*y^* + y^*,
	\end{empheq}
\end{subequations}	
from which four pairs of fixed points $(x^*,y^*)$ may be derived as
\begin{subequations}\label{eq: fixed points version 6}
	\begin{empheq}[left=\empheqlbrace]{align}
		&E_{0}=(0,0)\label{eq:fixed point E0 version 6}\\
		&E_{1}=(0,1)\label{eq:fixed point E1 version 6}\\
		&E_{2}=(1,0)\label{eq:fixed point E2version 6}\\
		&E_{3}=\bigg(\frac{ \nu_{0} (\mu_{1}-\mu_{0}) } {-\mu_{0}\nu_{0}+\mu_{1}\nu_{1}},\frac{\mu_{0}(\nu_{1}-\nu_{0})}{-\mu_{0}\nu_{0}+\mu_{1}\nu_{1}}\bigg),\label{eq:fixed point E3 version 6}
	\end{empheq}
\end{subequations}
provided that the denominator in Eq.($\ref{eq:fixed point E3 version 6}$) is not zero. On the contrary, however, when $\mu_{0}\nu_{0}=\mu_{1}\nu_{1}$, Eq.($\ref{eq:version 6}$) (Type 1) has fixed points $E_{0}$, $E_{1}$,and $E_{2}$.

Keeping \textbf{Lemma \ref{lemma}} in Sect. $\ref{subs:stability}$ in mind, we may be able to examine the topological type of each fixed point in Eq.($\ref{eq: fixed points version 6}$) as in Theorem $\ref{fixed points thm for version 6}$:
\begin{theorem}\label{fixed points thm for version 6}
	The topological types of fixed points in Eq.($\ref{eq: fixed points version 6}$) are 
	\begin{enumerate}
		\item For $E_{0}=(0,0)$,
		\begin{subequations}
			\begin{empheq}[left=\empheqlbrace]{align}\nonumber
				\Omega(1)&=-\mu_{0}\nu_{0}\\\nonumber
				\Omega(-1)&=(2+\mu_{0})(2-\nu_{0})\\\nonumber
				\mathrm{det(J)}&=(1+\mu_{0})(1-\nu_{0})\\\nonumber
				\mathrm{Tr(J)}&=2+\mu_{0}-\nu_{0}.
			\end{empheq}
		\end{subequations}
		Because $\Omega(1)<0$, $E_{0}$ is always a saddle.
		\item For $E_{1}=(0,1)$,
		\begin{subequations}
			\begin{empheq}[left=\empheqlbrace]{align}\nonumber
				\Omega(1)&=\nu_{0}(\mu_{0}-\mu_{1})\\\nonumber
				\Omega(-1)&=(2+\nu_{0})(2+\mu_{0}-\mu_{1})\\\nonumber
				\mathrm{det(J)}&=(1+\nu_{0})(1+\mu_{0}-\mu_{1})\\\nonumber
				\mathrm{Tr(J)}&=2+\mu_{0}-\mu_{1}+\nu_{0}.
			\end{empheq}
		\end{subequations}
		Therefore,
		\begin{enumerate}
			\item the fixed point is a source if $\mu_{0}>\mu_{1}$
			\item the fixed point is a saddle if $\mu_{0}<\mu_{1}$.
   			\item the fixed point is a non-hyperbole if $\mu_{0}=\mu_{1}$
		\end{enumerate}
		\item For $E_{2}=(1,0)$,
		\begin{subequations}
			\begin{empheq}[left=\empheqlbrace]{align}\nonumber
				\Omega(1)&=-\mu_{0}(\nu_{1}-\nu_{0})\\\nonumber
				\Omega(-1)&=(2-\mu_{0})(2+\nu_{1}-\nu_{0})\\\nonumber
				\mathrm{det(J)}&=(1-\mu_{0})(1+\nu_{1}-\nu_{0})\\\nonumber
				\mathrm{Tr(J)}&=2-\mu_{0}+(\nu_{1}-\nu_{0}).
			\end{empheq}
		\end{subequations}
		In this case,
		\begin{enumerate}
			\item the fixed point is a sink if one of the following sustains:
                    \begin{enumerate}
                        \item $\mu_{0}<2$, and $\nu_{1}<\nu_{0}<\nu_{1}+1$
                        \item $\mu_{0}<2$, and $\nu_{0}=\nu_{1}+1$
                        \item $\mu_{0}<2$, and $\nu_{1}+1<\nu_{0}<\nu_{1}+2$
                    \end{enumerate}
			\item if $\mu_{0}>2$ and $\nu_{0}>\nu_{1}+2$, then $E_{2}$ is a source;
                \item the fixed point is a saddle if one of the following sustains:
                    \begin{enumerate}
                        \item $\mu_{0}>2$ and $\nu_{1}<\nu_{0}<\nu_{1}+2$
                        \item $\mu_{0}<2$ and $\nu_{0}>\nu_{1}+2$
                        \item $\nu_{0}<\nu_{1}$
                    \end{enumerate}
			\item the fixed point is a non-hyperbole because of Lemma \textbf{\ref{lemma}}-$1(d)$ if one of the following sustains: 
                    \begin{enumerate}
                        \item $\mu_{0}=2$ and $\nu_{1}<\nu_{0}<\nu_{1}+2$
                        \item $\mu_{0}\leq2$, $\nu_{0}=\nu_{1}+2$
                        \item $\mu_{0}>2$ and $\nu_{0}=\nu_{1}+2$
                        \item $\mu_{0}=2$ and $\nu_{0}>\nu_{1}+2$
                    \end{enumerate}
   	\end{enumerate}
		\item For $E_{3}=\bigg(\frac{ \nu_{0} (\mu_{1}-\mu_{0}) } {-\mu_{0}\nu_{0}+\mu_{1}\nu_{1}},\frac{\mu_{0}(\nu_{1}-\nu_{0})}{-\mu_{0}\nu_{0}+\mu_{1}\nu_{1}}\bigg)$,
		\begin{subequations}
			\begin{empheq}[left=\empheqlbrace]{align}\nonumber
				\Omega(1)&=-\frac{\mu_{0}\nu_{0}(\nu_{0}-\nu_{1})(\mu_{0}-\mu_{1})}{\mu_{0}\nu_{0}-\mu_{1}\nu_{1}}\\\nonumber
				\Omega(-1)&=\frac{-\nu_{0}(\nu_{0}-\nu_{1}+2)\mu_{0}^{2}+\nu_{0}(\mu_{1}+2)(\nu_{0}-\nu_{1}+2)\mu_{0}-4\mu_{1}\nu_{1}}{\mu_{0}\nu_{0}-\mu_{1}\nu_{1}}\\\nonumber
				\mathrm{det(J)}&=\frac{-\nu_{0}(\nu_{0}-\nu_{1}+1)\mu_{0}^{2}+\nu_{0}(\mu_{1}+1)(\nu_{0}-\nu_{1}+1)\mu_{0}-\mu_{1}\nu_{1}}{\mu_{0}\nu_{0}-\mu_{1}\nu_{1}}\\\nonumber
				\mathrm{Tr(J)}&=\frac{  -\mu_{0}^{2}\nu_{0}+\nu_{0}(\nu_{0}+\mu_{1}-\nu_{1}+2)\mu_{0}-2\mu_{1}\nu_{1}}{\mu_{0}\nu_{0}-\mu_{1}\nu_{1}}.
			\end{empheq}
		\end{subequations}
		There are a lot of combinations for criteria of a sink from \textbf{Lemma \ref{lemma}}-$1(a)$, a source from \textbf{Lemma \ref{lemma}}-$1(b)$, and a saddle from \textbf{Lemma \ref{lemma}}-$1(c)$. We only provide conditions for a saddle from \textbf{Lemma \ref{lemma}}-$3$ and a non-hyperbole.
        \begin{enumerate}
            \item the fixed point is a saddle because of \textbf{Lemma \ref{lemma}}-$3$ if one of the following holds:
                \begin{enumerate}
                    \item $\mu_{0}=1$, $\mu_{1}<1$, $\nu_{1}<\nu_{0}$, and $\frac{\nu_{0}}{\nu_{1}}<\mu_{1}$
                    \item $\mu_{0}=1$, $\nu_{1}<\nu_{0}$, and $\frac{\nu_{0}}{\nu_{1}}<\mu_{1}$
                    \item $\mu_{0}=1$, $\mu_{1}<1$, $\nu_{0}<\nu_{1}$, and $\frac{\nu_{0}}{\nu_{1}}<\mu_{1}$
                    \item $\mu_{0}>1$, $\mu_{1}<\mu_{0}$, and $\nu_{1}<\nu_{0}$
                    \item $\mu_{0}>1$, $\nu_{1}<\nu_{0}$, and $\frac{\mu_{0}\nu_{0}}{\nu_{1}}<\mu_{1}$
                    \item $\mu_{0}>1$, $\mu_{1}<\mu_{0}$, $\nu_{0}<\nu_{1}$, and $\frac{\mu_{0}\nu_{0}}{\nu_{1}}<\mu_{1}$
                \end{enumerate}
            \item the fixed points is a non-hyperbole because of \textbf{Lemma \ref{lemma}}-$2$ if one of the following holds:
                \begin{enumerate}
                    \item $\mu_{0}=1$, $\mu_{1}=1$, and $\nu_{1}<\nu_{0}$
                    \item $\mu_{0}=1$, and $\mu_{1}\le1$
                    \item $\mu_{0}=1$, $\nu_{1}=\nu_{0}$, and $\mu_{1}>1$
                    \item $\mu_{0}=1$, $\mu_{1}=1$, and $\nu_{0}<\nu_{1}$
                    \item $\mu_{1}=\mu_{0}$, $\mu_{0}>1$, and $\nu_{1}<\nu_{0}$
                    \item $\nu_{1}=\nu_{0}$, $\mu_{0}>1$, and $\mu_{1}<\mu_{0}$
                    \item $\mu_{1}=\mu_{0}$, $\nu_{1}=\nu_{0}$, and $\mu_{0}>1$
                    \item $\nu_{1}=\nu_{0}$, $\mu_{0}>1$, and $\mu_{0}<\mu_{1}$
                    \item $\mu_{1}=\mu_{0}$, $\mu_{0}>1$, and $\nu_{0}<\nu_{1}$
                \end{enumerate}
        \item the fixed points is a non-hyperbole because of \textbf{Lemma \ref{lemma}}-$1(d)$ if one of the following holds:
                \begin{enumerate}
                    \item $\mu_{0}=1$, $\mu_{1}=-\frac{\nu_{0}(-\nu_{1}+\nu_{0}+2)}{\nu_{2}^{2}-\nu_{1}\nu_{0}+2\nu_{0}-4\nu_{1}}$, $\nu_{0}<2$, $\nu_{0}<\nu_{1}$, and $\nu_{1}<\nu_{0}+2$
                    \item $\mu_{1}=\frac{\mu_{0\nu_{0}}(\mu_{0}\nu_{0}-\mu_{0}\nu_{1}+2\mu_{0}-2\nu{0}+2\nu_{1}-4)}{\mu_{0}\nu_{0}^{2}-\nu_{1}\mu_{0}\nu_{0}+2\mu_{0}\nu_{0}-4\nu_{1}}$, $1<\mu_{0}<2$, $\nu_{0}<2$, $\nu_{0}<\nu_{1}$, and $\nu_{1}<\nu_{0}+2$
                    \item $\mu_{1}=\frac{\mu_{0\nu_{0}}(\mu_{0}\nu_{0}-\mu_{0}\nu_{1}+2\mu_{0}-2\nu{0}+2\nu_{1}-4)}{\mu_{0}\nu_{0}^{2}-\nu_{1}\mu_{0}\nu_{0}+2\mu_{0}\nu_{0}-4\nu_{1}}$, $\mu_{0}>2$, $\nu_{0}<2$, $\nu_{0}<\nu_{1}$, and $\nu_{1}<\frac{\mu_{0}\nu_{0}(\nu_{0}+2)}{\mu_{0}\nu_{0}+4}$
                    \item $\mu_{1}=\frac{\mu_{0\nu_{0}}(\mu_{0}\nu_{0}-\mu_{0}\nu_{1}+2\mu_{0}-2\nu{0}+2\nu_{1}-4)}{\mu_{0}\nu_{0}^{2}-\nu_{1}\mu_{0}\nu_{0}+2\mu_{0}\nu_{0}-4\nu_{1}}$, $\mu_{0}>2$, , $\mu_{0}>2$, $\nu_{0}<2$, $\nu_{1}<\frac{\mu_{0}(\nu_{0}+2)}{\mu_{0}-2}$, and $\nu_{0}+2<\nu_{1}$
                    \item $\mu_{0}=1$, $\mu_{1}=-\frac{\nu_{1}-4}{3\nu_{1}-4}$, $\nu_{0}=2$, and $2<\nu_{1}<4$
                    \item $\mu_{1}=\frac{\mu_{0}(\mu_{0}\nu_{1}-4\mu_{0}-2\nu_{1}+8)}{\mu_{0}\nu_{1}-4\mu_{0}+2\nu_{1}}$, $\nu_{0}=2$, $1<\mu_{0}<2$, and $2<\nu_{1}<4$
                    \item $\mu_{1}=\frac{\mu_{0}(\mu_{0}\nu_{1}-4\mu_{0}-2\nu_{1}+8)}{\mu_{0}\nu_{1}-4\mu_{0}+2\nu_{1}}$, $\nu_{0}=2$, $\mu_{0}>2$, and $2<\nu_{1}<\frac{4\mu_{0}}{\mu_{0}+2}$
                    \item $\mu_{1}=\frac{\mu_{0}(\mu_{0}\nu_{1}-4\mu_{0}-2\nu_{1}+8)}{\mu_{0}\nu_{1}-4\mu_{0}+2\nu_{1}}$, $\nu_{0}=2$, $\mu_{0}>2$, and $4<\nu_{1}<\frac{4\mu_{0}}{\mu_{0}-2}$
                    \item $\mu_{1}=1$, $\mu_{1}=-\frac{\nu_{0}(-\nu_{1}+\nu_{0}+2)}{\nu_{0}^{2}-\nu_{1}\nu_{0}+2\nu_{0}-4\nu_{1}}$, and $2<\nu_{0}<\nu_{1}<\nu_{0}+2$
                    \item $\mu_{1}=\frac{\mu_{0}\nu_{0}(\mu_{0}\nu_{0}-\mu_{0}\nu_{1}+2\mu_{0}-2\nu_{0}+2\nu_{1}-4)}{\mu_{0}\nu_{0}^{2}-\nu{1}\mu_{0}\nu_{0}+2\mu_{0}\nu_{0}-4\nu_{1}}$, $1<\mu_{0}<2$, and $2<\nu_{0}<\nu_{1}<\nu_{0}+2$
                    \item $\mu_{1}=\frac{\mu_{0}\nu_{0}(\mu_{0}\nu_{0}-\mu_{0}\nu_{1}+2\mu_{0}-2\nu_{0}+2\nu_{1}-4)}{\mu_{0}\nu_{0}^{2}-\nu_{1}\mu_{0}\nu_{0}+2\mu_{0}\nu_{0}-4\nu_{1}}$, $2<\mu_{0}$, and $2<\nu_{0}<\nu_{1}<\frac{\mu_{0}\nu_{0}(\nu_{0}+2)}{\mu_{0}\nu_{0}+4}$
                    \item $\mu_{1}=\frac{\mu_{0}\nu_{0}(\mu_{0}\nu_{0}-\mu_{0}\nu_{1}+2\mu_{0}-2\nu_{0}+2\nu_{1}-4)}{\mu_{0}\nu_{0}^{2}-\nu_{1}\mu_{0}\nu_{0}+2\mu_{0}\nu_{0}-4\nu_{1}}$, $2<\mu_{0}$, $2<\nu_{0}$, and $\nu_{0}+2<\nu_{1}<\frac{\mu_{0}(\nu_{0}+2)}{\mu_{0}-2}$ 
                \end{enumerate}
        \end{enumerate}
	\end{enumerate}
\end{theorem}

By \Cref{hyp:b}, we further assume proportionality of parameters $\mu_{1}=\alpha\mu_{0}$, $\nu_{0}=\beta\mu_{0}$, and $\nu_{1}=\gamma\mu_{0}$, where $\alpha$, $\beta$, and $\gamma$ are other designated parameters. Under this circumstance, the nontrivial fixed point $E_{3}$ becomes
\begin{equation}\nonumber
	E_{3}=\bigg(\frac{ \beta(\alpha-1) }{\alpha\gamma-\beta} ,\frac{\gamma-\beta}{\alpha\gamma-\beta}\bigg),
\end{equation}
which is independent of the growth rate parameter $\mu_{0}$. It should be noted that for $E_{3}$ to remain inside the square enclosed by the vortexes $(0,0),(0,1),(1,1),(1,0)$, $\alpha$, $\beta$, and $\gamma$ must satisfy either $\{\alpha<1$, and $\gamma<\beta\}$, or $\{\alpha>1$, and $\gamma>\beta\}$. The Jacobian eigenvalues at $E_{3}$ in Eq.($\ref{eq:Jacobian of version 6}$) is 
\begin{subequations}\label{eq:evals version 6 alpha beta gamma}
	\begin{empheq}[left=\empheqlbrace]{align}
		\omega_{0}&= \frac{1}{2\alpha\gamma-2\beta}\bigg(\Xi_{0}+\frac{\alpha\gamma-\beta}{|\alpha\gamma-\beta|}\Xi_{1}\bigg)\\ 
		\omega_{1}&= \frac{1}{2\alpha\gamma-2\beta}\bigg(\Xi_{0}-\frac{\alpha\gamma-\beta}{|\alpha\gamma-\beta|}\Xi_{1}\bigg), 
	\end{empheq}
\end{subequations}
where 
\begin{subequations}
	\begin{empheq}[left=\empheqlbrace]{align}
		\Xi_{0}&=2\alpha\gamma-\beta^{2}\mu_{0}-\big(2+(\alpha-\gamma-1)\mu_{0}\big)\beta\\
		\Xi_{1}&=\mu_{0}\Bigg[\beta  
		\bigg(  
		(4\beta\gamma-4\gamma^{2}+\beta)\alpha^{2}-
		(2\beta^{2}+2\beta\gamma-4\gamma^{2}+2\beta)\alpha
		+\beta(\beta-\gamma+1)^{2}
		\bigg)
		\Bigg]^{\frac{1}{2}}		 
	\end{empheq}
\end{subequations}

\subsection{Properties of Eq.($\ref{eq:2D logistic eq}$)  (Type 2)}\label{subs:version 5}
Similar to Eq.($\ref{eq:2D mapping version 6}$), the two-dimensional logistic equations in Eq.($\ref{eq:2D logistic eq}$)  (Type 2) have the mappings
\begin{subequations}\label{eq:2D logistic map}
	\begin{empheq}[left=\empheqlbrace]{align}
		f(x,y)&=\mu_{0}x(1-x)-\mu_{1}xy;\label{eq: logistic map for prey}\\	
		g(x,y)&=-\nu_{0}y(1-y) + \nu_{1}xy.\label{eq: logistic map for predator}
	\end{empheq}
\end{subequations}
Eq.($\ref{eq:2D logistic map}$) has Jacobian that is slightly different from Eq.($\ref{eq:Jacobian of version 6}$)
\begin{equation}\label{eq:Jacobian of version 5} 
	J =	\begin{pmatrix}
		\mu_{0}(1-2x) - \mu_{1}y & -\mu_{1}x \\ 
		\nu_{1}y & \nu_{0}(-1+2y) + \nu_{1}x
	\end{pmatrix},
\end{equation}			   		
with eigenvalues $\omega_{0}$ and $\omega_{1}$ being, respectively,
\begin{subequations}\label{eq:evals version 5}
	\begin{empheq}[left=\empheqlbrace]{align}
		\omega_{0}&= -\mu_{0}x + \nu_{0}y + \frac{1}{2}(x\nu_{1} - y\mu_{1}) + \frac{1}{2}(\mu_{0} - \nu_{0}) + \frac{\omega}{2}\\ 
		\omega_{1}&= -\mu_{0}x + \nu_{0}y + \frac{1}{2}(x\nu_{1} - y\mu_{1}) + \frac{1}{2}(\mu_{0} - \nu_{0}) - \frac{\omega}{2} 
	\end{empheq}
\end{subequations}
where $\omega$ is the same as in Eq.($\ref{eq:omega}$).
Fixed points for Eq.($\ref{eq:2D logistic eq}$)  (Type 2) are
\begin{subequations}\label{eq: fixed points version 5}
	\begin{empheq}[left=\empheqlbrace]{align}
		&E_{0}^{\prime}=(0,0)\label{eq:fixed point E0}\\
		&E_{1}^{\prime}=(0,1+\frac{1}{\nu_{0}})\label{eq:fixed point E1}\\
		&E_{2}^{\prime}=(1-\frac{1}{\mu_{0}},0)\label{eq:fixed point E2}\\
		&E_{3}^{\prime}=\bigg(\frac{-\mu_{0}\nu_{0}+\mu_{1}\nu_{0}+\mu_{1}+\nu_{0}}{-\mu_{0}\nu_{0}+\mu_{1}\nu_{1}},\frac{-\mu_{0}\nu_{0}+\mu_{0}\nu_{1}-\mu_{0}-\nu_{1}}{-\mu_{0}\nu_{0}+\mu_{1}\nu_{1}}\bigg)\label{eq:fixed point E3}
	\end{empheq}
\end{subequations}
However, when $\mu_{0}\nu_{0}=\mu_{1}\nu_{1}$, Eq.($\ref{eq:2D logistic eq}$)  (Type 2) has fixed points $E_{0}^{\prime}$, $E_{1}^{\prime}$, and $E_{2}^{\prime}$.

We may also make use of \textbf{Lemma \ref{lemma}} in Sect. $\ref{subs:stability}$ to examine the topological type of each fixed point in Eq.($\ref{eq:2D logistic eq}$)  (Type 2) as in Theorem $\ref{fixed points thm for version 5}$:
\begin{theorem}\label{fixed points thm for version 5}
	The topological types of fixed points in Eq.($\ref{eq: fixed points version 5}$) are 
	\begin{enumerate}
		\item For $E_{0}^{\prime}=(0,0)$,
		\begin{subequations}
			\begin{empheq}[left=\empheqlbrace]{align}\nonumber
				\Omega(1)&=-(\mu_{0}-1)(\nu_{0}+1)\\\nonumber
				\Omega(-1)&=-(\mu_{0}+1)(\nu_{0}-1)\\\nonumber
				\mathrm{det(J)}&=-\mu_{0}\nu_{0}\\\nonumber
				\mathrm{Tr(J)}&=\mu_{0}-\nu_{0}.
			\end{empheq}
		\end{subequations}
		In this case,
		\begin{enumerate}
			\item if $\mu_{0}>1$, then $E_{0}^{\prime}$ is a source.
			\item if $\mu_{0}=1$, then $E_{0}^{\prime}$ is a non-hyperbole.
		\end{enumerate}
		\item For $E_{1}^{\prime}=(0,1+\frac{1}{\nu_{0}})$,
		\begin{subequations}
			\begin{empheq}[left=\empheqlbrace]{align}\nonumber
				\Omega(1)&=\frac{(\nu_{0}+1)((\mu_{0}-\mu_{1}-1)\nu_{0}-\mu_{1})}{\nu_{0}}\\\nonumber
				\Omega(-1)&=\frac{(\nu_{0}+3)((\mu_{0}-\mu_{1}+1)\nu_{0}-\mu_{1})}{\nu_{0}}\\\nonumber
				\mathrm{det(J)}&=\frac{(\nu_{0}+2)((\mu_{0}-\mu_{1})\nu_{0}-\mu_{1})}{\nu_{0}}\\\nonumber
				\mathrm{Tr(J)}&=\frac{\nu_{0}^{2}+(\mu_{0}-\mu_{1}+2)\nu_{0}-\mu_{1}}{\nu_{0}}.
			\end{empheq}
		\end{subequations}
		Thus,
		\begin{enumerate}
			\item if $\mu_{0}>\frac{\mu_{1}\nu_{0}+\mu_{1}+\nu_{0}}{\nu_{0}}$, then $E_{1}^{\prime}$ is a source.
			\item if $\mu_{0}<\frac{\mu_{1}\nu_{0}+\mu_{1}+\nu_{0}}{\nu_{0}}$, then $E_{1}^{\prime}$ is a saddle.
			\item if $\mu_{0}=\frac{\mu_{1}\nu_{0}+\mu_{1}+\nu_{0}}{\nu_{0}}$, then $E_{1}^{\prime}$ is a non-hyperbole.
		\end{enumerate}
		\item For $E_{2}^{\prime}=(1-\frac{1}{\mu_{0}},0)$,
		\begin{subequations}
			\begin{empheq}[left=\empheqlbrace]{align}\nonumber
				\Omega(1)&=\frac{(\mu_{0}-1)((\nu_{0}-\nu_{1}+1)\mu_{0}+\nu_{1})}{\mu_{0}}\\\nonumber
				\Omega(-1)&=\frac{(\mu_{0}-3)((\nu_{0}-\nu_{1}-1)\mu_{0}+\nu_{1})}{\mu_{0}}\\\nonumber
				\mathrm{det(J)}&=\frac{(\mu_{0}-2)((\nu_{0}-\nu_{1})\mu_{0}+\nu_{1})}{\mu_{0}}\\\nonumber
				\mathrm{Tr(J)}&=-\frac{\mu_{0}^{2}+(\nu_{0}-\nu_{1}-2)\mu_{0}+\nu_{1}}{\mu_{0}}.
			\end{empheq}
		\end{subequations}
		Afterward,
		\begin{enumerate}
			\item the fixed point is a sink if one of the following holds:
                    \begin{enumerate}
                        \item $1<\mu_{0}<2$ and $\frac{\nu_{1}\mu_{0}-\mu_{0}-\nu_{1}}{\mu_{0}}<\nu_{0}<\frac{\nu_{1}\mu_{0}+\mu_{0}-\nu_{1}}{\mu_{0}}$
                        \item $\mu_{0}=2$ and $\frac{\nu_{1}}{2}-1<\nu_{0}<\frac{\nu_{1}}{2}+1$ with $\nu_{1}>2$
                        \item $2<\mu_{0}<3$ and $\frac{\nu_{1}\mu_{0}-\mu_{0}-\nu_{1}}{\mu_{0}}<\nu_{0}<\frac{\nu_{1}\mu_{0}+\mu_{0}-\nu_{1}}{\mu_{0}}$
                    \end{enumerate}
                \item the fixed point is a source if $\mu_{0}>3$ and $\nu_{0}>\frac{\nu_{1}\mu_{0}+\mu_{0}-\nu_{1}}{\mu_{0}}$
                \item the fixed point is a saddle if one of the following holds:
                    \begin{enumerate}
                        \item $1\le\mu_{0}<3$ and $\nu_{0}>\frac{\nu_{1}\mu_{0}+\mu_{0}-\nu_{1}}{\mu_{0}}$
                        \item $\mu_{0}>3$ and $\frac{\nu_{1}\mu_{0}-\mu_{0}-\nu_{1}}{\mu_{0}}<\nu_{0}<\frac{\nu_{1}\mu_{0}+\mu_{0}-\nu_{1}}{\mu_{0}}$
                        \item $1<\mu_{0}$ and  $\nu_{0}<\frac{\nu_{1}\mu_{0}-\mu_{0}-\nu_{1}}{\mu_{0}}$
                    \end{enumerate}
   			\item the fixed point is a non-hyperbole if one of the following holds:
                    \begin{enumerate}
                        \item $\mu_{1}=1$
                        \item $\nu_{0}=\frac{\nu_{1}\mu_{0}-\mu_{0}-\nu_{1}}{\mu_{0}}$ 
                        \item $\nu_{0}=\frac{\nu_{1}\mu_{0}+\mu_{0}-\nu_{1}}{\mu_{0}}$ and $\mu_{0}\ne1$
                        \item $\mu_{0}=3$ and $\nu_{0}\ne\frac{2\nu_{1}}{3}+1$ and $\nu_{0}>\frac{2\nu_{1}}{3}-1$
                    \end{enumerate}
		\end{enumerate}
		
		\item For $E_{3}^{\prime}=\bigg(\frac{-\mu_{0}\nu_{0}+\mu_{1}\nu_{0}+\mu_{1}+\nu_{0}}{-\mu_{0}\nu_{0}+\mu_{1}\nu_{1}},\frac{-\mu_{0}\nu_{0}+\mu_{0}\nu_{1}-\mu_{0}-\nu_{1}}{-\mu_{0}\nu_{0}+\mu_{1}\nu_{1}}\bigg)$,
		\begin{subequations}
			\begin{empheq}[left=\empheqlbrace]{align}\nonumber
				\Omega(1)&=-\frac{\mu_{0}\nu_{0}(\nu_{0}-\nu_{1})(\mu_{0}-\mu_{1})}{\mu_{0}\nu_{0}-\mu_{1}\nu_{1}}\\\nonumber
				\Omega(-1)&=\frac{-\nu_{0}(\nu_{0}-\nu_{1}+2)\mu_{0}^{2}+\nu_{0}(\mu_{1}+2)(\nu_{0}-\nu_{1}+2)\mu_{0}-4\mu_{1}\nu_{1}}{\mu_{0}\nu_{0}-\mu_{1}\nu_{1}}\\\nonumber
				\mathrm{det(J)}&=\frac{-\nu_{0}(\nu_{0}-\nu_{1}+1)\mu_{0}^{2}+\nu_{0}(\mu_{1}+1)(\nu_{0}-\nu_{1}+1)\mu_{0}-\mu_{1}\nu_{1}}{\mu_{0}\nu_{0}-\mu_{1}\nu_{1}}\\\nonumber
				\mathrm{Tr(J)}&=\frac{-\mu_{0}^{2}\nu_{0}+\nu_{0}(\nu_{0}+\mu_{1}-\nu_{1}+2)\mu_{0}-2\mu_{1}\nu_{1}}{\mu_{0}\nu_{0}-\mu_{1}\nu_{1}}.
			\end{empheq}
		\end{subequations}
		As before, conditions for a saddle from \textbf{Lemma \ref{lemma}}-$3$ and a non-hyperbole are given as follows:
        \begin{enumerate}
            \item the fixed point is a saddle if one of the following holds:
            \begin{enumerate}
                \item $\mu_{0}=1$ and $\frac{\nu_{0}}{\nu_{1}}<\mu_{1}$
                \item $\mu_{0}>1$, $\mu_{1}<\frac{\nu_{0}\mu_{0}-1}{\nu_{0}+1}$, and $\nu_{1}<\frac{(\nu_{1}+1)\mu_{0}}{\mu_{0}-1}$
                \item $\mu_{0}>1$, $\nu_{1}<\frac{(\nu_{0}+1)\mu_{0}}{\mu_{0}-1}$, and $\frac{\mu_{0}\nu_{0}}{\nu_{1}}<\mu_{1}$
                \item $1<\mu_{0}$, $\frac{\mu_{0}\nu_{0}}{\nu_{1}}<\mu_{1}<\frac{\nu_{0}(\mu_{0}-1)}{\nu_{0}+1}$, and $\frac{(\nu_{0}+1)\mu_{0}}{\mu{0}-1}<\nu_{1}$
            \end{enumerate}
            \item the fixed point is a non-hyperbole if one of the following holds for \textbf{Lemma \ref{lemma}}-$2$:
            \begin{enumerate}
                \item $\mu_{1}=\frac{\nu_{0}(\mu_{0}-1)}{\nu_{0}+1}$, $\mu_{0}>1$ and $\nu_{1}<\frac{(\nu_{0}+1)\mu_{0}}{\mu_{0}-1}$
                \item $\nu_{1}=\frac{(\nu_{0}+1)\mu_{0}}{\mu_{0}-1}$, $\mu_{0}>1$, and $\mu_{1}<\frac{\nu_{0}(\mu_{0}-1)}{\nu_{0}+1}$
                \item $\mu_{1}=\frac{\nu_{0}(\mu_{0}-1)}{\nu_{0}+1}$, $\nu_{1}=\frac{(\nu_{0}+1)\mu_{0}}{\mu_{0}-1}$, and $\mu_{0}>1$
                \item $\nu_{1}=\frac{(\nu_{0}+1)\mu_{0}}{\mu_{0}-1}$, $\mu_{0}>1$, and $\mu_{1}>\frac{\nu_{0}(\mu_{0}-1)}{\nu_{0}+1}$
                \item $\mu_{1}=\frac{\nu_{0}(\mu_{0}-1)}{\nu_{0}+1}$, $\mu_{0}>1$, and $\nu_{1}>\frac{(\nu_{0}+1)\mu_{0}}{\mu_{0}-1}$
            \end{enumerate}
            \item the fixed point is a non-hyperbole if one of the following holds for \textbf{Lemma \ref{lemma}}-$1(d)$:
            \begin{enumerate}
                \item $1<\mu_{0}<3$, $\mu_{1}=\frac{\nu_{0}(\mu_{0}^{2}\nu_{0}-\mu_{0}^{2}\nu_{1}+3\mu_{0}^{2}-3\mu_{0}\nu_{0}+4\mu_{0}\nu_{1}-9\mu_{0}-3\nu_{1})}{\mu_{0}\nu_{0}^{2}-\nu_{1}\mu_{0}\nu_{0}+4\mu_{0}\nu_{0}-\nu_{1}\mu_{0}+\nu_{1}\nu_{0}+3\mu_{0}-3\nu_{1}}$, $\nu_{0}<1$, and $\frac{\mu_{0}(\nu_{0}+1)}{\mu_{0}-1}<\nu_{1}<\frac{\mu_{0}(\nu_{0}+3)}{\mu_{0}-1}$
                \item $\mu_{0}>3$, $\mu_{1}=\frac{\nu_{0}(\mu_{0}^{2}\nu_{0}-\mu_{0}^{2}\nu_{1}+3\mu_{0}^{2}-3\mu_{0}\nu_{0}+4\mu_{0}\nu_{1}-9\mu_{0}-3\nu_{1})}{\mu_{0}\nu_{0}^{2}-\nu_{1}\mu_{0}\nu_{0}+4\mu_{0}\nu_{0}-\nu_{1}\mu_{0}+\nu_{1}\nu_{0}+3\mu_{0}-3\nu_{1}}$, $\nu_{0}<1$, and $\frac{\mu_{0}(\nu_{0}+1)}{\mu_{0}-1}<\nu_{1}<\frac{\mu_{0}(\nu_{0}^{2}+4\nu_{0}+3)}{\mu_{0}\nu_{0}+\mu_{0}-\nu_{0}+3}$
                \item $\mu_{0}>3$, $\mu_{1}=\frac{\nu_{0}(\mu_{0}^{2}\nu_{0}-\mu_{0}^{2}\nu_{1}+3\mu_{0}^{2}-3\mu_{0}\nu_{0}+4\mu_{0}\nu_{1}-9\mu_{0}-3\nu_{1})}{\mu_{0}\nu_{0}^{2}-\nu_{1}\mu_{0}\nu_{0}+4\mu_{0}\nu_{0}-\nu_{1}\mu_{0}+\nu_{1}\nu_{0}+3\mu_{0}-3\nu_{1}}$, $\nu_{0}<1$, and $\frac{\mu_{0}(\nu_{0}+3)}{\mu_{0}-1}<\nu_{1}<\frac{\mu_{0}(\nu_{0}+3)}{\mu_{0}-3}$
                \item $1<\mu_{0}<3$, $\mu_{1}=\frac{\mu_{0}^{2}\nu_{1}-4\mu_{0}^{2}-4\mu_{0}\nu_{1}+12\mu_{0}+3\nu_{1}}{2(\mu_{0}\nu_{1}-4\mu_{0}+\nu_{1})}$, $\nu_{0}=1$, and $\frac{2\mu_{0}}{\mu_{0}-1}<\nu_{1}<\frac{4\mu_{0}}{\mu_{0}-1}$
                \item $\mu_{0}>3$, $\mu_{1}=\frac{\mu_{0}^{2}\nu_{1}-4\mu_{0}^{2}-4\mu_{0}\nu_{1}+12\mu_{0}+3\nu_{1}}{2(\mu_{0}\nu_{1}-4\mu_{0}+\nu_{1})}$, $\nu_{0}=1$, and $\frac{2\mu_{0}}{\mu_{0}-1}<\nu_{1}<\frac{4\mu_{0}}{\mu_{0}-1}$
                \item $\mu_{0}>3$, $\mu_{1}=\frac{\mu_{0}^{2}\nu_{1}-4\mu_{0}^{2}-4\mu_{0}\nu_{1}+12\mu_{0}+3\nu_{1}}{2(\mu_{0}\nu_{1}-4\mu_{0}+\nu_{1})}$, $\nu_{0}=1$, and $\frac{4\mu_{0}}{\mu_{0}-1}<\nu_{1}<\frac{4\mu_{0}}{\mu_{0}-3}$
                \item $1<\mu_{0}<3$, $\mu_{1}=\frac{\nu_{0}(\mu_{0}^{2}\nu_{0}-\mu_{0}^{2}\nu_{1}+3\mu_{0}^{2}-3\mu_{0}\nu_{0}+4\mu_{0}\nu_{1}-9\mu_{0}-3\nu_{1})}{\mu_{0}\nu_{0}^{2}-\nu_{1}\mu_{0}\nu_{0}+4\mu_{0}\nu_{0}-\nu_{1}\mu_{0}+\nu_{1}\nu_{0}+3\mu_{0}-3\nu_{1}}$, $1<\nu_{0}<3$, and $\frac{\mu_{0}(\nu_{0}+1)}{\mu_{0}-1}<\nu_{1}<\frac{\mu_{0}(\nu_{0}+3)}{\mu_{0}-1}$
                \item $\mu_{0}>3$, $\mu_{1}=\frac{\nu_{0}(\mu_{0}^{2}\nu_{0}-\mu_{0}^{2}\nu_{1}+3\mu_{0}^{2}-3\mu_{0}\nu_{0}+4\mu_{0}\nu_{1}-9\mu_{0}-3\nu_{1})}{\mu_{0}\nu_{0}^{2}-\nu_{1}\mu_{0}\nu_{0}+4\mu_{0}\nu_{0}-\nu_{1}\mu_{0}+\nu_{1}\nu_{0}+3\mu_{0}-3\nu_{1}}$, $1<\nu_{0}<3$, and $\frac{\mu_{0}(\nu_{0}+1)}{\mu_{0}-1}<\nu_{1}<\frac{\mu_{0}(\nu_{0}^{2}+4\nu_{0}+3)}{\mu_{0}\nu_{0}+\mu_{0}-\nu_{0}+3}$
                \item $\mu_{0}>3$, $\mu_{1}=\frac{\nu_{0}(\mu_{0}^{2}\nu_{0}-\mu_{0}^{2}\nu_{1}+3\mu_{0}^{2}-3\mu_{0}\nu_{0}+4\mu_{0}\nu_{1}-9\mu_{0}-3\nu_{1})}{\mu_{0}\nu_{0}^{2}-\nu_{1}\mu_{0}\nu_{0}+4\mu_{0}\nu_{0}-\nu_{1}\mu_{0}+\nu_{1}\nu_{0}+3\mu_{0}-3\nu_{1}}$, $1<\nu_{0}<3$, and $\frac{\mu_{0}(\nu_{0}+3)}{\mu_{0}-1}<\nu_{1}<\frac{\mu_{0}(\nu_{0}+3)}{\mu_{0}-3}$
                \item $3>\mu_{0}>1$, $\mu_{1}=\frac{3(\mu_{0}^{2}\nu_{1}-6\mu_{0}^{2}-4\mu_{0}\nu_{1}+18\mu_{0}+3\nu_{1})}{4\mu_{0}(\nu_{1}-6)}$, $\nu_{0}=3$, and $\frac{4\mu_{0}}{\mu_{0}-1}<\nu_{1}<\frac{6\mu_{0}}{\mu_{0}-1}$
                \item $\mu_{0}>3$, $\mu_{1}=\frac{3(\mu_{0}^{2}\nu_{1}-6\mu_{0}^{2}-4\mu_{0}\nu_{1}+18\mu_{0}+3\nu_{1})}{4\mu_{0}(\nu_{1}-6)}$, $\nu_{0}=3$, and $\frac{4\mu_{0}}{\mu_{0}-1}<\nu_{1}<6$
                \item $\mu_{0}>3$, $\mu_{1}=\frac{3(\mu_{0}^{2}\nu_{1}-6\mu_{0}^{2}-4\mu_{0}\nu_{1}+18\mu_{0}+3\nu_{1})}{4\mu_{0}(\nu_{1}-6)}$, $\nu_{0}=3$, and $\frac{6\mu_{0}}{\mu_{0}-1}<\nu_{1}<\frac{6\mu_{0}}{\mu_{0}-3}$
                \item $1<\mu_{0}<3$, $\mu_{1}=\frac{\nu_{0}(\mu_{0}^{2}\nu_{0}-\mu_{0}^{2}\nu_{1}+3\mu_{0}^{2}-3\mu_{0}\nu_{0}+4\mu_{0}\nu_{1}-9\mu_{0}-3\nu_{1})}{\mu_{0}\nu_{0}^{2}-\nu_{1}\mu_{0}\nu_{0}+4\mu_{0}\nu_{0}-\nu_{1}\mu_{0}+\nu_{1}\nu_{0}+3\mu_{0}-3\nu_{1}}$, $\nu_{0}>3$, and $\frac{\mu_{0}(\nu_{0}+1)}{\mu_{0}-1}<\nu_{1}<\frac{\mu_{0}(\nu_{0}+3)}{\mu_{0}-1}$
                \item $1<\mu_{0}<3$, $\mu_{1}=\frac{\nu_{0}(\mu_{0}^{2}\nu_{0}-\mu_{0}^{2}\nu_{1}+3\mu_{0}^{2}-3\mu_{0}\nu_{0}+4\mu_{0}\nu_{1}-9\mu_{0}-3\nu_{1})}{\mu_{0}\nu_{0}^{2}-\nu_{1}\mu_{0}\nu_{0}+4\mu_{0}\nu_{0}-\nu_{1}\mu_{0}+\nu_{1}\nu_{0}+3\mu_{0}-3\nu_{1}}$, $\nu_{0}>3$, and $\frac{\mu_{0}(\nu_{0}+1)}{\mu_{0}-1}<\nu_{1}<\frac{\mu_{0}(\nu_{0}^{2}+4\nu_{0}+3)}{\mu_{0}\nu_{0}+\mu_{0}-\nu_{0}+3}$
                \item $\mu_{0}>3$, $\mu_{1}=\frac{\nu_{0}(\mu_{0}^{2}\nu_{0}-\mu_{0}^{2}\nu_{1}+3\mu_{0}^{2}-3\mu_{0}\nu_{0}+4\mu_{0}\nu_{1}-9\mu_{0}-3\nu_{1})}{\mu_{0}\nu_{0}^{2}-\nu_{1}\mu_{0}\nu_{0}+4\mu_{0}\nu_{0}-\nu_{1}\mu_{0}+\nu_{1}\nu_{0}+3\mu_{0}-3\nu_{1}}$, $\nu_{0}>3$, and $\frac{\mu_{0}(\nu_{0}+3)}{\mu_{0}-1}<\nu_{1}<\frac{\mu_{0}(\nu_{0}+3)}{\mu_{0}-3}$
            \end{enumerate}
        \end{enumerate}
	\end{enumerate}
\end{theorem}

Also by \Cref{hyp:b}, in terms of $\alpha$, $\beta$, and $\gamma$, $E^{\prime}_{3}$ in Eq.($\ref{eq:fixed point E3}$) is \(E^{\prime}_{3}=\bigg(\frac{\alpha\beta\mu_{0}-\beta\mu_{0}+\alpha+\beta}{\mu_{0}(\alpha\gamma-\beta)},\frac{-\beta\mu_{0}+\gamma\mu_{0}-\gamma-1}{\mu_{0}(\alpha\gamma-\beta)}\bigg)\). Jacobian eigenvalues of $E^{\prime}_{3}$ in Eq.($\ref{eq:Jacobian of version 5}$) is 
\begin{subequations}\label{eq:evals version 5 alpha beta gamma}
	\begin{empheq}[left=\empheqlbrace]{align}
		\omega_{0}&= \frac{1}{2\alpha\gamma-2\beta}\bigg(\Xi^{\prime}_{0}+\frac{\alpha\gamma-\beta}{|\alpha\gamma-\beta|}\Xi^{\prime}_{1}\bigg)\\ 
		\omega_{1}&= \frac{1}{2\alpha\gamma-2\beta}\bigg(\Xi^{\prime}_{0}-\frac{\alpha\gamma-\beta}{|\alpha\gamma-\beta|}\Xi^{\prime}_{1}\bigg), 
	\end{empheq}
\end{subequations}
where 
\begin{subequations}
	\begin{empheq}[left=\empheqlbrace]{align}\label{eq:Xi0}
		\Xi^{\prime}_{0}&=(2\gamma-1)\alpha-\beta^{2}\mu_{0}-\big(\alpha\mu_{0}+(-\mu_{0}+1)\gamma-\mu_{0}+4\big)\beta\\
		\Xi^{\prime}_{1}&=\bigg( \left( \mu_{{0}}-1 \right)  \left(  \left( -4\beta\mu_{0}-4 \right){\alpha}^{2}+4\beta\left(\mu_{0}-1 \right) \alpha+{\beta}^{2} \left( \mu_{0}-1\right)  \right) {\gamma}^{2}\\\nonumber
		&+\left( 4\left( \beta\mu_{0}+1 \right) ^{2}{\alpha}^{2}-2\beta\left( \mu_{0}-1\right) \left( \beta\mu_{0}+1 \right) \alpha-2{\beta}^{2}\mu_{0}\left( \mu_{0}-1 \right)  \left( \beta+1 \right) \right) \gamma\\\nonumber
		&+\big(  \left( \beta\mu_{0}+1 \right) \alpha-\beta\mu_{0} \left( \beta+1 \right)  \big)^{2}\bigg)^{\frac{1}{2}}\\\label{eq:Xi1}
	\end{empheq}
\end{subequations}
Eq.($\ref{eq:evals version 5 alpha beta gamma}$) shows that whenever $\Xi^{\prime\hspace{2pt}2}_{1}$ is negative, $\omega_{0}$ and $\omega_{1}$ are complex conjugates. Unlike the previous case, the fixed points $E^{\prime}_{1}$, $E^{\prime}_{2}$, and $E^{\prime}_{3}$ now depend on the growth rate $\mu_{0}$.
\subsection{Corresponding relationships between Eq.($\ref{eq:version 6}$) (Type 1) and Eq.($\ref{eq:2D logistic eq}$)  (Type 2)}
In this subsection we would like to demonstrate that, after linear transformation, Eq.($\ref{eq:version 6}$) (Type 1) is similar to Eq.($\ref{eq:2D logistic eq}$)  (Type 2). The only difference is that we need to find appropriate parameters for different sets of equations. First, rewrite Eq.($\ref{eq:2D mapping version 6}$) as
\begin{subequations}\label{eq:2D mapping version 6 prime}
	\begin{empheq}[left=\empheqlbrace]{align}
		\overline{f}({x},{y}) &= \mu_{0}' {x} - \mu_{0}' {x} ^ 2 - \mu_1' {x}{y} + x = \mu_0' {x}{(1-x)} - \mu_1' {x}{y} + x;\\	
		\overline{g}({x},{y}) &= -\nu_{0}' {y} + \nu_{0}' {y} ^ 2 - \nu_{1}' {x}{y} + y = -\nu_{0}' {y}{(1-y)} + \nu_1' {x}{y} + y.
	\end{empheq}
\end{subequations}
Also Eq.($\ref{eq:2D logistic map}$) is written as 
\begin{subequations}\label{eq:2D logistic map XY}
	\begin{empheq}[left=\empheqlbrace]{align}
		{f}({X},{Y}) &= \mu_0 {X} - \mu_0 {X} ^ 2 - \mu_1 {X}{Y} = \mu_0 {X}{(1-X)} - \mu_1 {X}{Y};\\	
		{g}({X},{Y}) &= -\nu_0 {Y} + \nu_0 {Y} ^ 2 + \nu_1 {X}{Y} = -\nu_0 {Y}{(1-Y)} + \nu_1 {X}{Y}.
	\end{empheq}
\end{subequations}
We consider the linear transformation as follows
\begin{subequations}\label{eq:transformation}
  \begin{empheq}[left=\empheqlbrace]{align}
  x = \alpha_XX+\alpha_YY+\alpha_0;\\
  y = \beta_XX+\beta_YY+\beta_0,
  \end{empheq}
\end{subequations}
where $\alpha_X$, $\alpha_Y$, $\alpha_0$, $\beta_X$, $\beta_Y$, and $\beta_0$ are about to be determined with comparisons.
Substituting into Eq.($\ref{eq:2D mapping version 6 prime}$), we obtain
\begin{subequations}
\begin{empheq}{align}
    \overline{f}({X},{Y})\nonumber
    &= \mu_{0}'(\alpha_XX+\alpha_YY+\alpha_0) - \mu_{0}'(\alpha_XX+\alpha_YY+\alpha_0)^2 \\\nonumber
    &\phantom{=} - \mu_{1}'(\alpha_XX+\alpha_YY+\alpha_0)(\beta_XX+\beta_YY+\beta_0) + (\alpha_XX+\alpha_YY+\alpha_0)\\\nonumber
    &= \mu_{0}'\alpha_XX+\mu_{0}'\alpha_YY + \mu_{0}'\alpha_0\nonumber \\
    &\phantom{=} - \mu_{0}'({\alpha_{X}}^2X^2 + {\alpha_{Y}}^2Y^2 + {\alpha_0}^2 + 2\alpha_{X}\alpha_{Y}XY+ 2\alpha_{0}\alpha_{X}X + 2\alpha_{0}\alpha_{Y}Y)\nonumber \\
    &\phantom{=} - \mu_{1}'\big[\alpha_X\beta_XX^2+\alpha_Y\beta_YY^2 + \alpha_0\beta_0 \nonumber\\
    &\phantom{====} + (\alpha_X\beta_Y + \alpha_Y\beta_X)XY + (\alpha_X\beta_0 + \alpha_0\beta_X)X + (\alpha_Y\beta_0 + \beta_Y\alpha_0)Y\big]\nonumber \\
    &\phantom{=} + (\alpha_XX +\beta_YY +\alpha_0)\nonumber \\
    &= \big[\mu_0'\alpha_X - \mu_0' \cdot 2\alpha_0\alpha_X - \mu_1' \cdot (\alpha_X\beta_0+ \alpha_0\beta_X)+\alpha_X\big]X\label{eq:eqn1}\\
    &\phantom{=} - (\mu_0'{\alpha_X}^2+\mu_1'\alpha_X\beta_X)X^2\label{eq:eqn2}\\
    &\phantom{=} +\big(-\mu_{0}'\cdot 2\alpha_X\alpha_Y-\mu_{1}'(\alpha_X\beta_Y+\alpha_Y\beta_X)\big)XY\label{eq:eqn9}\\
    &\phantom{=} -\big(-\mu_{0}'\alpha_Y+\mu_{0}'\cdot 2\alpha_0\alpha_Y+\mu_{1}'(\alpha_Y\beta_0+\alpha_0\beta_Y)-\beta_Y\big)Y\label{eq:eqn3}\\
    &\phantom{=} +(-\mu_{0}'{\alpha_{Y}}^2+\mu_{1}'\alpha_Y\beta_Y)Y^2\label{eq:eqn4}\\
    &\phantom{=} +(\mu_{0}'\alpha_{0}-\mu_{1}'\alpha_0\beta_0-\mu_{0}'{\alpha_0}^2+\alpha_0)\label{eq:eqn11}
\end{empheq}
\end{subequations}
Similarly, 
\begin{subequations}
\begin{empheq}{align}
    \overline{g}({X},{Y})\nonumber  
    &= -\nu_{0}'(\beta_XX+\beta_YY+\beta_0)+\nu_{0}'(\beta_XX+\beta_YY+\beta_0)^2 \\\nonumber 
    &\phantom{=} + \nu_{1}'(\alpha_XX+\alpha_YY+\alpha_0)(\beta_XX+\beta_YY+\beta_0)+(\beta_XX+\beta_YY+\beta_0)\\\nonumber 
    &=-\nu_{0}'\beta_XX-\nu_{0}'\beta_YY-\nu_{0}'\beta_0\\\nonumber 
    &\phantom{=} +\nu_{0}'(\beta_X^2{X^2}+\beta_Y^2{Y^2}+\beta_{0}^2+2\beta_X\beta_YXY+2\beta_0\beta_XX+2\beta_0\beta_YY)\\\nonumber 
    &\phantom{=}+\nu_{1}'\big[\alpha_X\beta_XX^2+\alpha_Y\beta_YY^2+\alpha_0\beta_0\\\nonumber 
    &\phantom{====}+(\alpha_X\beta_Y+\alpha_Y\beta_X)XY+(\alpha_X\beta_0+\alpha_0\beta_X)X+(\alpha_Y\beta_0+\alpha_0\beta_Y)Y\big]\\\nonumber 
    &\phantom{=} +(\beta_XX +\beta_YY+\beta_0)\\
    &=\big[-\nu_0'\beta_X + \nu_0' \cdot 2\beta_0\beta_X + \nu_1'(\alpha_X \beta_0 + \alpha_0 \beta_X) + \beta_X\big]X\label{eq:eqn5}\\
    &\phantom{=} - (-\nu_0' \beta_X^2-\nu_1'\alpha_X\beta_X)X^2\label{eq:eqn6}\\
    &\phantom{=} - \big(-\nu_0' \cdot 2\beta_X\beta_Y - \nu_1'(\alpha_X\beta_Y+\alpha_Y\beta_X)\big) XY\label{eq:eqn10}\\
    &\phantom{=} - \big(+\nu_0'\beta_Y - \nu_0' \cdot 2\beta_0\beta_Y - \nu_1'(\alpha_Y\beta_0+\alpha_0\beta_Y) - \beta_Y\big) Y\label{eq:eqn7}\\
    &\phantom{=} + (+\nu_0'\beta_Y^2 + \nu_1'\alpha_Y\beta_Y) Y^2\label{eq:eqn8}\\
    &\phantom{=} + (-\nu_0'\beta_0 + \nu_0'\beta_0^2 + \nu_1'\alpha_0\beta_0 + \beta_0)\label{eq:eqn12}
\end{empheq}
\end{subequations}
Transformation from Eq.($\ref{eq:2D mapping version 6 prime}$) to Eq.($\ref{eq:2D logistic map XY}$) may be accomplished if we choose coefficients in Eq.($\ref{eq:eqn1}$), Eq.($\ref{eq:eqn2}$), Eq.($\ref{eq:eqn11}$), Eq.($\ref{eq:eqn7}$), Eq.($\ref{eq:eqn8}$), and Eq.($\ref{eq:eqn12}$) to be all zero, coefficient in Eq.($\ref{eq:eqn9}$) equal to $\nu_{1}$, coefficients in Eq.($\ref{eq:eqn3}$) and Eq.($\ref{eq:eqn4}$) both equal to $\nu_{0}$, coefficients in Eq.($\ref{eq:eqn5}$) and Eq.($\ref{eq:eqn6}$) both equal to $\mu_{0}$, and coefficient in  Eq.($\ref{eq:eqn10}$) equal to $\mu_{1}$. Subsequently, 
\begin{subequations}
    \begin{empheq}[left=\empheqlbrace]{align}
         \overline{f}({X},{Y})&=-\nu_{0}Y+\nu_{0}Y^{2}+\nu_{1}XY={g}({X},{Y}),\\
         \overline{g}({X},{Y})&=\mu_{0}X-\mu_{0}X^{2}-\mu_{1}XY={f}({X},{Y}).
    \end{empheq}
\end{subequations}
Non-trivial solutions for coefficients in Eq.($\ref{eq:transformation}$) are, 
\begin{subequations}
    \begin{empheq}[left=\empheqlbrace]{align}
        \alpha_{X} &= \frac{(-\nu_{0}\mu_{0}+\nu_{1}\mu_{0}+\mu_{0}+\nu_{1})\mu_{1}}{(-\nu_{0}\mu_{0}+\nu_{1}\mu_{1})\mu_{0}}\\
        \alpha_{Y} &= \frac{\nu_{0}\mu_{0}-\mu_{1}\nu_{0}+\mu_{1}+2\nu_{0}+\nu_{1}}{-\nu_{0}\mu_{0}+\nu_{1}\mu_{1}}\\
        \beta_{X} &= \frac{\nu_{0}\mu_{0}-\nu_{1}\mu_{0}-\mu_{0}-\nu_{1}}{-\nu_{0}\mu_{0}+\nu_{1}\mu_{1}}\\
        \beta_{Y} &= \frac{-\nu_{1}(\nu_{0}\mu_{0}-\mu_{1}\nu_{0}+\mu_{1}+2\nu_{0}+\nu_{1})}{(-\nu_{0}\mu_{0}+\nu_{1}\mu_{1})\nu_{0}}\\
        \alpha_{0} &= \frac{-\nu_{0}\mu_{0}+\mu_{1}\nu_{0}-\mu_{1}-\nu_{0}}{-\nu_{0}\mu_{0}+\nu_{1}\mu_{1}}\\
        \beta_{0} &= \frac{-\nu_{0}\mu_{0}+\nu_{1}\mu_{0}+\mu_{0}+\nu_{1}}{-\nu_{0}\mu_{0}+\nu_{1}\mu_{1}},
    \end{empheq}
\end{subequations}
which give us the relationships of parameters between Eq.($\ref{eq:2D mapping version 6 prime}$) and Eq.($\ref{eq:2D logistic map XY}$) as follows
\begin{subequations}
    \begin{empheq}[left=\empheqlbrace]{align}
    -\frac{\big(\mu_{0}'(\nu_{0}'-\nu_{1}'-1)-\nu_{1}'\big)^2}{\mu_{0}'(\mu_0'\nu_0'-\mu_1'\nu_1')} &= \mu_0\\
    -\frac{\big(\nu_{0}'(\mu_{0}'-\mu_{1}'+2)+\mu_{1}'+\nu_1'\big)^2}{\nu_{0}'(\mu_0'\nu_0'-\mu_1'\nu_1')} &= \nu_0 = \beta\mu_0\\
    \frac{\nu_1'\big((\nu_{0}'-\nu_{1}'-1)\mu_0'-\nu_1'\big)\big(\mu_0'\nu_0'+\nu_0'(2-\mu_1')+\mu_1'+\nu_1'\big)}{\mu_0'\nu_0'(\mu_0'\nu_0'-\mu_1'\nu_1')} &= \mu_1 = \alpha\mu_0\\
    \frac{\mu_1'\big((\nu_{0}'-\nu_{1}'-1)\mu_0'-\nu_1'\big)\big(\mu_0'\nu_0'+\nu_0'(2-\mu_1')+\mu_1'+\nu_1'\big)}{\mu_0'\nu_0'(\mu_0'\nu_0'-\mu_1'\nu_1')} &= \nu_1 = \gamma\mu_0
    \end{empheq}
\end{subequations}
Interestingly, we note that $\frac{\mu_{1}'\nu_{1}'}{\mu_{0}'\nu_{0}'} = \frac{\mu_{1}\nu_{1}}{\mu_{0}\nu_{0}}$ and $\mu_{1}\mu_{1}'=\nu_{1}\nu_{1}'$. Unfortunately, general solutions of $\mu_{0}',$ $\mu_{1}',$  $\nu_{0}',$ and $\nu_{1}'$ in terms of $\mu_{0},$ $\mu_{1},$  $\nu_{0},$ and $\nu_{1}$ are difficult to find. However, if we notice that since 
\begin{subequations}
    \begin{empheq}[left=\empheqlbrace]{align}
    \nu_{1}'^{2} = \frac{\mu_{1}^2}{\mu_0\nu_0}\mu_{0}'\nu_0';\\
    \mu_{1}'^{2} = \frac{\nu_{1}^2}{\mu_0\nu_0}\mu_{0}'\nu_0',
    \end{empheq}
\end{subequations}
we then further acquire parameters of the system in Eq.($\ref{eq:2D mapping version 6 prime}$), in the counterpart of that in Eq.($\ref{eq:2D logistic map XY}$), 
\begin{subequations}
    \begin{empheq}[left=\empheqlbrace]{align}
    \alpha'&\equiv\frac{\mu_{1}'}{\mu_{0}'}=\frac{\nu_{1}}{\sqrt{\mu_{0}\nu_{0}}}\sqrt{\frac{\nu_{0}'}{\mu_{0}'}}=\frac{\gamma\mu_{0}\sqrt{\beta'}}{\sqrt{\mu_{0}\beta\mu_{0}}}=\frac{\gamma}{\sqrt{\beta}}\sqrt{\beta'};\\
    \beta'&\equiv\frac{\nu_{0}'}{\mu_{0}'}\\
    \gamma'&\equiv\frac{\nu_{1}'}{\mu_{0}'}=\frac{\mu_{1}}{\sqrt{\mu_{0}\nu_{0}}}\sqrt{\frac{\nu_{0}'}{\mu_{0}'}}=\frac{\alpha\mu_{0}\sqrt{\beta'}}{\sqrt{\mu_{0}\beta\mu_{0}}}=\frac{\alpha}{\sqrt{\beta}}\sqrt{\beta'},
    \end{empheq}
\end{subequations}
Therefore, once $\alpha$, $\beta$, and $\gamma$ in the system described in Eq.($\ref{eq:2D logistic map XY}$) (Eq.($\ref{eq:2D logistic map}$)) in Sect. $\ref{subs:version 5}$ are determined, we should be able to reconstruct a similar bifurcation diagram in the system in Eq.($\ref{eq:2D mapping version 6 prime}$) (Eq.($\ref{eq:2D mapping version 6}$)) in Sect. $\ref{subs:version 6}$  with the parameters $\alpha'$, $\beta'$, and $\gamma'$. Therefore, in this study, we show only bifurcation diagrams for Eq.($\ref{eq:2D logistic eq}$) (Type 2). 

\subsection{Lyapunov exponents}
In addition, the chaotic behavior may be better examined by introducing Lyapunov exponents, which are defined as, base 2 being chosen to conform to Wolf et al.$\cite{Wolf et al}$,
\begin{subequations}\label{eq:Lyapunov eq}
	\begin{empheq}[left=\empheqlbrace]{align}
		\lambda_{x}\equiv\log_{2}|w_{0}|&=\frac{\ln|w_{0}|}{\ln2}\label{eq:Lyapunov eq 1}\\
		\lambda_{y}\equiv\log_{2}|w_{1}|&=\frac{\ln|w_{1}|}{\ln2}\label{eq:Lyapunov eq 2}.
	\end{empheq}
\end{subequations}
The positive value of $\lambda_{x}$ or $\lambda_{y}$, together with the negative value of the total sum of the Lyapunov exponent, either $\sum\lambda_{x}<0$ or $\sum\lambda_{y}<0$, are strong inferences of chaos for the prey or predator$\cite{Escot and Galan}$. For comparison, Lyapunov exponents of time series data of prey and predator populations were also calculated by both the algorithms of Rosenstein$\cite{Rosenstein}$ and Eckmann et al.$\cite{Eckmann}$ with the package of NOnLinear measures for Dynamical Systems (nolds)$\cite{nolds}$. For the Rosenstein algorithm, the embedding dimension for delay embedding was $\mathrm{emb\_dim}=10$, and the step size between time series data points was set at $\tau=1$ seconds. While the number of data points ($\mathrm{trajectory\_len}$) was set to $20$ and used for the distance trajectories between two neighboring points, the mean period of the time series data, obtained from the fast Fourier transform, was used as the minimal temporal separation($\mathrm{min\_tsep}$) between two neighbors. The search for the appropriate delay was terminated when several potential neighbors of a vector were found to be smaller than the minimal neighbors, which were set as $\mathrm{min\_neighbors}=20$. Finally, the RANSAC fitting was used for the line fitting.
For the algorithm proposed by Eckmann et al., the matrix dimension was set to $2$, and the embedding dimension was set to $10$ as in the Rosenstein algorithm. Furthermore, $\tau=1$ s, the minimum number of neighbors ($\mathrm{min\_nb}$) was $4$, and $\mathrm{min\_tsep}=0$ were used in the algorithm.

There are at least four disadvantages for the above algorithms, as mentioned by Escot and Galan$\cite{Escot and Galan}$: lack of the ability to estimate the full Lyapunov spectrum, not resilient to noise in time-series data, poor detection performance in nonlinearity with an adequate sample size, and no theoretical derivations for the algorithms about their consistency and asymptotic distributions, making it impossible to statistically infer with respect to chaos.
\section{Results and Discussions}
 In the present study, we focus only on drawings of the equations in Sect. $\ref{subs:version 5}$. We observed that there could be four different bifurcation diagrams with various kinds of combinations of parameters.  The first category is \textbf{Normal}, referring to normal competitive behavior in the increasing and decreasing number of species between the prey and the predator. The second category is \textbf{Standard}, referring to the standard bifurcation diagram as shown in the well-known $1$D logistic equation in the prey in the absence of the predator. The third category is \textbf{Extinction}, where a small value $\beta$  plays an important role in contributing to this phenomenon, which is different from previous study$\cite{Danca et al}$, connoting to extinction of the predator when the prey becomes chaotic. The fourth category is named \textbf{Vorticella} because its topological shape resembles the vorticella. The categories and parameters are summarized in Table $\ref{tab:parameters}$. initX and initY indicate the initial values of the prey and the predator, respectively. Special attention should be paid to the cases of Normal and Extinction, where the inters-pecies parameters are deliberately made the same, but the initial population was different: for Normal, the two species have the same initial population, whereas for Extinction, the predator has $10$ times more population than the prey. The discrepancy in the initial population in these two cases shows completely different evolution consequences. Furthermore, to minimize variations between parameters, initX, initY, $\alpha$,  and $\beta$ are chosen to be the same for \textbf{Standard}, \textbf{Vorticella}, and \textbf{Extinction} to demonstrate the topological dependence on $\gamma$. Detailed discussions of the topology dependence on $\gamma$ are given in Sect. $\ref{subs:topology dependence on gamma}$. 
 
 Regarding the Lyapunov exponents, Equation, Rosenstein, Eckmann X, and Eckmann Y in the legend of $\lambda_{x}$ and $\lambda_{y}$ refer to calculations directly from Eq.($\ref{eq:Lyapunov eq}$), from the time-series algorithm of Rosenstein, Eckmann et al. of the prey, and Eckmann et al. of the predator, respectively. Comparisons of sum of Lyapunov exponents for every algorithm are summarized in Table $\ref{tab:lyapunov exponents for every algorithm}$. The codes, together with animations on population iterations, phase portraits, and phase diagrams under different growth rates, can be retrieved via Ref($\cite{codes and animations}$). 
 
\begin{table}[!htbp]
	\centering
	\begin{tabular}{lllccc}
		\hline 
		                  &   initX   &  initY   &  $\alpha$  &  $\beta$   & $\gamma$   \\ \hline
		Normal              &   0.200   &  0.200   &  1.000     &   0.001    &  0.500     \\
		Standard\textsuperscript{1}           &   0.010   &  0.100   &  1.000     &   0.001    &  0.197     \\
        Extinction          &   0.010   &  0.100   &  1.000     &   0.001    &  0.500     \\
		Vorticella          &   0.010   &  0.100   &  1.000     &   0.001    &  0.912     \\\hline
	\end{tabular}
	\caption{Parameters used for simulations used in Sect. $\ref{subs:version 5}$.}
	\label{tab:parameters}
	\small\textsuperscript{1}{Parameters of 0.100, 0.500, 1.000, 0.100, 0.500 also produce Standard, which made a perfect 0, in the sense of computer error, for $y$. These parameters were used to make both Rosenstein and Eckmann et al. algorithms work. The small value $\beta$ also makes contributions to the first term in Eq.($\ref{eq:version 5 for predator}$), and it can be confirmed by examining its order of magnitudes. }
\end{table}

\begin{table}[!htbp]
	\centering
	\begin{tabular}{lllcccc}
		\hline 
	$\sum\lambda$	&  Eq.($\ref{eq:Lyapunov eq 1}$) $\lambda{x}$ & Eq.($\ref{eq:Lyapunov eq 2}$) $\lambda{y}$ & Rosenstein $\lambda{x}$  & Rosenstein $\lambda{y}$ & Eckmann et al. $\lambda{x}$   & Eckmann et al. $\lambda{y}$\\ \hline
		Normal          &  -357.012    &  -365.094   &  -76.198      &   -162.843    &  -148.710     &   -262.202 \\
		Standard        &   -321.133   &  -698.992   &  -115.988     &  -443.234     &  Not reliable & Not reliable     \\
        Extinction      &   -326.102   &  -336.777   &  -103.789     &   -204.999    &  -135.607     &  -269.497   \\
		Vorticella      &   -332.520   &  -201.130   &  -58.598      &   -125.890    &  -59.522      &  -142.697   \\\hline
	\end{tabular}
	\caption{Comparison of sum of Lyapunov for every algorithm with same parameters in Table $\ref{tab:parameters}$.}
	\label{tab:lyapunov exponents for every algorithm}
\end{table}

\subsection{$\gamma$ dependence of the topological transformation in the bifurcation diagram}\label{subs:topology dependence on gamma}
\begin{figure}[!htbp]
		\begin{subfigure}[b]{0.5\textwidth}
			\centering
			\includegraphics[width=1.0\linewidth]{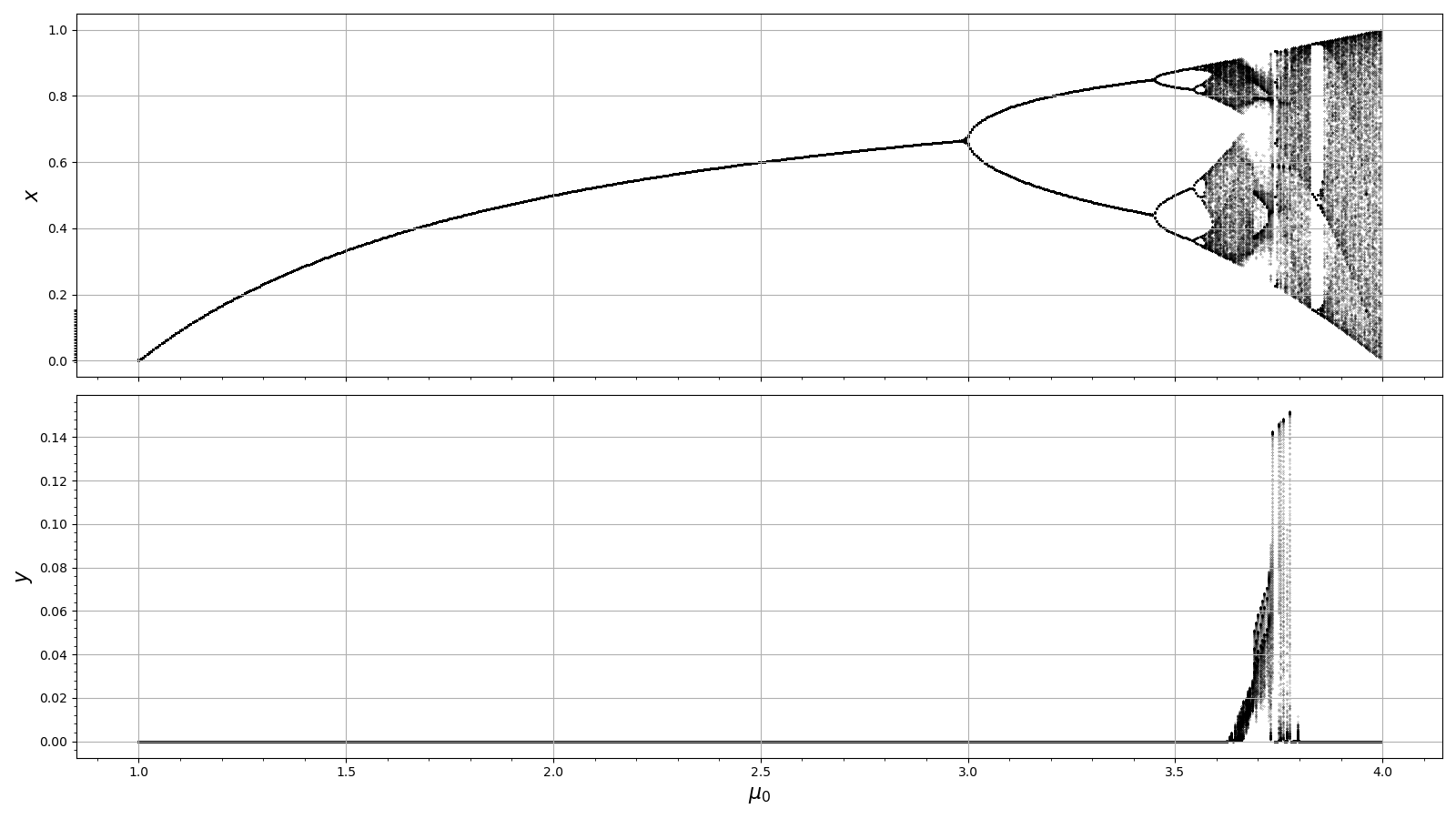}
                \includegraphics[width=1.0\linewidth]{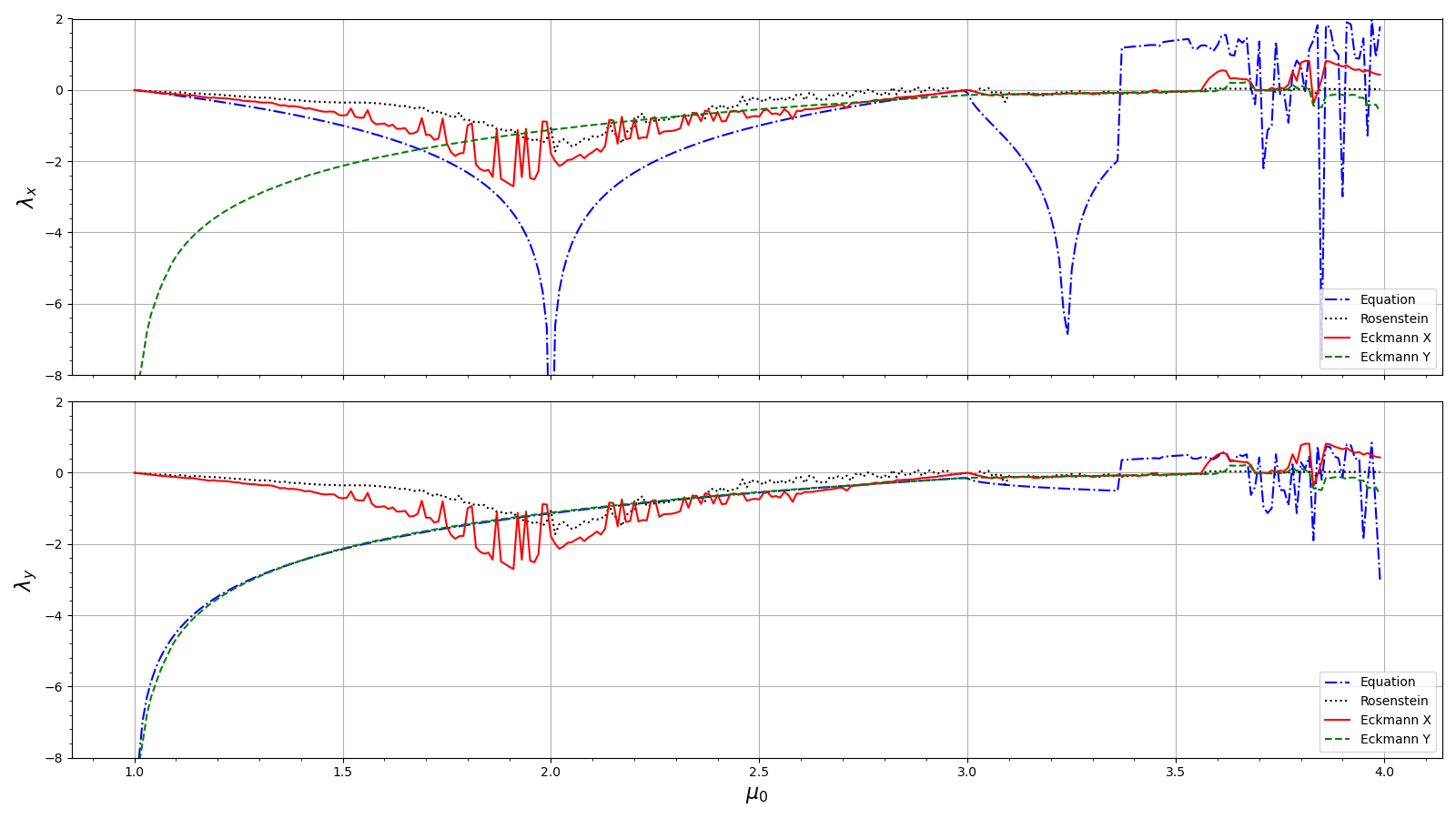}
			\caption{}
			\label{fig:gammaDep0.455}
		\end{subfigure}
		\begin{subfigure}{0.5\textwidth}
			\centering
			\includegraphics[width=1.0\linewidth]{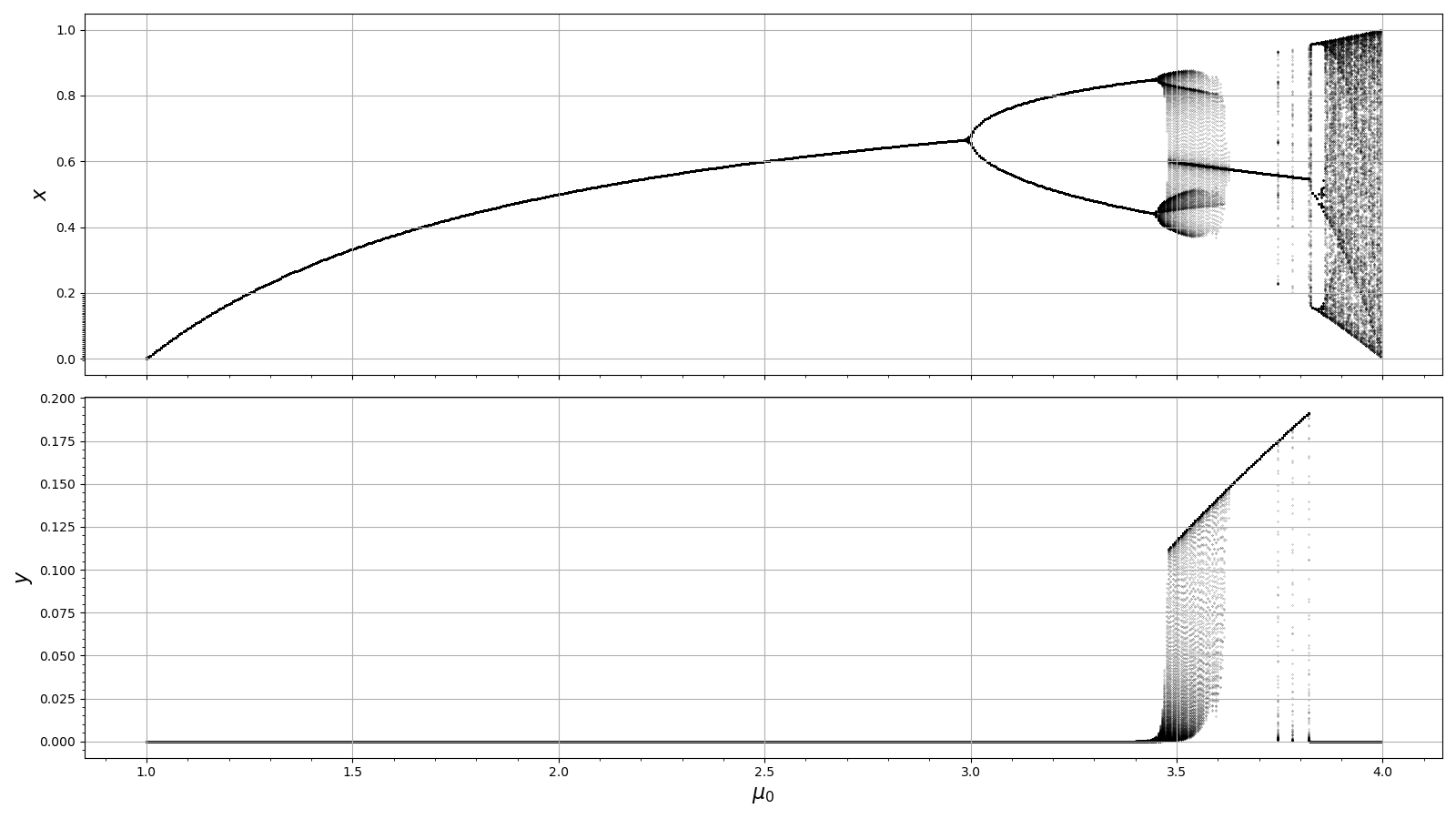}
                \includegraphics[width=1.0\linewidth]{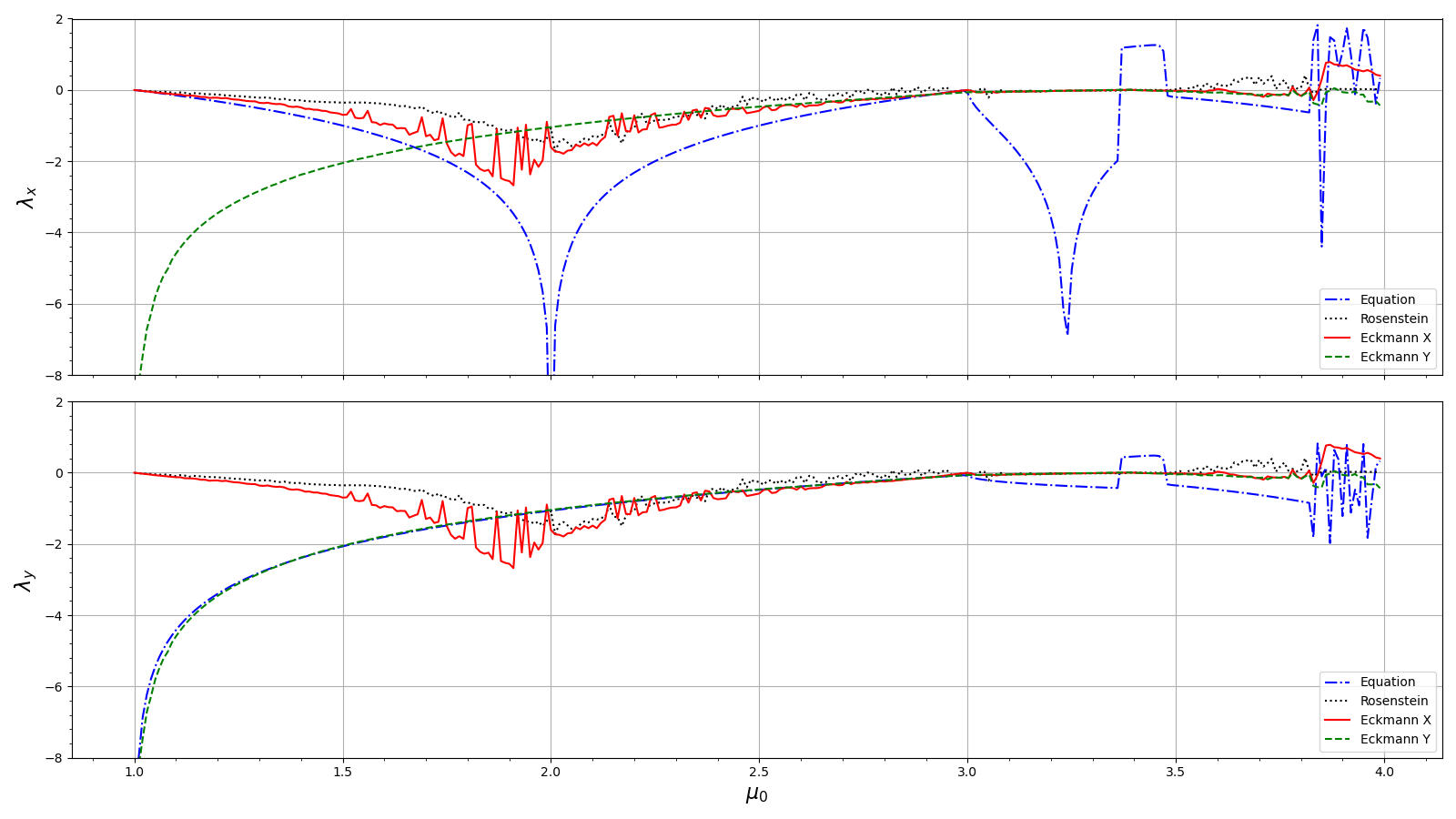}
			\caption{}
			\label{fig:gammaDep0.480}
		\end{subfigure}
		\begin{subfigure}{0.5\textwidth}
			\centering
			\includegraphics[width=1.0\linewidth]{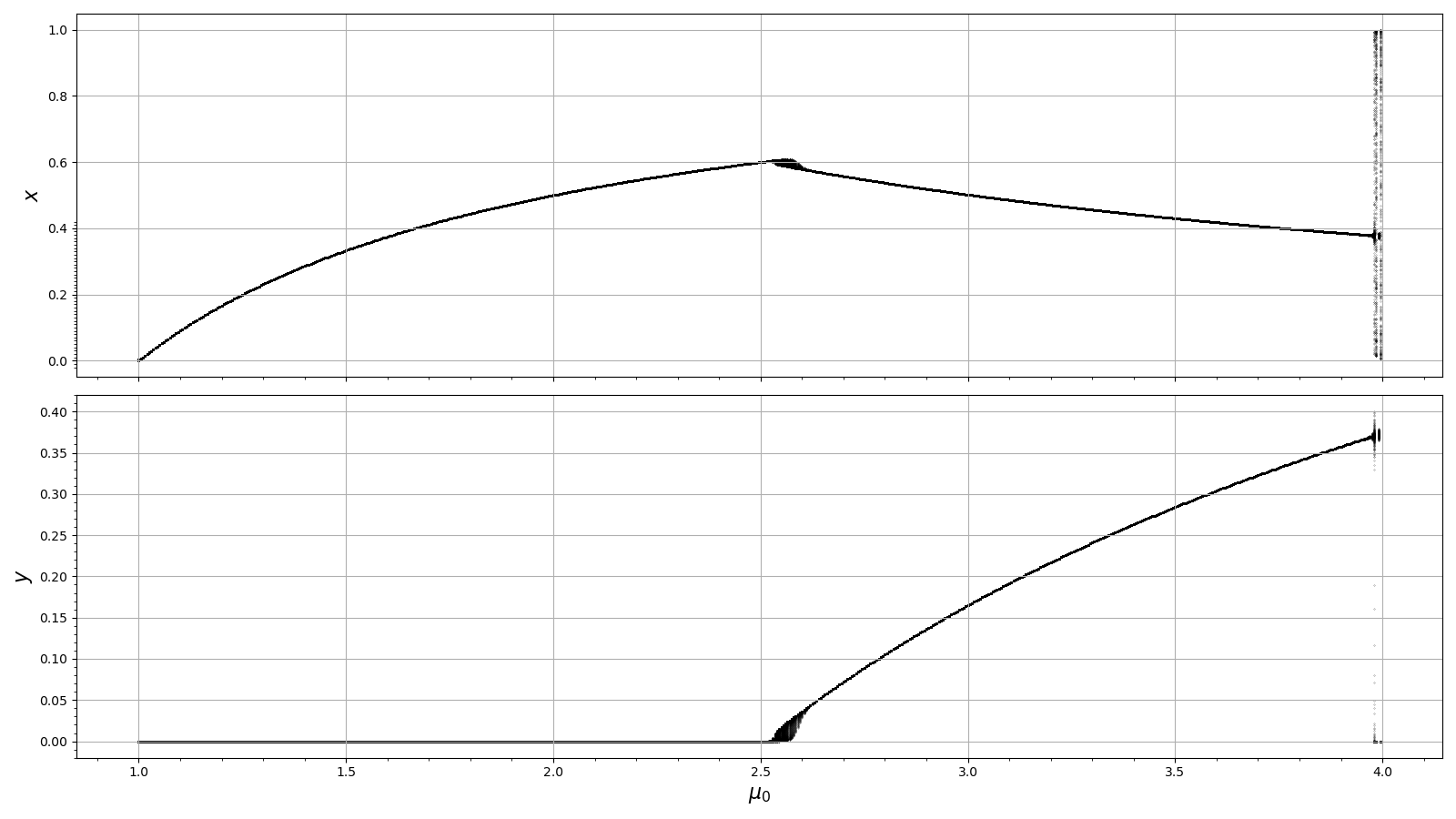}
                \includegraphics[width=1.0\linewidth]{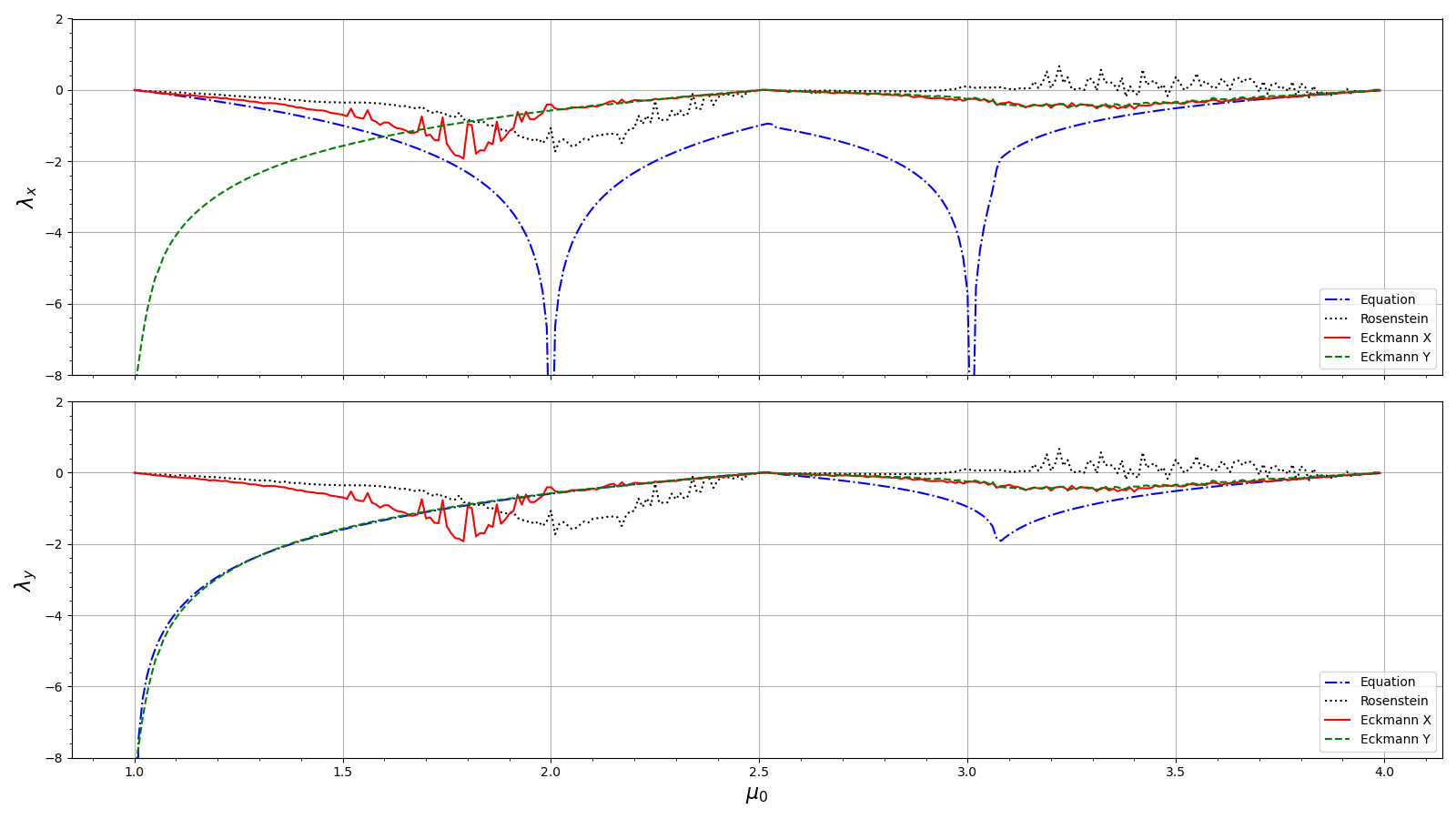}
			\caption{}
			\label{fig:gammaDep0.666}
		\end{subfigure}
  		\begin{subfigure}[b]{0.5\textwidth}
			\centering
			\includegraphics[width=1.0\linewidth]{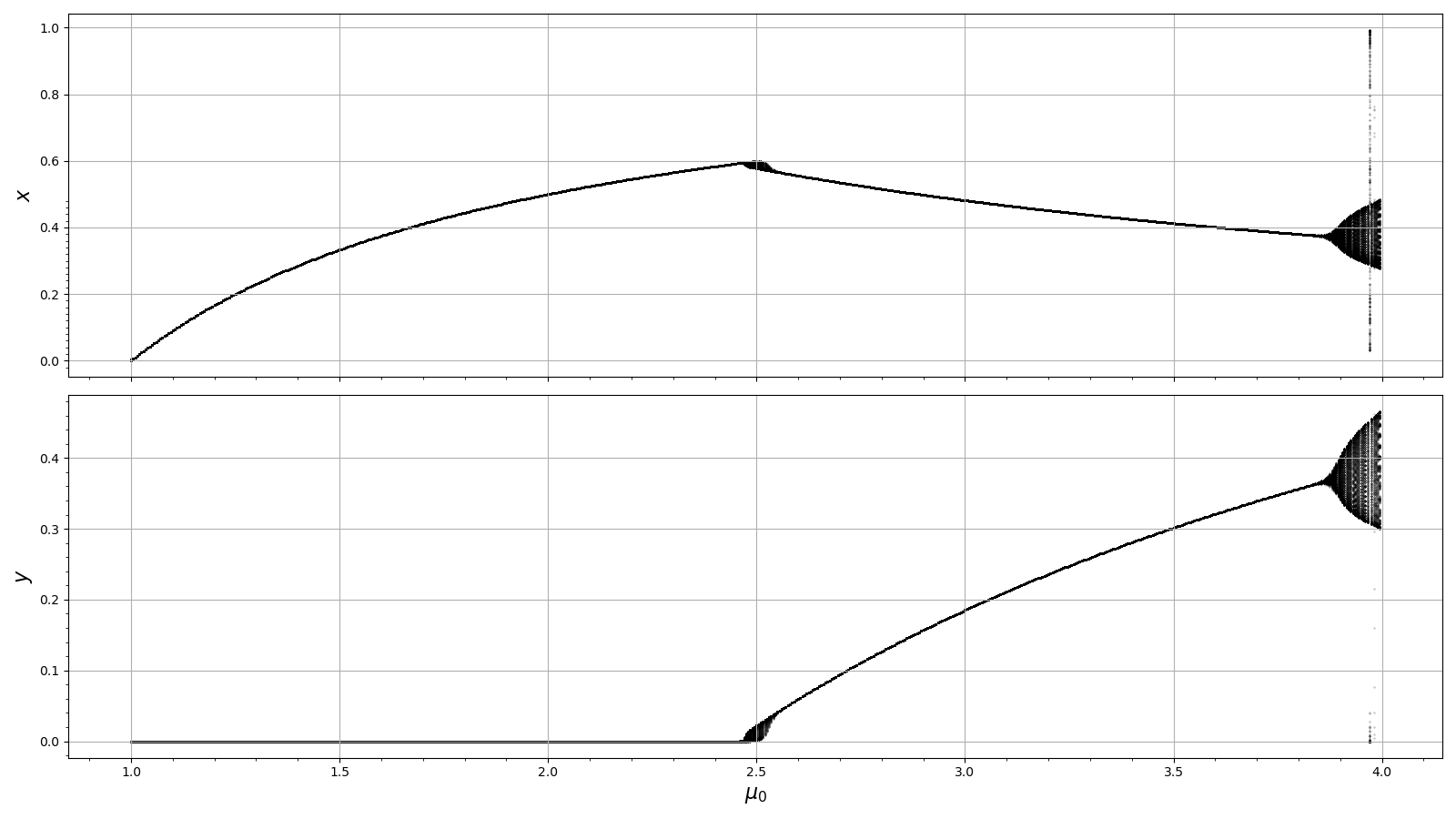}
                \includegraphics[width=1.0\linewidth]{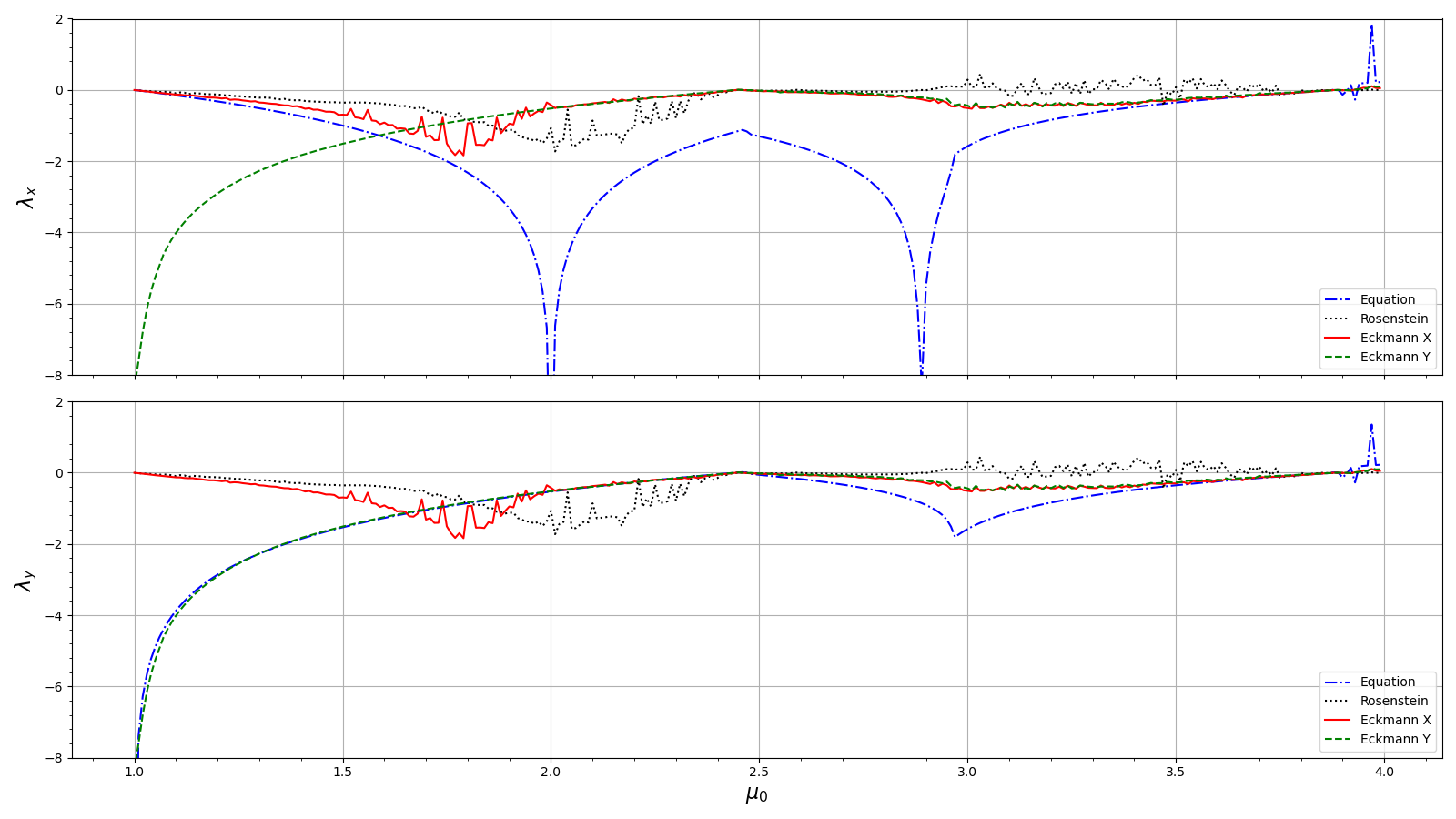}
			\caption{}
			\label{fig:gammaDep0.694}
		\end{subfigure}
		\caption{Topological Dependence on $\gamma$ of bifurcation diagrams, together with corresponding Lyapunov exponents calculated from Eq.($\ref{eq:Lyapunov eq}$), are compared. $\gamma$ for  ($\ref{fig:gammaDep0.455}$), ($\ref{fig:gammaDep0.480}$), ($\ref{fig:gammaDep0.666}$) and ($\ref{fig:gammaDep0.694}$) are $0.455$, $0.480$, $0.666$, and $0.694$, respectively.}
		\label{fig:topological Dependence}
\end{figure}

Before discussing about each bifurcation diagram, first we show the topological variations of bifurcation diagram depending on $\gamma$ when all the other parameters are kept the same for \textbf{Standard}, \textbf{Extinction}, and \textbf{Vorticella} given in Table $\ref{tab:parameters}$. In Figure $\ref{fig:topological Dependence}$, simulations show that as $\gamma<0.431$, the bifurcation diagram is classified as \textbf{Standard} with negligible $y$. The detained discussions on \textbf{Standard} are given in Sect. $\ref{subs:standard}$. However, significant $y$ begins to appear around $\mu_{0}=3.7$ after $\gamma=0.445$, \textit{breaking} the structure of \textbf{Standard} for $x$, as shown in Figure $\ref{fig:gammaDep0.455}$, while gradually transforming the topology into \textbf{Extinction}. When $0.480<\gamma<0.666$, the bifurcation diagram falls into the topology of \textbf{Extinction}, as in Figure $\ref{fig:gammaDep0.480}$. More detailed discussions on \textbf{Extinction} are given in Sect. $\ref{subs:extinction}$. Meanwhile, the predictable values $x$ and $y$ as in \textbf{Normal} appear within $3.5<\mu_{0}<3.7$ (see Figure $\ref{fig:gammaDep0.480}$). Subsequently, \textbf{Vorticella} grows out of the disappearance of \textbf{Extinction}, see Figure $\ref{fig:gammaDep0.666}$. After $\gamma>0.694$, the bifurcation diagram becomes \textbf{Vorticella}, as shown in Figure $\ref{fig:gammaDep0.694}$. The detained discussions on \textbf{Vorticella} are also given in Sect. $\ref{subs:vorticella}$. In the range $0.431<\gamma<0.480$, both \textbf{Standard} and \textbf{Extinction} coexist (see Figure $\ref{fig:gammaDep0.455}$). On the other hand, in the range $0.666<\gamma<0.694$, the bifurcation diagram is considered as \textbf{Normal}, regardless of some small black regions around $\mu_{0}=2.5$ and $\mu_{0}=4.0$, the latter exhibits chaos that may be confirmed by the Lyapunov exponents shown below. Within the range, \textbf{Extinction} and \textbf{Vorticella} appear simultaneously (see Figure $\ref{fig:gammaDep0.666}$). Another pure \textbf{Normal} bifurcation without any chaos region is given in Sect. $\ref{subs:normal}$. 

\begin{figure}[!htbp]
		\begin{subfigure}[b]{0.5\textwidth}
			\centering
			\includegraphics[width=1.0\linewidth]{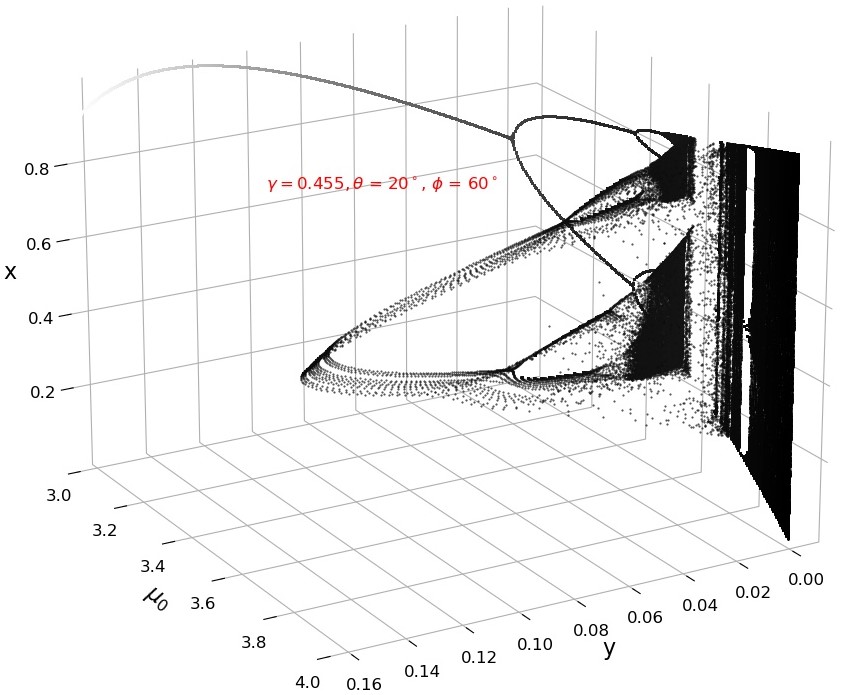}
			\caption{}
			\label{fig:3DgammaDep0.455}
		\end{subfigure}
		\begin{subfigure}{0.5\textwidth}
			\centering
			\includegraphics[width=1.0\linewidth]{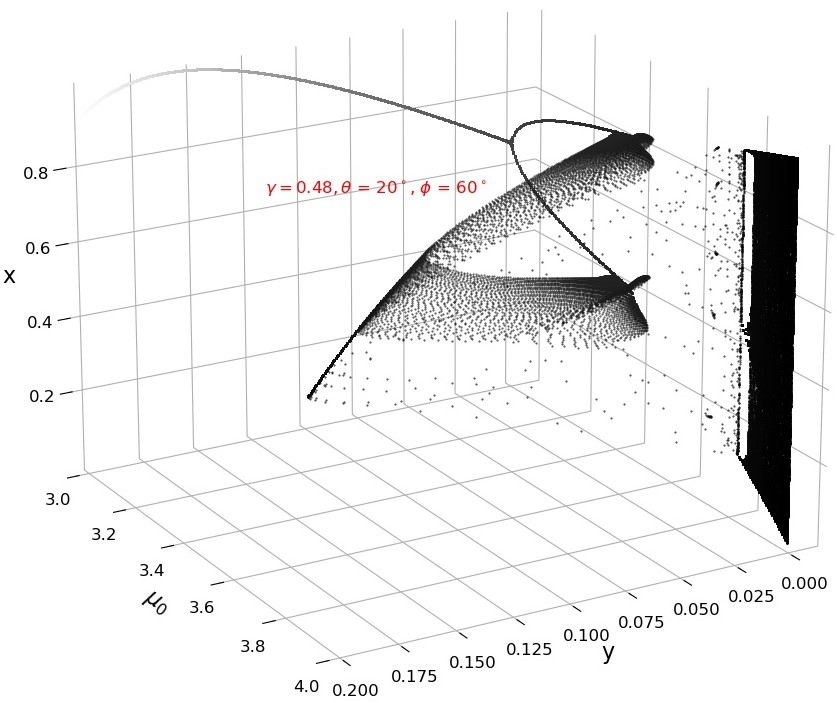}
			\caption{}
			\label{fig:3DgammaDep0.480}
		\end{subfigure}
		\begin{subfigure}{0.5\textwidth}
			\centering
			\includegraphics[width=1.0\linewidth]{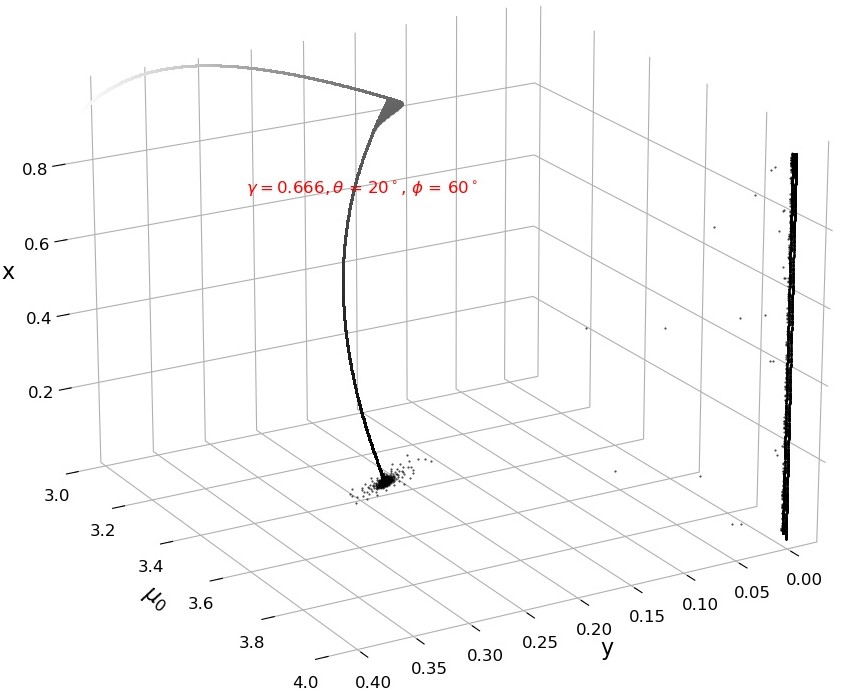}
			\caption{}
			\label{fig:3DgammaDep0.666}
		\end{subfigure}
  		\begin{subfigure}[b]{0.5\textwidth}
			\centering
			\includegraphics[width=1.0\linewidth]{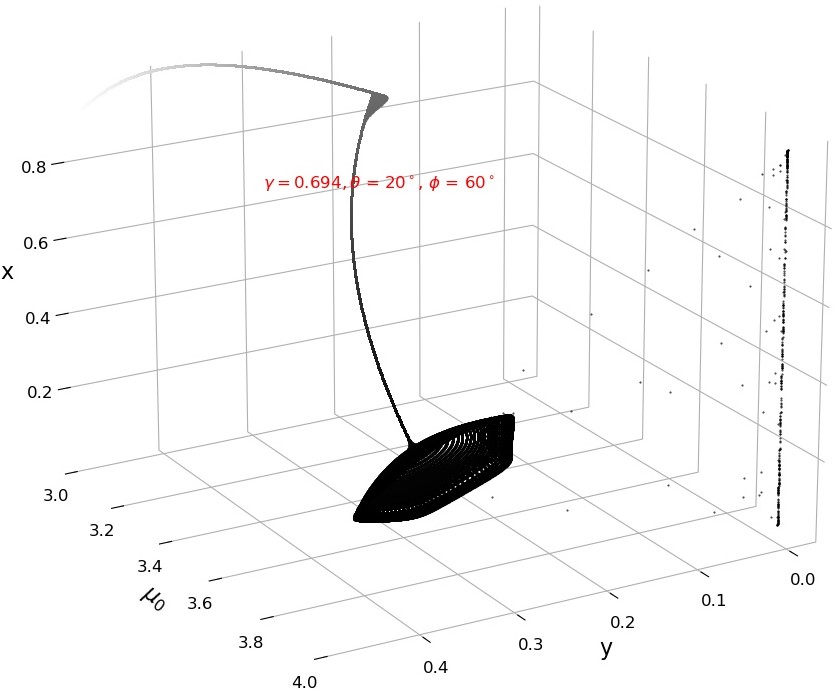}
			\caption{}
			\label{fig:3DgammaDep0.694}
		\end{subfigure}
		\caption{$3$D phase portrait for various $\gamma$ values viewed at an elevation angle $20^{\circ}$ and an azimuthal angle $60^{\circ}$}
		\label{fig:topological Dependence 3d}
\end{figure}

It is much more interesting to plot bifurcation diagrams in three dimensions, as shown in Figure $\ref{fig:topological Dependence 3d}$ with an elevation angle $20^{\circ}$ and an azimuthal angle $60^{\circ}$. Figure $\ref{fig:gammaDep0.455}$ shows the case with $\gamma=0.455$, where stable orbitals originally lying in $x-\mu_{0}$ within the range $3.6<\mu_{0}<3.8$ now become blurred and instead grow in the $x-y$ plane, forming a structure similar to \textbf{Standard} within the range $3.0<\mu_{0}<3.6$. Figure $\ref{fig:3DgammaDep0.480}$ with $\gamma=0.480$ shows a butterfly structure of \textbf{Extinction} as in Figure $\ref{fig:Extinction_phasePortrait3D_040_020}$. Figure $\ref{fig:3DgammaDep0.666}$ shows the $3$D bifurcation diagram of \textbf{Normal} with $\gamma=0.666$. In addition to normal competing behavior in the middle, there is a blacked-in area at the top and a spiral-in attractor at the bottom of the $y-\mu_{0}$ plane. Meanwhile, a stripe appears around $\mu_{0}\approx4.0$, showing the last occurrence of \textbf{Extinction}. Figure $\ref{fig:3DgammaDep0.694}$ shows the three-dimensional bifurcation diagram at the beginning of \textbf{ Viracella} with $\gamma=0.694$, with the chaotic blacked region shown at the bottom. 

\newpage
\subsection{Normal}\label{subs:normal}
\begin{center}
	\begin{figure}[!htbp]
		\begin{subfigure}{1.0\textwidth}
			\centering
			\includegraphics[width=1.0\linewidth,height=1.0\textheight,keepaspectratio]{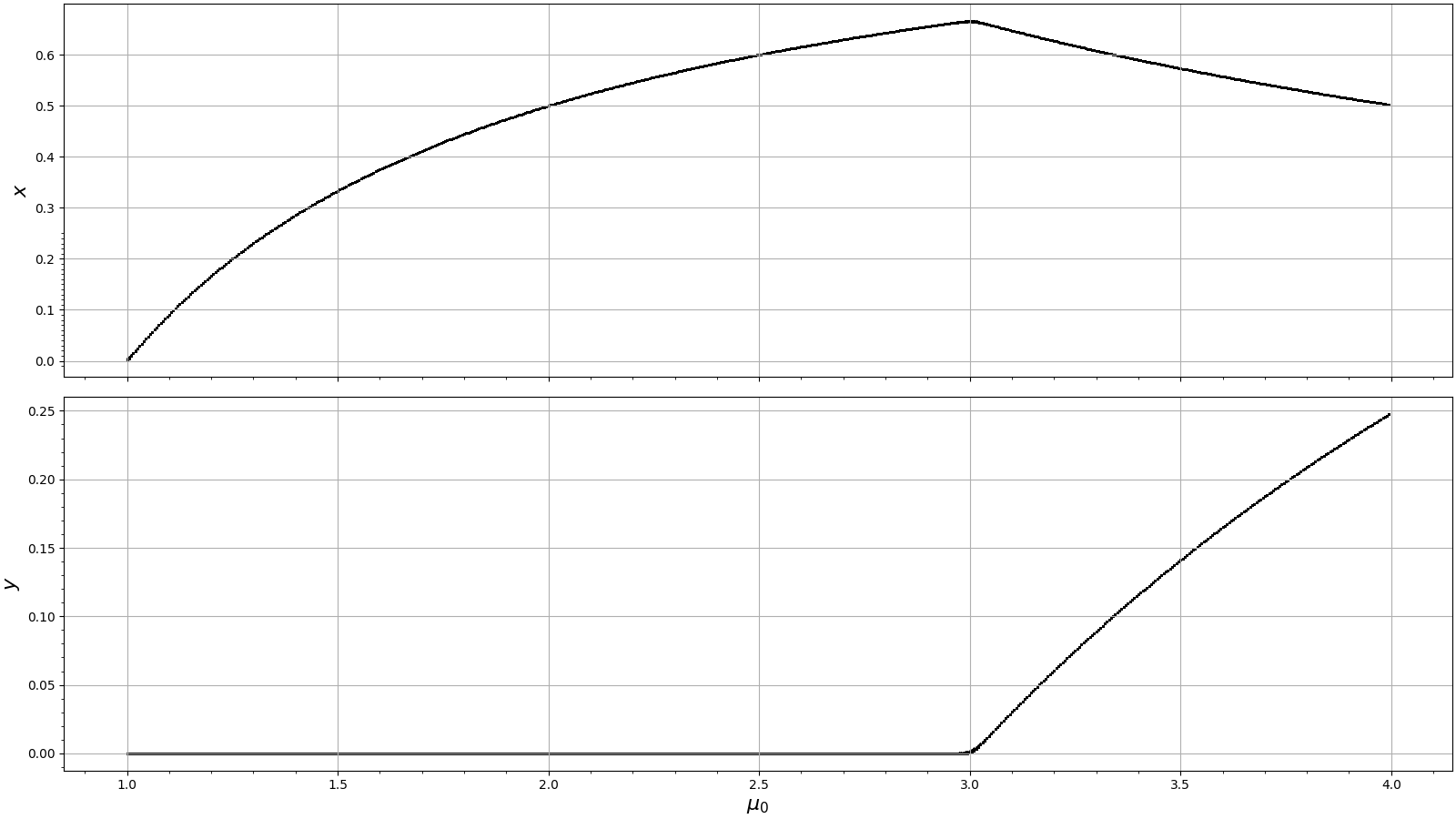}
			\caption{}
			\label{fig:Normal_bifurcation}
		\end{subfigure}
		\begin{subfigure}{1.0\textwidth}
			\centering
			\includegraphics[width=1.0\linewidth,height=1.0\textheight,keepaspectratio]{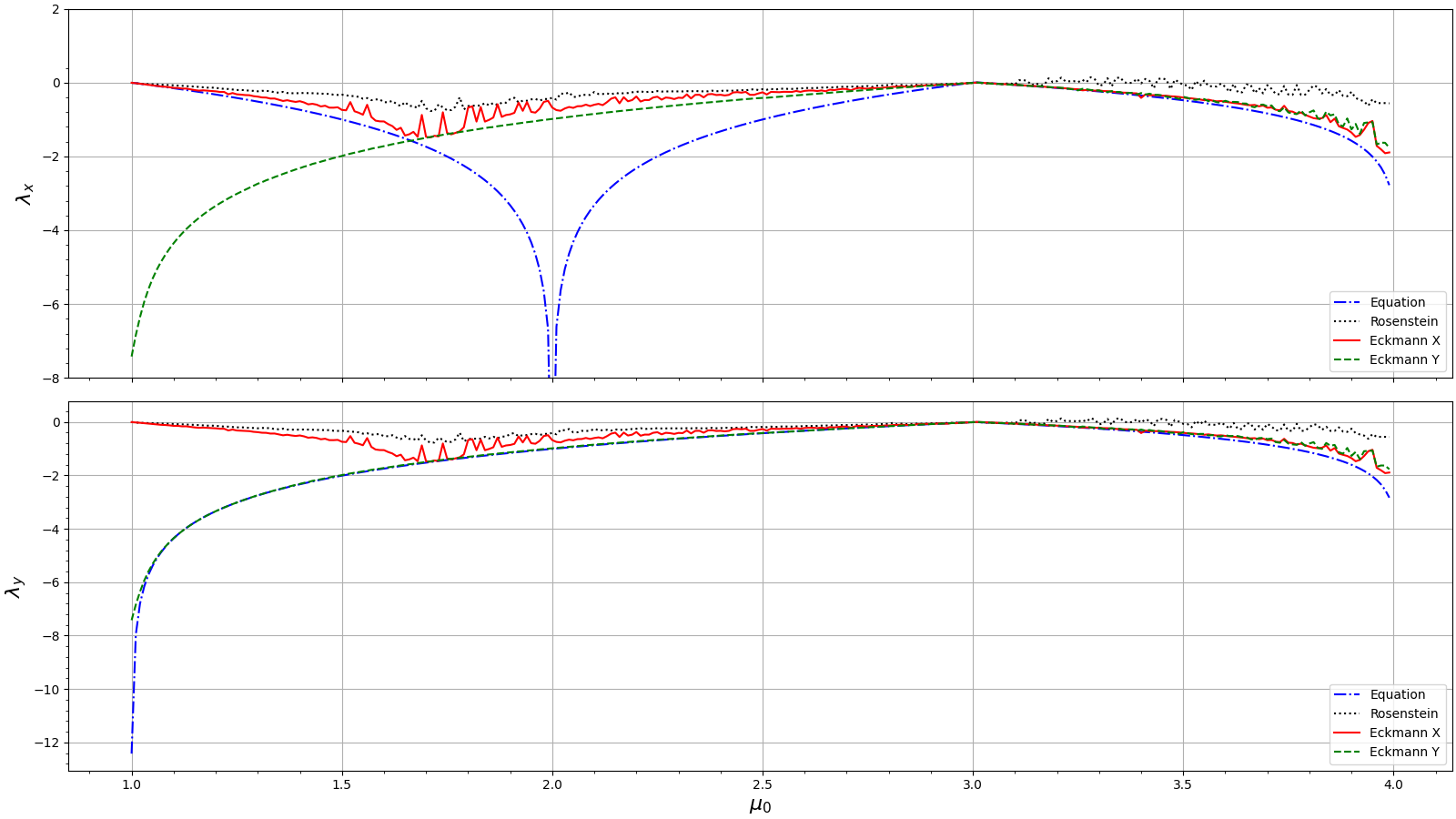}
			\caption{}
			\label{fig:Normal_lyapunovExponents}
		\end{subfigure}
		\caption{Competitive behavior and Lyapunov exponents of Normal.}
		\label{fig:Normal}
	\end{figure}
\end{center}

Our dynamical system may describe normal competitiveness without chaos, compared with Figure $\ref{fig:gammaDep0.666}$ showing slight chaos regins at $\mu_{0}$ around $2.5$ and $4.0$, between two species. Under the circumstance of an equal initial population, Figure $\ref{fig:Normal_bifurcation}$ shows a steadily increasing population of $x$ in the absence of predator when $\mu_{0}<3.0$. At the appearance of $y$ after $\mu_{0}>3.0$, the prey gradually decreases with the number of predators.

Figure $\ref{fig:Normal_lyapunovExponents}$ ensures that in this scenario there is no chaos, as the Lyapunou exponents calculated by every algorithm are negative. However, the results are different from those of various algorithms. First, for $\lambda_{x}$, there is a trench around $\mu_{0}=2$ by Eq.($\ref{eq:Lyapunov eq}$), whereas all other algorithms fail to reproduce. Rosenstein, Eckmann X, and Eckmann Y only reproduced a shallow dip around $1.5<\mu_{0}<2.5$. Furthermore, we observe that Rosenstein and Eckmann X have quite similar results in the entire range of $\mu_{0}$, except that Rosenstein has a slightly higher value. Furthermore, Eckmann X and Eckmann Y have almost identical values when $\mu_{0}>1.66$, but Eckmann Y digresses a lot from the other three curves with a low growth rate below $\mu_{0}=1.66$. For $\lambda_{y}$, all algorithms show a close spectrum $\mu_{0}>1.692$, with larger values for Rosenstein. The four algorithms divide into two groups of results below $\mu_{0}=1.692$, with Rosenstein and Eckmann X showing an increasing tail that is different from the other two algorithms showing curves of drop. 

\newpage
\subsection{Standard}\label{subs:standard}
\begin{center}
	\begin{figure}[!htbp]
		\begin{subfigure}{1.0\textwidth}
			\centering
			\includegraphics[width=1.0\linewidth]{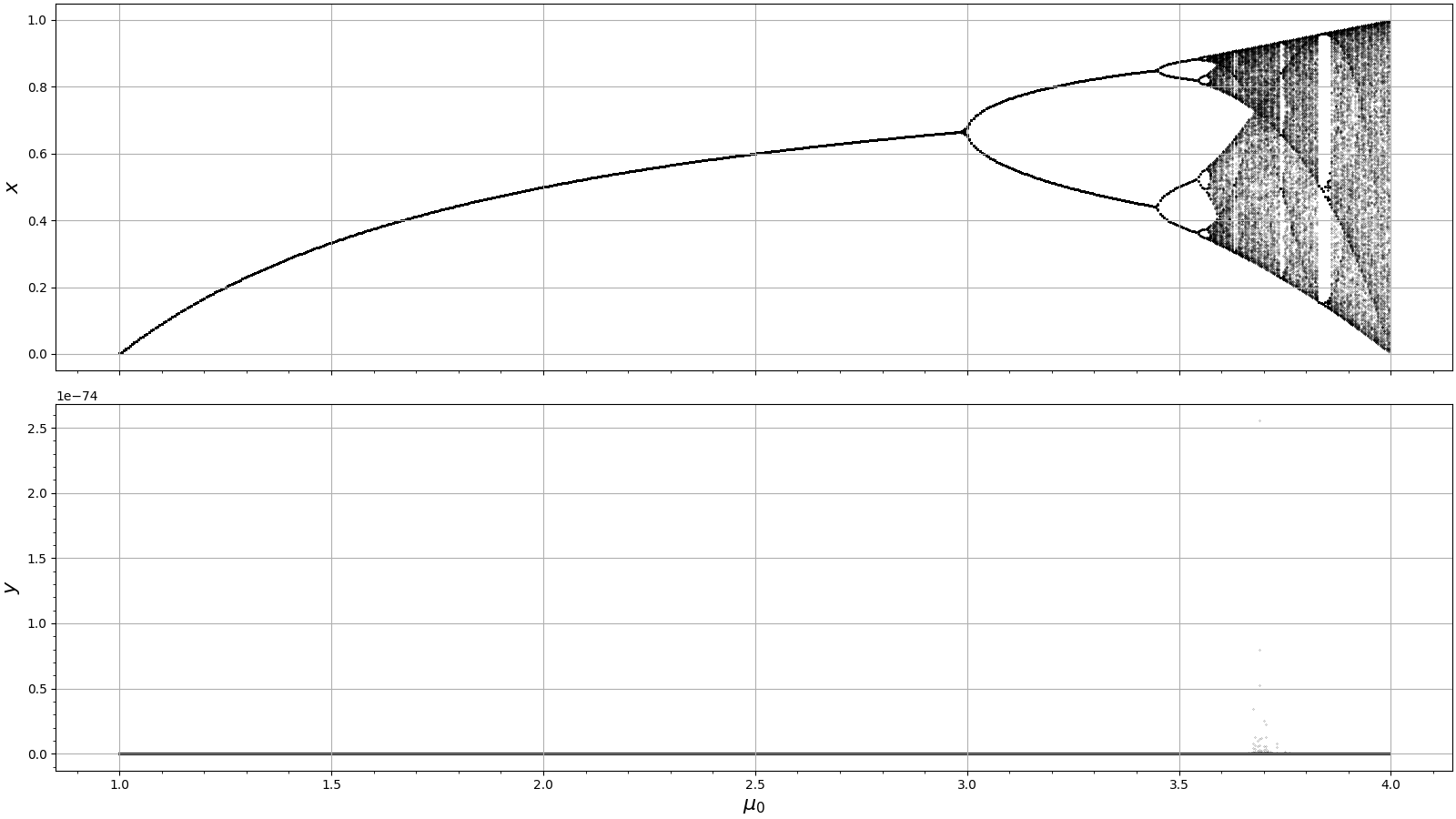}
			\caption{}
			\label{fig:Standard_bifurcation}
		\end{subfigure}
		\begin{subfigure}{1.0\textwidth}
			\centering
			\includegraphics[width=1.0\linewidth]{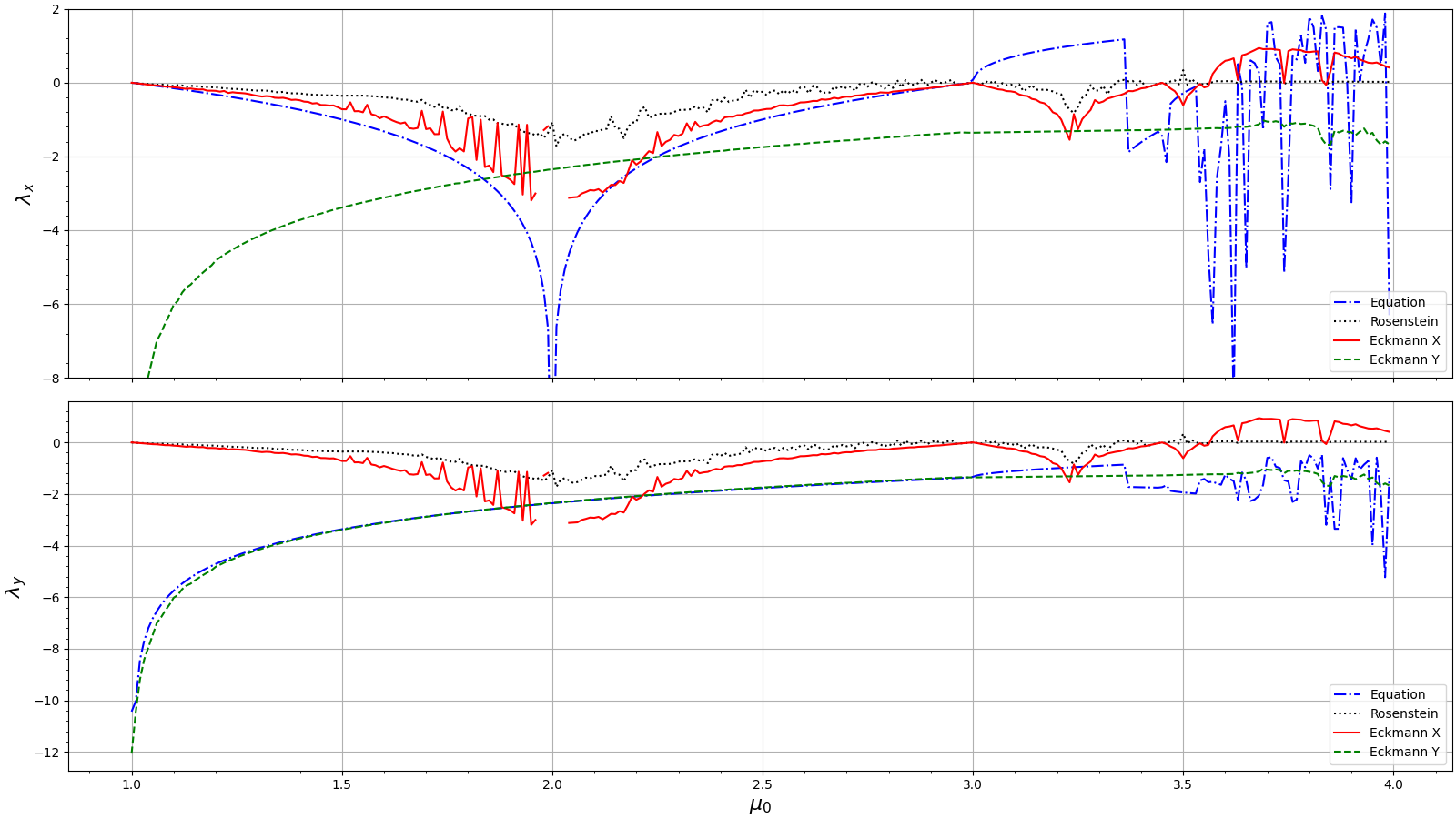}
			\caption{}
			\label{fig:Standard_lyapunovExponents}
		\end{subfigure}
		\caption{Bifurcation diagram and Lyapunov exponents of Standard.}
		\label{fig:Standard}
	\end{figure}
\end{center}

Figure $\ref{fig:Standard_bifurcation}$ shows the bifurcation diagram and the Lyapunov exponents of Standard. Our model shows that even with nonzero initial population and nonzero inter-species relationships of $\alpha$, $\beta$, and $\gamma$, we may still acquire a flip bifurcation for $1$D logistic equation$\cite{Thornton and Marion}$ for the prey in the absence of the predator. A flip bifurcation is a counterpart in the discrete dynamical system to describe the concept of periodic doubling in the continuous dynamic system$\cite{IGI Global}$. 

Figure $\ref{fig:Standard_lyapunovExponents}$ shows the Lyapunov exponents. We may see that for the general trend, all algorithms show that both $\lambda_{x}$ and $\lambda_{y}$ are negative for $\mu_{0}<3.0$, while $\lambda_{x}$ and $\lambda_{y}$ have positive and negative values for $\mu_{0}>3.5$. It is widely accepted$\cite{Goldstein}$ that the Lyapunov exponent values occur interchangeably between positive and negative inferring chaos, which is consistent with the shaded area in Figure $\ref{fig:Standard_bifurcation}$. Another inconsistency occurs with $3.0<\mu_{0}<3.5$ with $\lambda_{x}>0$ for Eq.($\ref{eq:Lyapunov eq}$) but $\lambda_{x}<0$ for all other algorithms, where $x$ exhibits a flip bifurcation from $2$-cycle into $4$-cycle. Nevertheless, this inconsistency may not be a problem for us to distinguish chaos from happening. There is a break around $\mu_{0}=2$ for Eckmann X in both $\lambda_{x}$ and $\lambda_{y}$, at which Eq.($\ref{eq:Lyapunov eq}$) shows a deep trench in $\lambda_{x}$. 

\begin{figure}[!htbp]
		\begin{subfigure}[b]{0.5\textwidth}
			\centering
			\includegraphics[width=1.0\linewidth]{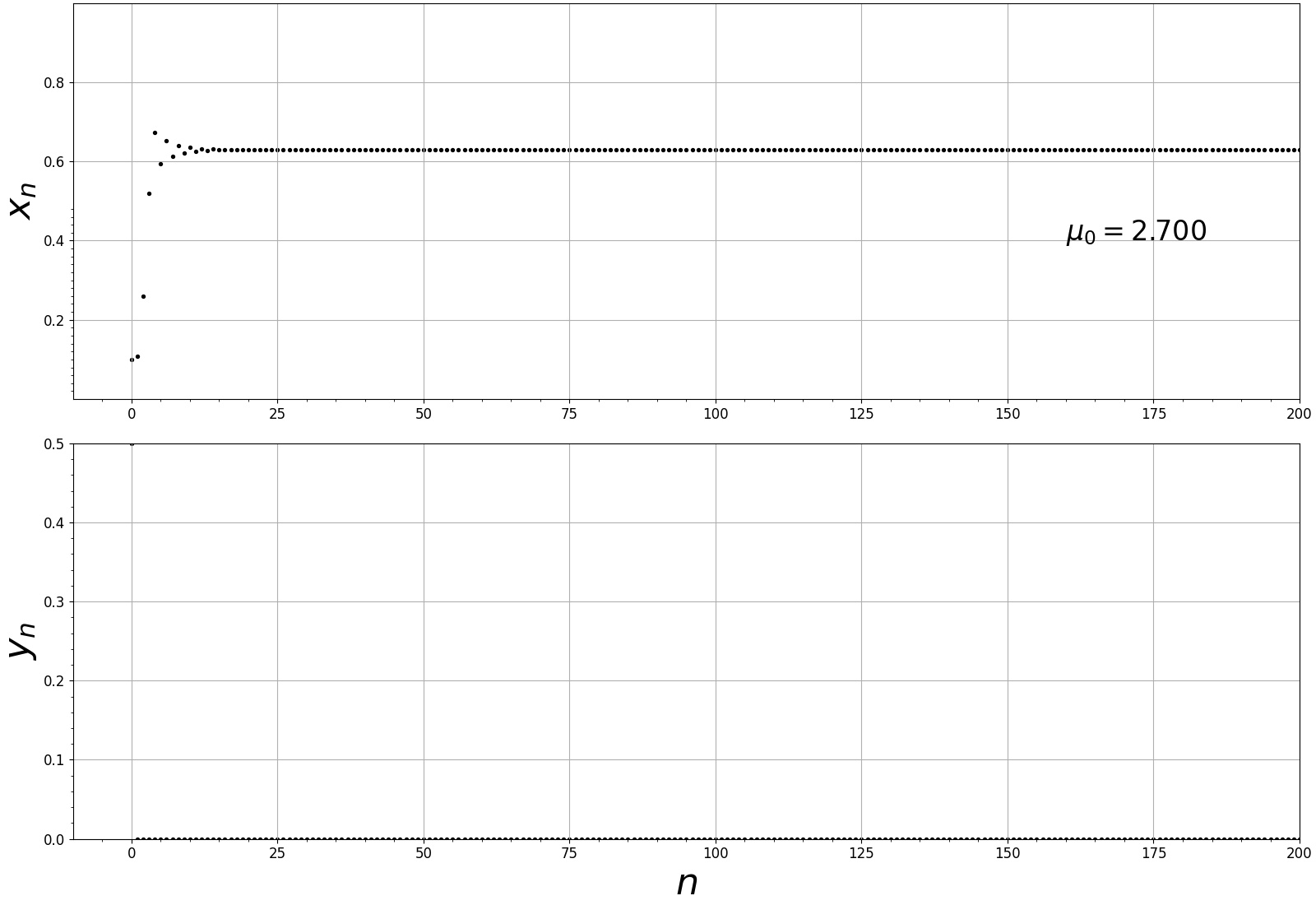}
			\caption{}
			\label{}
		\end{subfigure}
		\begin{subfigure}[b]{0.5\textwidth}
			\centering
			\includegraphics[width=1.0\linewidth]{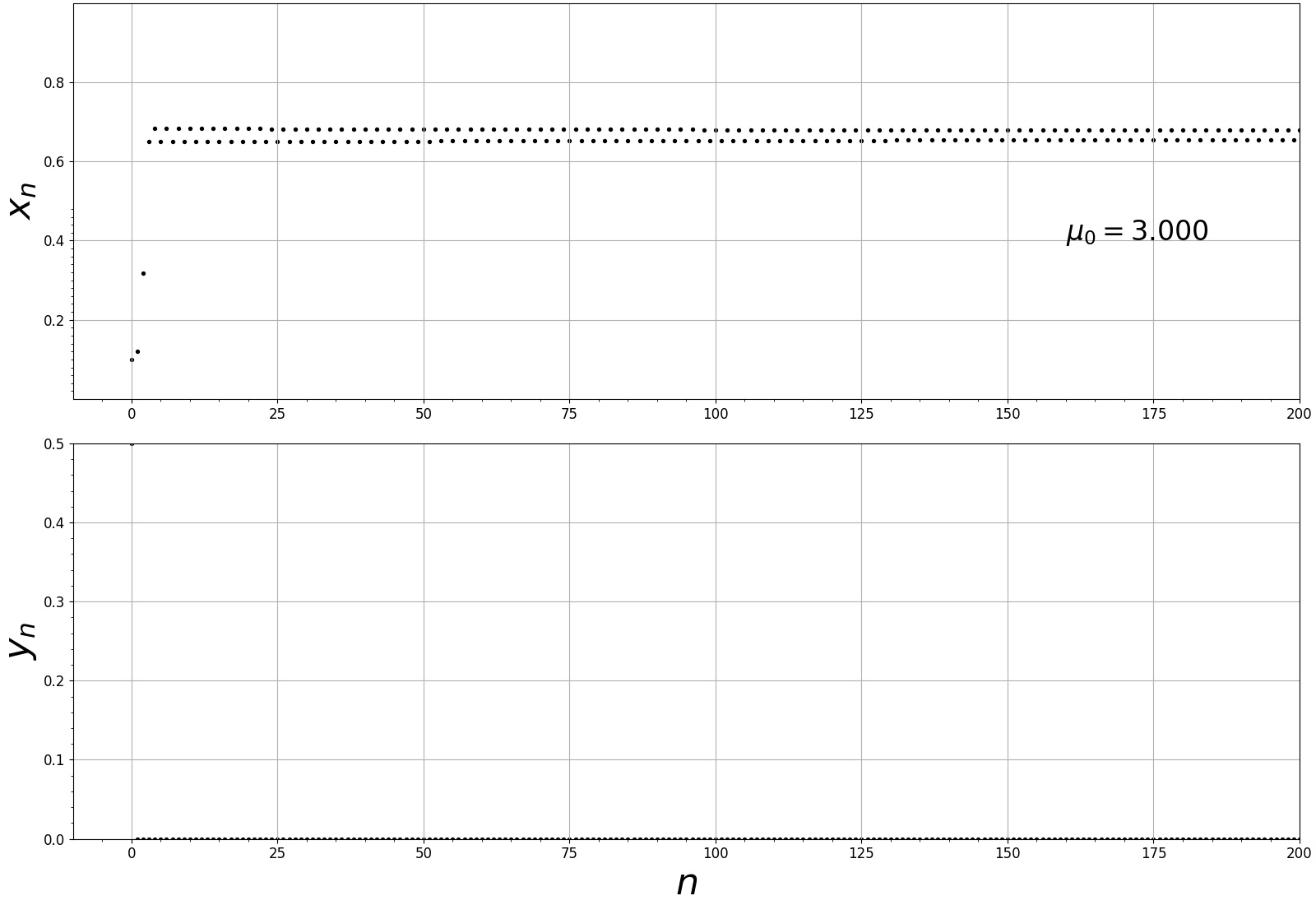}
			\caption{}
			\label{}
		\end{subfigure}
		\begin{subfigure}{0.5\textwidth}
			\centering
			\includegraphics[width=1.0\linewidth]{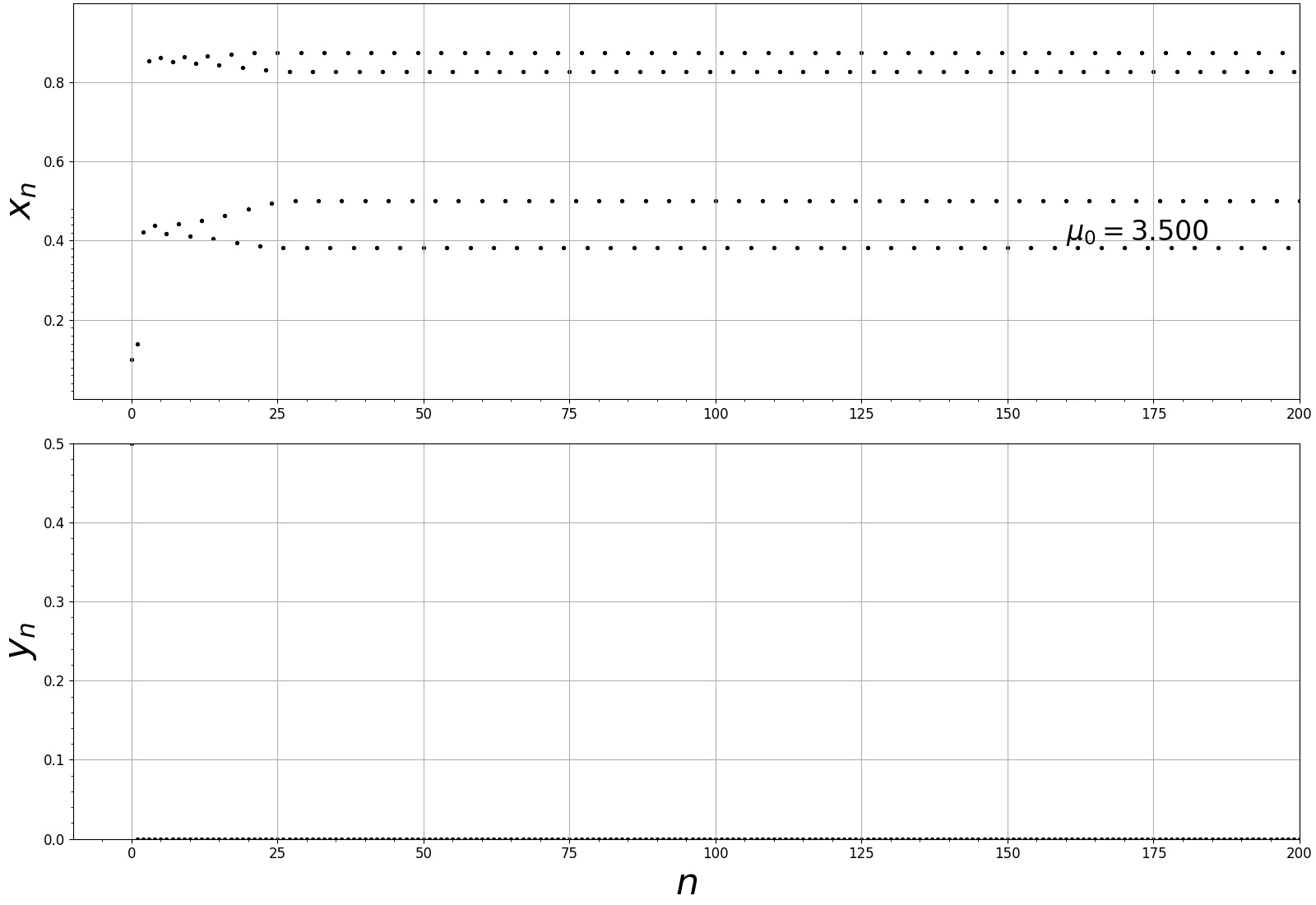}
			\caption{}
			\label{}
		\end{subfigure}
		\begin{subfigure}{0.5\textwidth}
			\centering
			\includegraphics[width=1.0\linewidth]{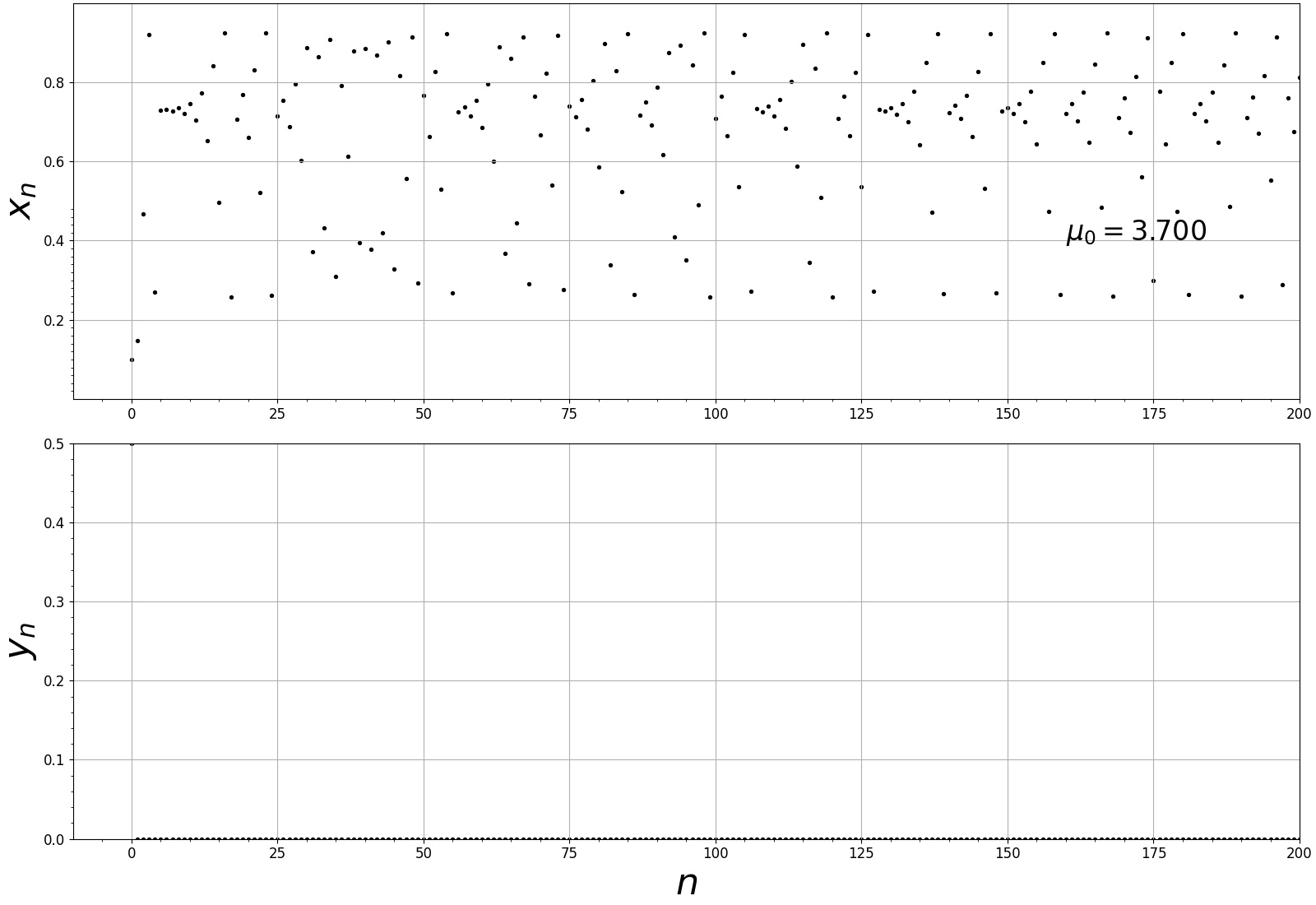}
			\caption{}
			\label{}
		\end{subfigure}
		\begin{subfigure}{0.5\textwidth}
			\centering
			\includegraphics[width=1.0\linewidth]{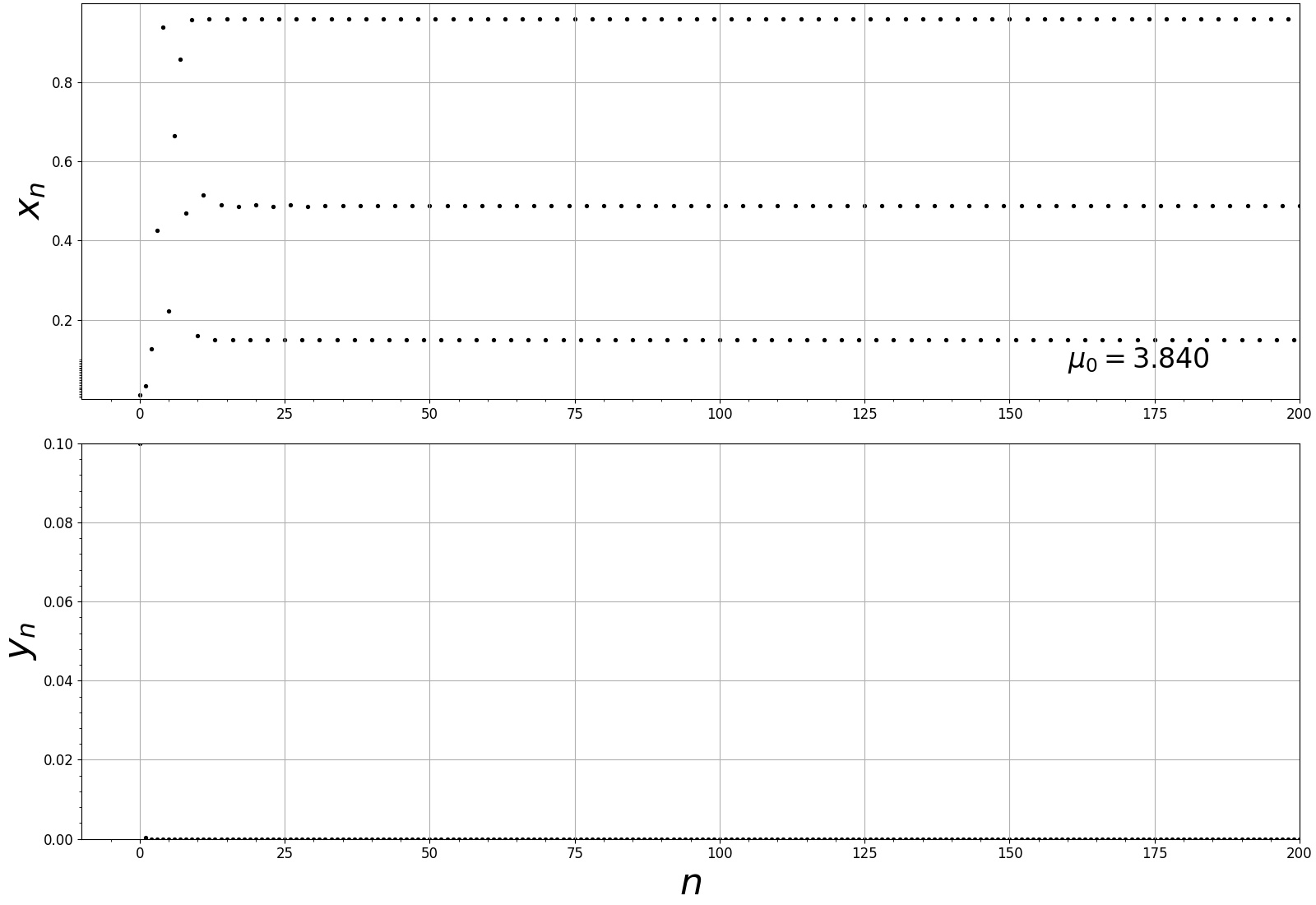}
			\caption{}
			\label{}
		\end{subfigure}
		\begin{subfigure}{0.5\textwidth}
			\centering
			\includegraphics[width=1.0\linewidth]{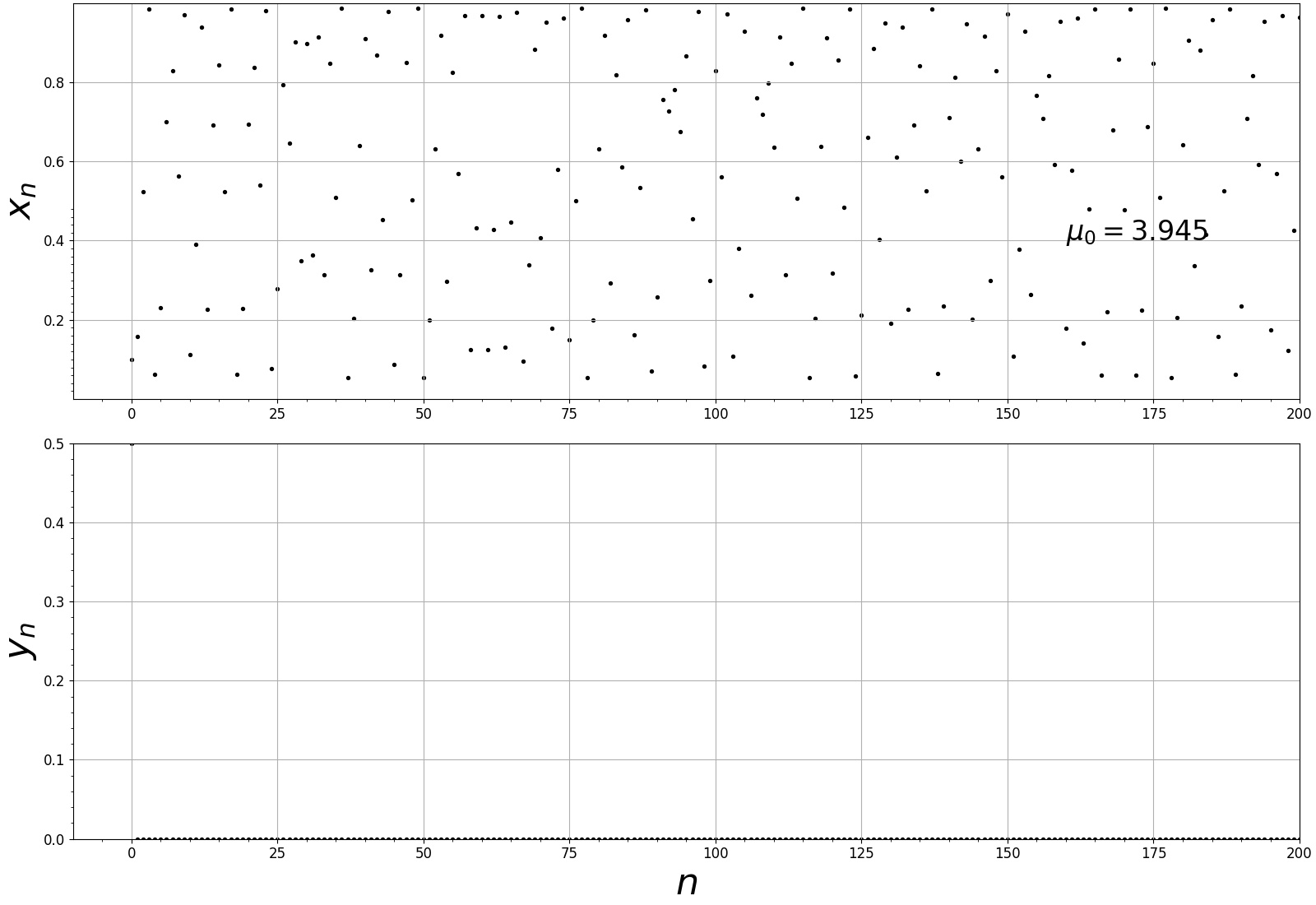}
			\caption{}
			\label{}
		\end{subfigure}
		\caption{Population vs. iteration of Standard.}
		\label{fig:Standard_PopVsN}
\end{figure}

Figure $\ref{fig:Standard_PopVsN}$ studies the population in the course of time (iteration) at $\mu_{0}$ equal to $2.700$ ($1$-cycle), $3.000$ ($2$-cycle where the flip bifurcation occurs) and $3.500$ ($4$-cycle), $3.700$ (at which the system goes into chaos), $3.840$ (the system returning to a more stable $3$-cycle) and $3.945$ (where the system returns to chaos again) in successive order. The zero predator population is obtained over the course of time.

\begin{figure}[!htbp]
        \centering
        \includegraphics[width=\linewidth,height=0.95\textheight,keepaspectratio, angle=0]{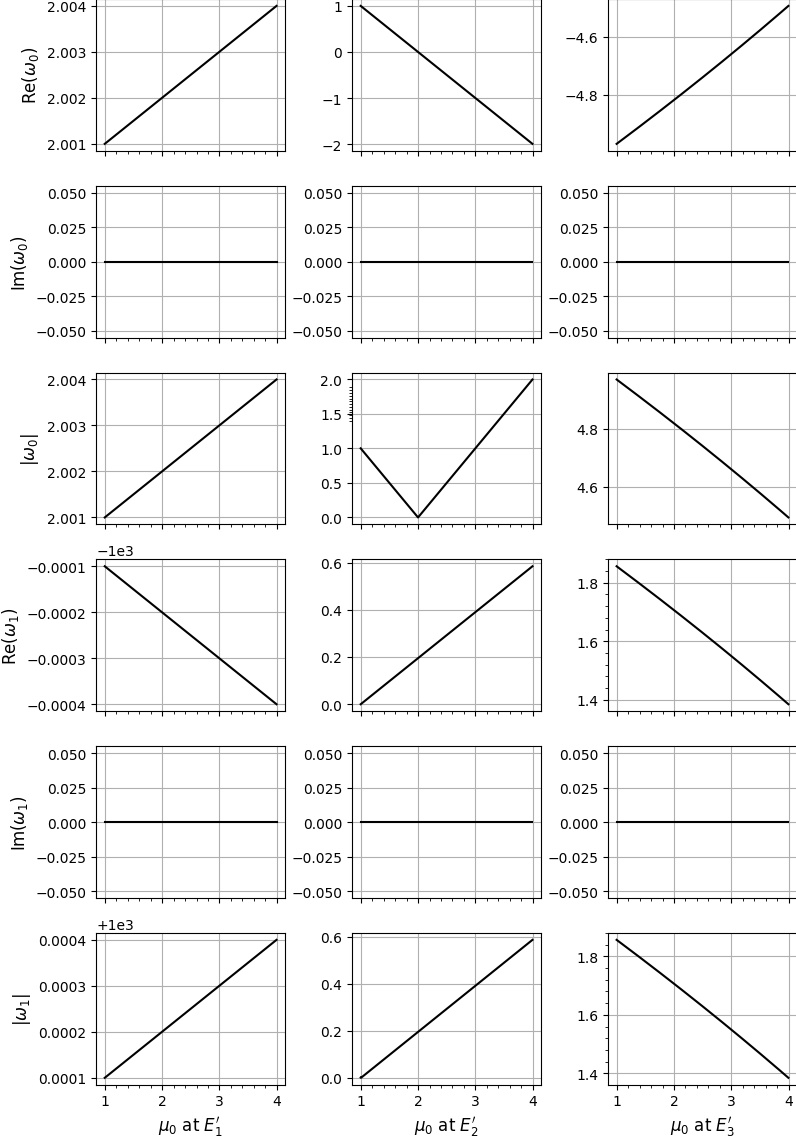}
	\caption{Absolute values of eigenvalues vs. growth rate at fixed points for Standard. We observe that $E^{\prime}_{2}$ is non-hyperbolic at $\mu_{0}=1$ and $3$, while $E^{\prime}_{3}$ is non-hyperbolic at $\mu_{0}=3.75$.} 
	\label{fig:Standard_absEval_vs_mu0}
    \end{figure}

Figure $\ref{fig:Standard_absEval_vs_mu0}$ shows the absolute values of the eigenvalues $\omega_{0}$ and $\omega_{1}$ at fixed points $E^{\prime}_{1}$, $E^{\prime}_{2}$ and $E^{\prime}_{3}$. The topological types of fixed points may be checked more straightforwardly with the figure. The first and third columns show that $E^{\prime}_{1}$ and $E^{\prime}_{3}$ are sources because $\omega_{0}$ and $\omega_{1}$ are always greater than $1$. The second column shows that $E^{\prime}_{2}$ is a sink when $1<\mu_{0}<3$ and at $3<\mu_{0}<4$, $E^{\prime}_{2}$ is a saddle. Non-hyperboles can also be examined at $\mu_{0}=1$ and $\mu_{0}=3$ for $E^{\prime}_{2}$ (in both cases $\omega_{0}=1$ and $\omega_{1}\ne1$). Finally, when $\mu_{0}>3$, $E^{\prime}_{2}$ is a source. 

\newpage
\subsection{Extinction}\label{subs:extinction}
\begin{center}
	\begin{figure}[!htbp]
		\begin{subfigure}[b]{1.0\textwidth}
			\centering
			\includegraphics[width=1.0\linewidth]{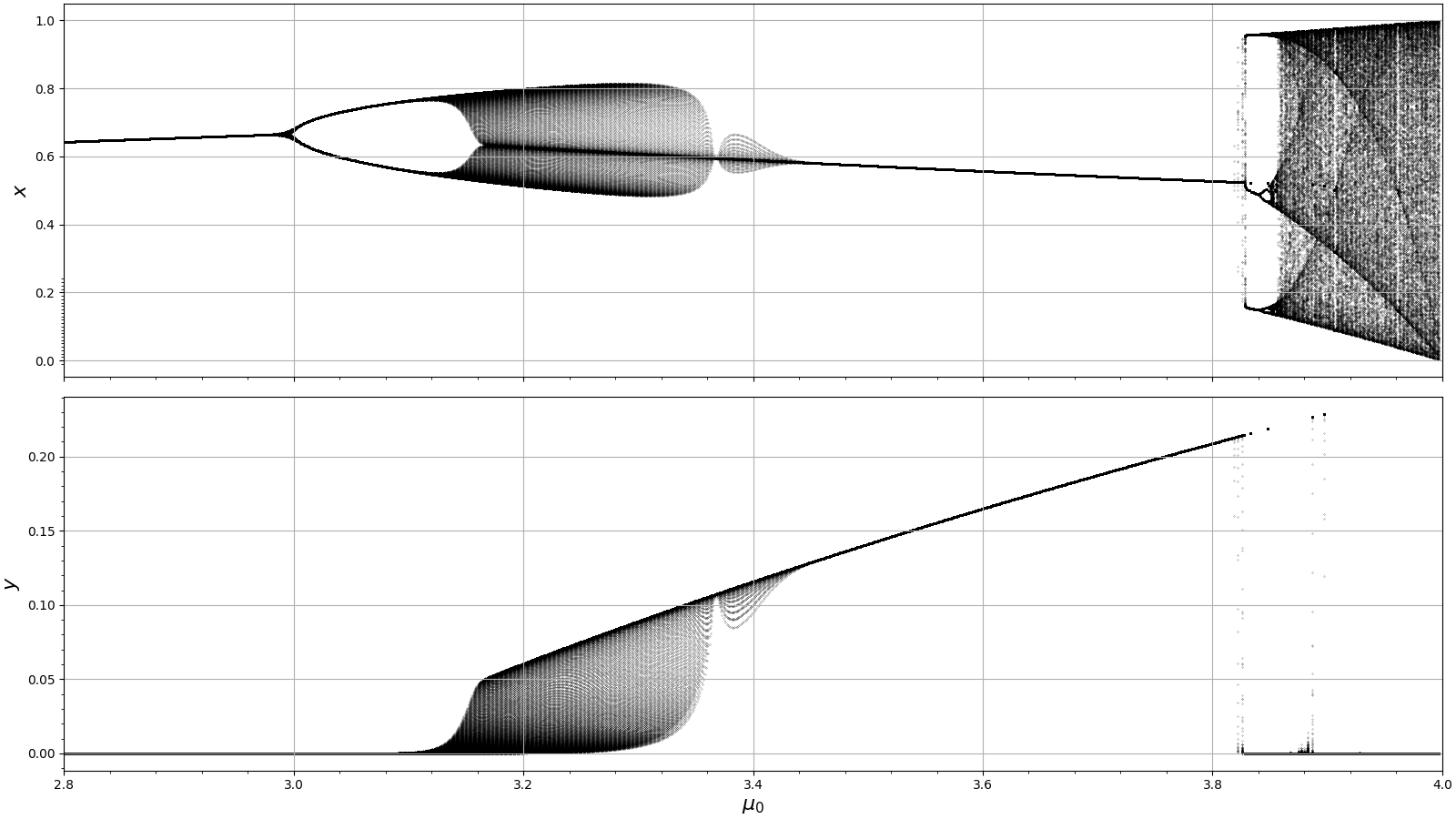}
			\caption{}
			\label{fig:Extinction_bifurcation}
		\end{subfigure}
		\begin{subfigure}[b]{1.0\textwidth}
			\centering
			\includegraphics[width=1.0\linewidth]{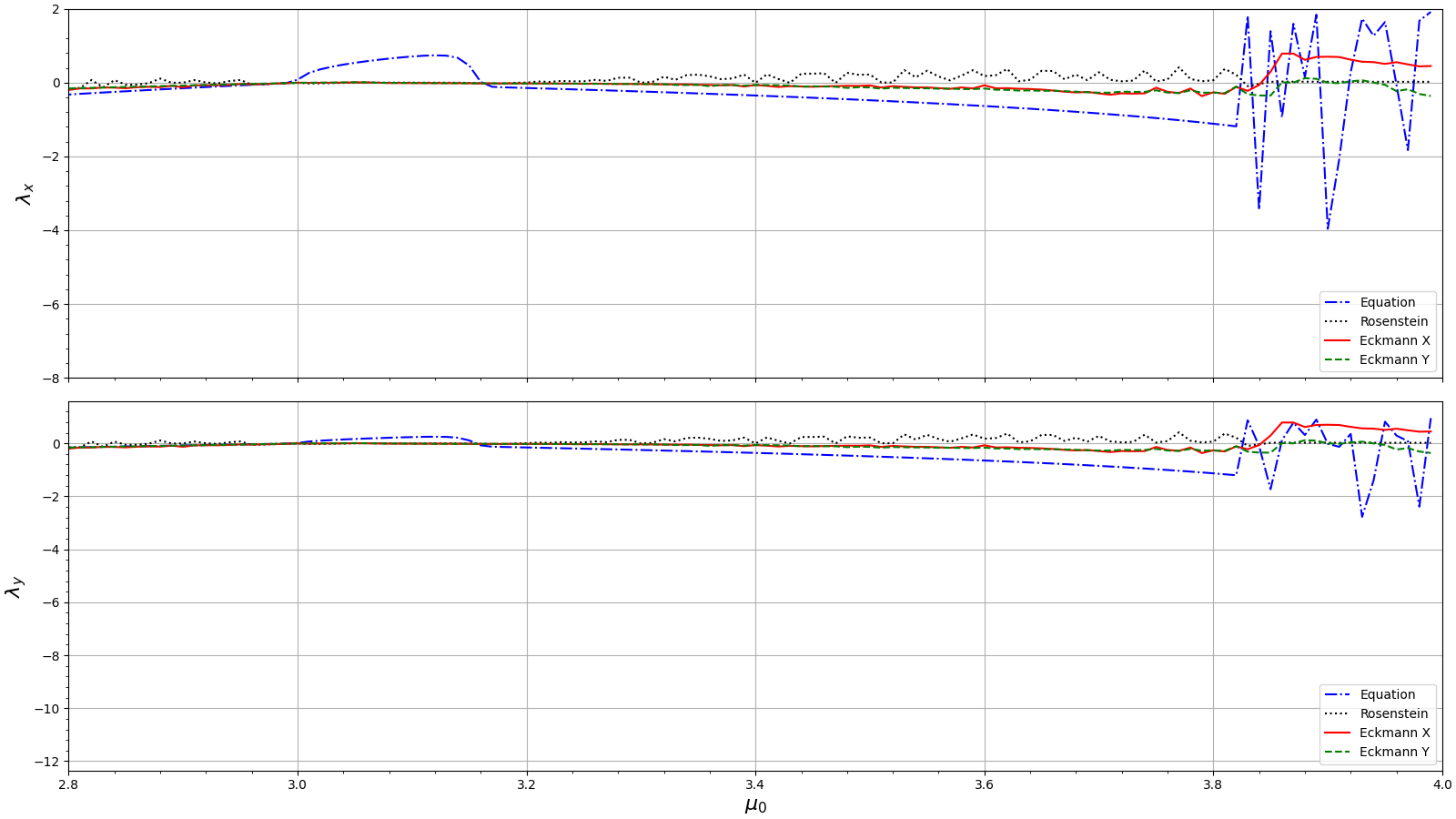}
			\caption{}
			\label{fig:Extinction_lyapunovExponents}
		\end{subfigure}
		\caption{Bifurcation diagram and Lyapunov exponents of Extinction.}
		\label{fig:Extinction}
	\end{figure}
\end{center}

Figure $\ref{fig:Extinction_bifurcation}$ shows the bifurcation diagram of \textbf{Extinction}. There is a flip bifurcation for $x$ around $\mu_{0}\approx2.99$. More explanation will be given for the shaded regions between $3.165<\mu_{0}<3.434$ for $x$ and $3.10<\mu_{0}<3.343$ for $y$, which are not chaos regions. Subsequently, the bifurcation diagram shows that the population returns to Normal for both $x$ and $y$, and represents pleasant conditions of predictable values before $x$ enters the $3$-cycle, at $\mu_{0}\approx3.828$ as in Standard, where $y$ drops to zero cliff fall at $\mu_{0}\approx3.824$, manifesting the occurrence of extinction for the predator while the prey is around the $3$-cycle state where the onset of chaos is around the corner. Fortunately, there could still be few small chances for the predator to survive at ($\mu_{0},y$) = ($3.845,0.219$), ($3.887,0.226$) and ($3.897,0.228$), where we can see that three isolated fixed points appear with vertical tails. When the prey becomes fully chaotic, the predator population reduces back to zero dramatically again and never has any further opportunity to rise back. This astonishing phenomenon, called \textit{chaotic extinction}, could be the most profound finding of the study, which states that prey in chaos generated by overpopulation of the predator would erase the entire predator species. 

Figure $\ref{fig:Extinction_lyapunovExponents}$ shows the Lyapunov exponents for \textbf{Extinction} that are similar to those in Figure $\ref{fig:Normal_lyapunovExponents}$, except for the regions at $3.0<\mu_{0}<3.18$ and $\mu_{0}>3.86$, the former showing a bum by Eq.($\ref{eq:Lyapunov eq}$), which is also the same region for the prey to be in the $2$-cycle, and the latter presenting chaos for $x$ and extinction for $y$. 

\begin{figure}[!htbp]
	\begin{subfigure}[b]{0.5\textwidth}
		\centering
		\includegraphics[width=1.0\linewidth]{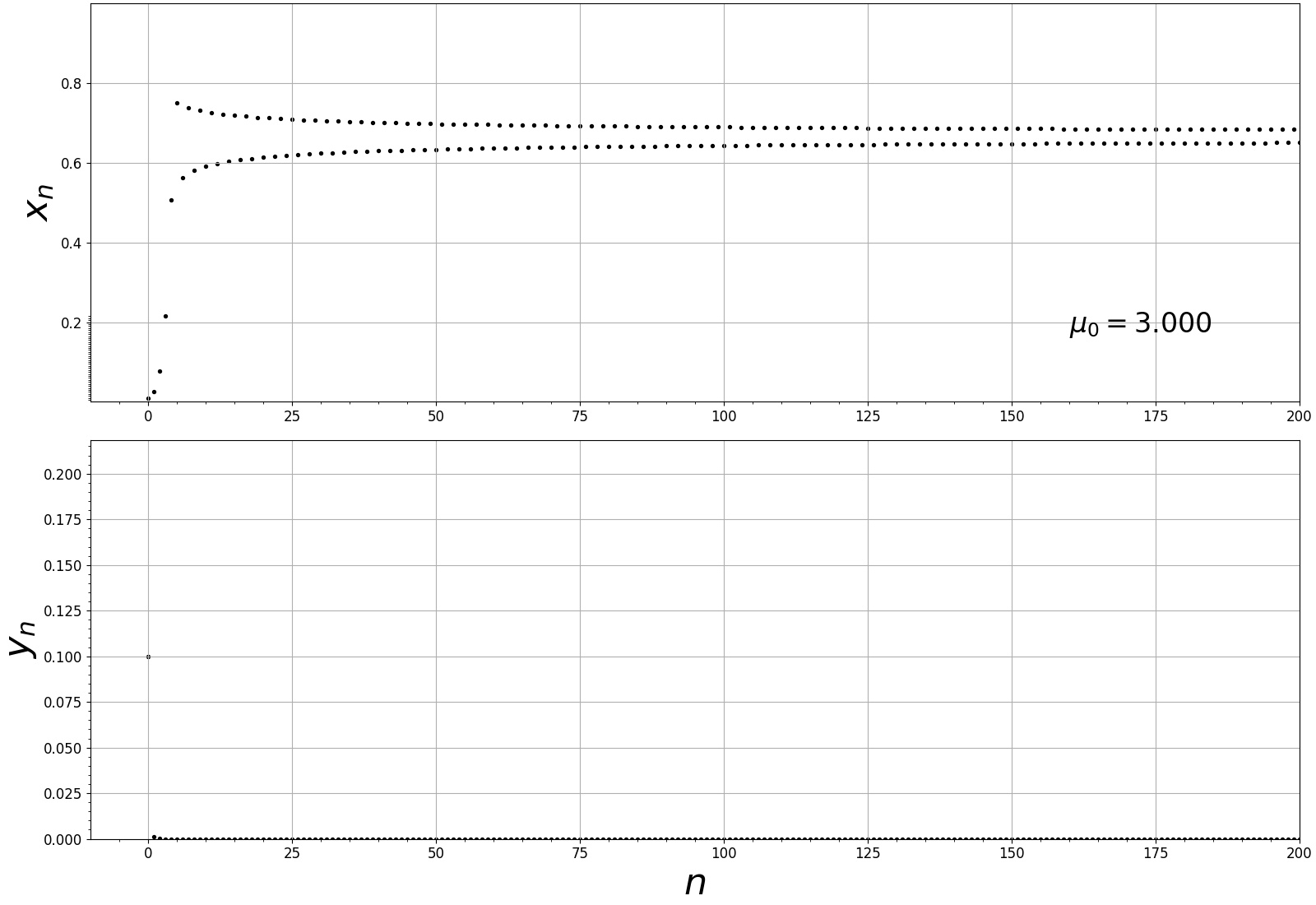}
		\caption{}
		\label{fig:Extinction_PopVsN_3.00000}
	\end{subfigure}
	\begin{subfigure}[b]{0.5\textwidth}
		\centering
		\includegraphics[width=1.0\linewidth]{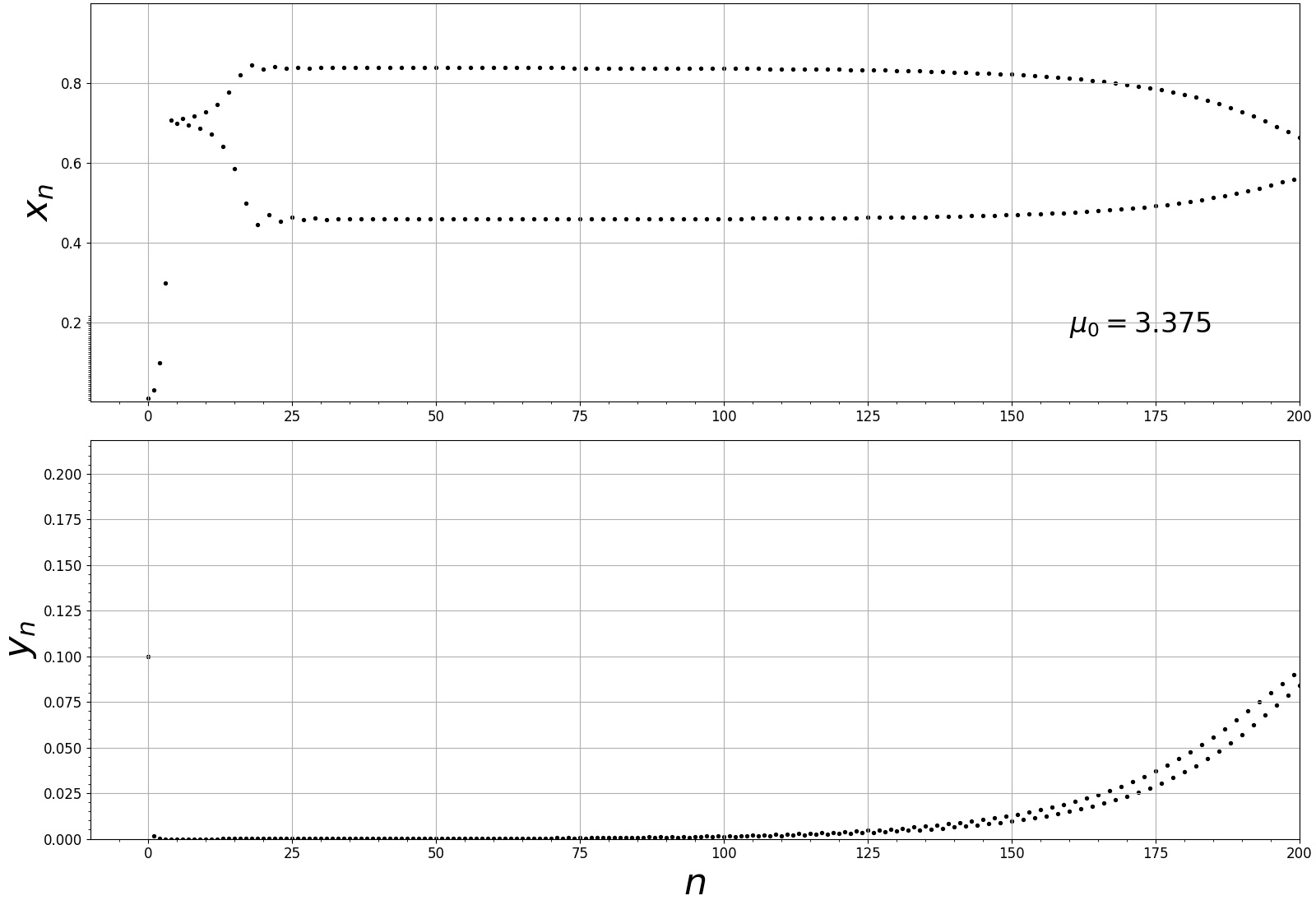}
		\caption{}
		\label{fig:Extinction_PopVsN_3.37500}
	\end{subfigure}
	\begin{subfigure}[b]{0.5\textwidth}
		\centering
		\includegraphics[width=1.0\linewidth]{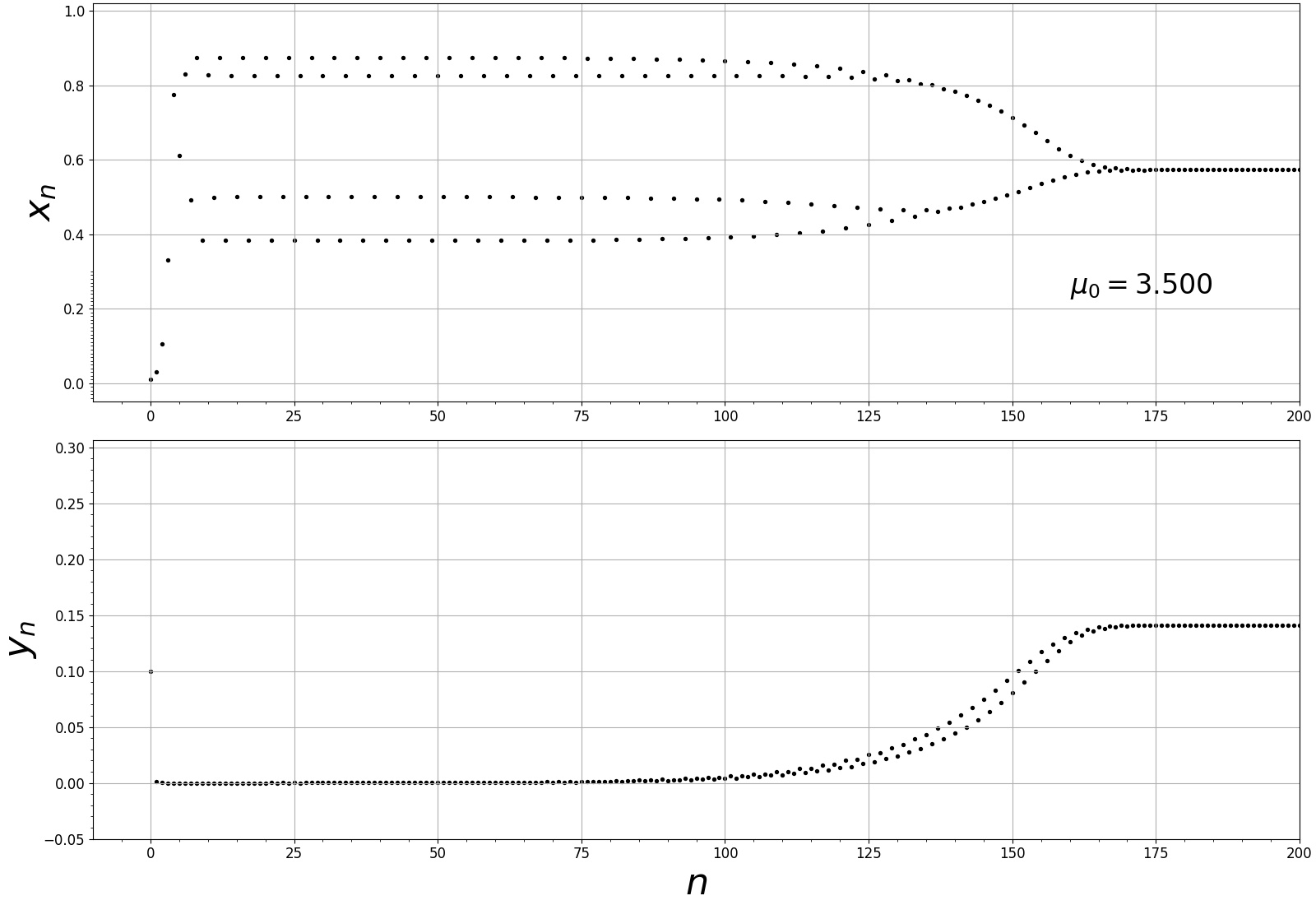}
		\caption{}
		\label{fig:Extinction_PopVsN_3.50000}			
	\end{subfigure}
	\begin{subfigure}[b]{0.5\textwidth}
		\centering
		\includegraphics[width=1.0\linewidth]{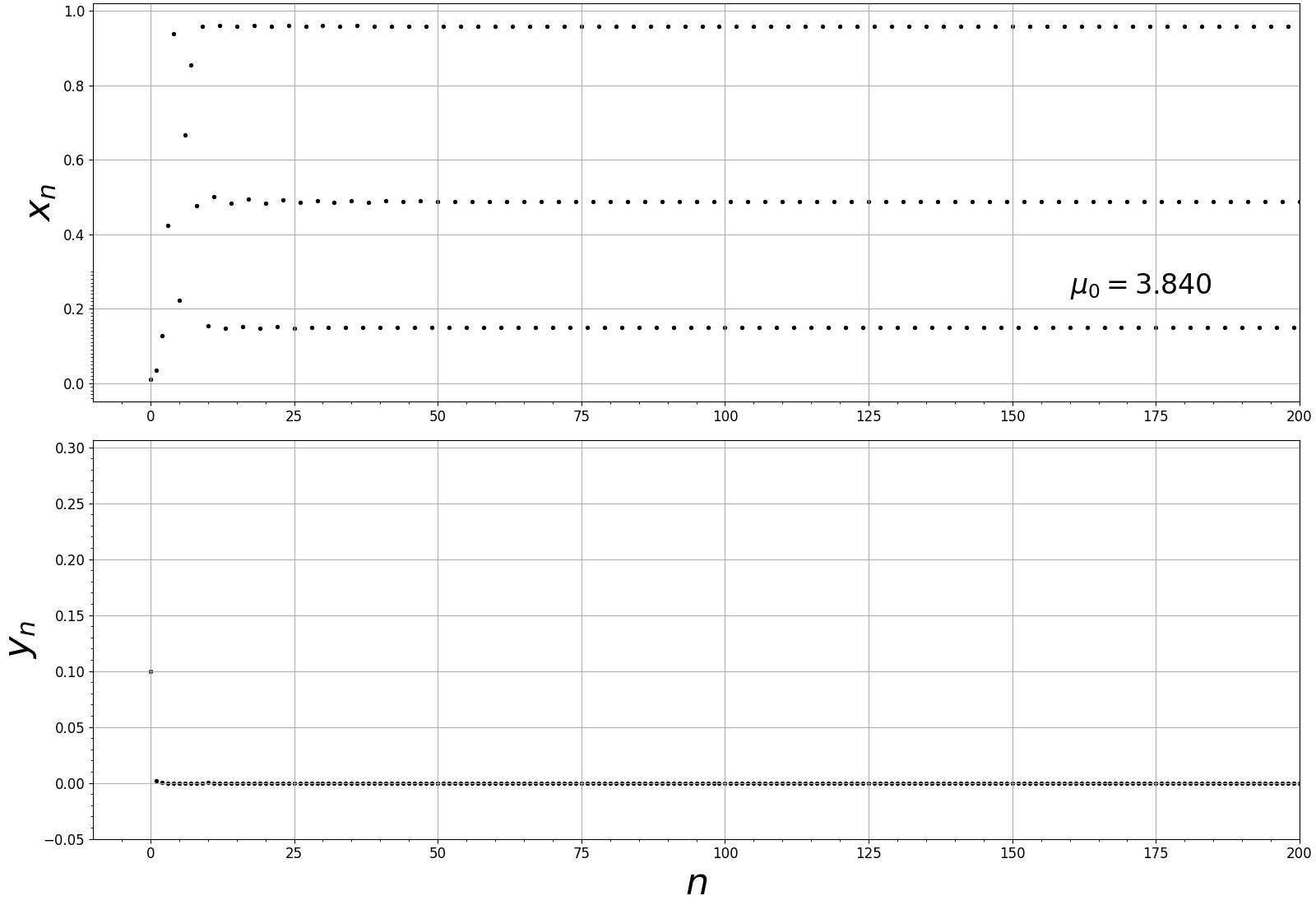}
		\caption{}
		\label{fig:Extinction_PopVsN_3.84000}
	\end{subfigure}
	\begin{subfigure}[b]{0.5\textwidth}
		\centering
		\includegraphics[width=1.0\linewidth]{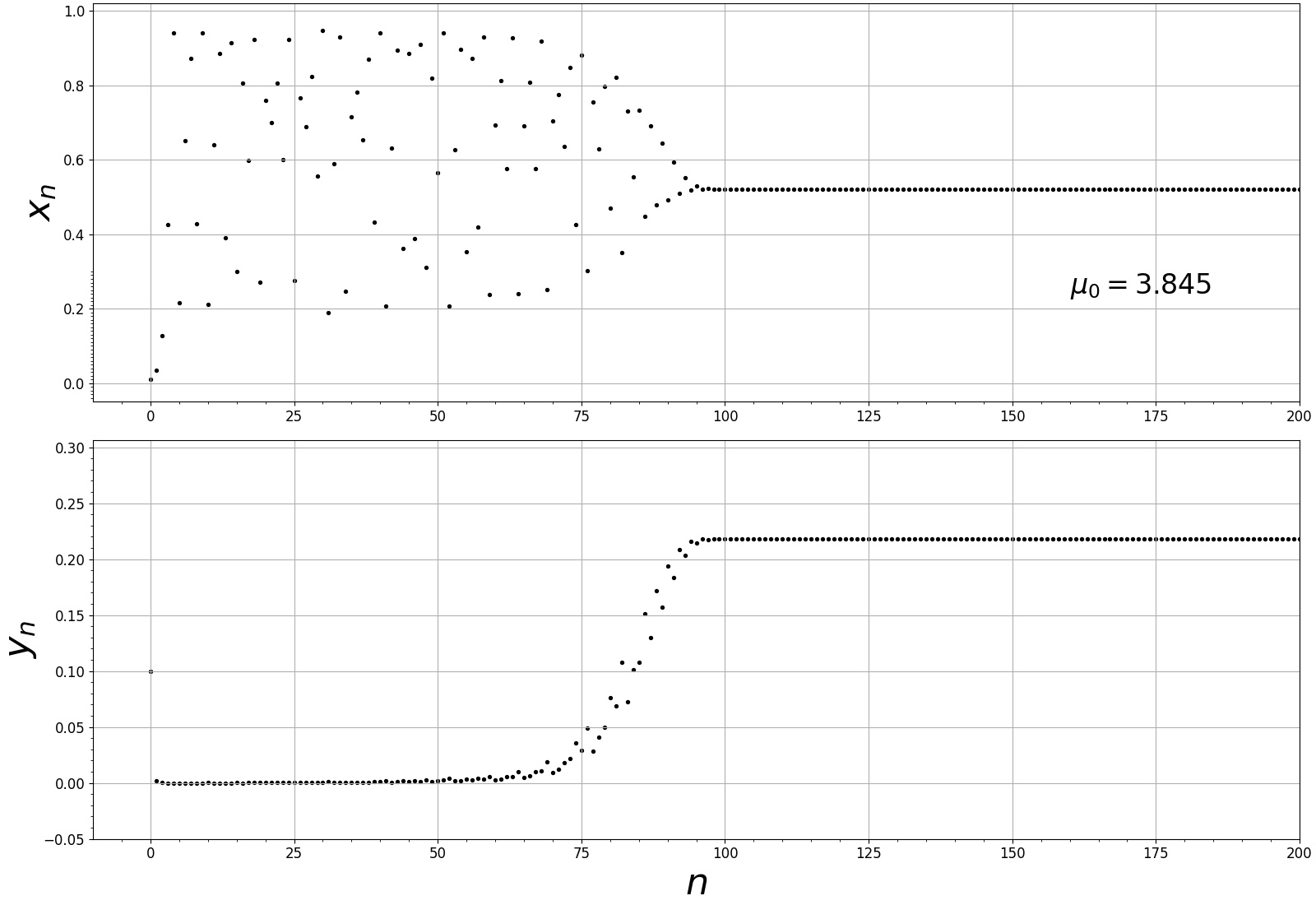}
		\caption{}
		\label{fig:Extinction_PopVsN_3.84500}
	\end{subfigure}
	\begin{subfigure}[b]{0.5\textwidth}
		\centering
		\includegraphics[width=1.0\linewidth]{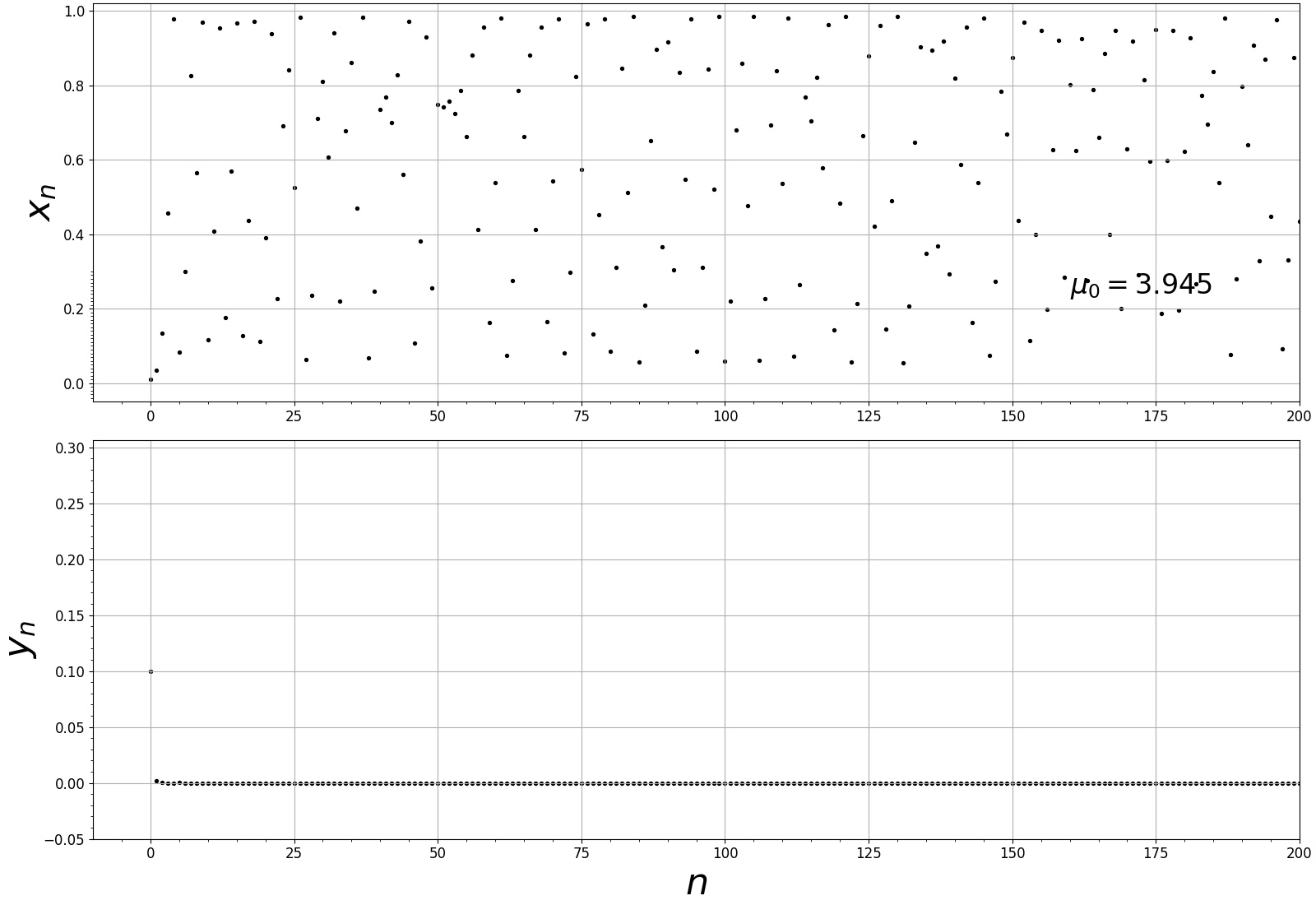}
		\caption{}
		\label{fig:Extinction_PopVsN_3.94500}
	\end{subfigure}
	\caption{Population vs. iteration of Extinction.}
	\label{fig:Extinction_PopVsN}
\end{figure}

Figure $\ref{fig:Extinction_PopVsN}$ shows the iteration of the \textbf{Extinction} population. As we can see, the flip bifurcation starts at $\mu_{0}=3.000$ as in Figure($\ref{fig:Extinction_PopVsN_3.00000}$), while in Figure($\ref{fig:Extinction_PopVsN_3.37500}$) the two fixed points collide and after transient states ($n>200$), they tend to merge into a single fixed point. We further demonstrate that at  $\mu_{0}=3.500$ in Figure($\ref{fig:Extinction_PopVsN_3.50000}$), after transient($n>175$), bifurcation collapses to the $1$-cycle. Furthermore, the $3$-cycle opens in $x$ at $\mu_{0}=3.84$, as indicated in Figure($\ref{fig:Extinction_PopVsN_3.84000}$), with extinction of $y$ at exactly the same time. However, the sunlight of survival for $y$ is shed in the window at $\mu_{0}=3.845$, as shown in Figure($\ref{fig:Extinction_PopVsN_3.84500}$), where the two species may still exist in a predictable population. 

One may argue that since the value $\beta=0.001$ is relatively small, our system could simply reproduce that of Danca et al.$\cite{Danca et al}$. However, this is not completely true. For one thing, the parameters $\mu_{1}$ and $\nu_{1}$ depend on $\mu_{0}$ with the help of Hypothesis \ref{hyp:b}. This hypothesis causes the predator to die out very quickly, under certain $\mu_{0}$, within only a few iterations, corresponding to a short time, and leads to an abrupt drop of the predator population shown in Figure $\ref{fig:Extinction_PopVsN_3.94500}$, which portends the extinction of predators in the chaos of prey in Figure $\ref{fig:Extinction_bifurcation}$. In contrast, Danca et al. also pointed out a similar scenario of extinction after thousands of iterations because their system omitted the term $-\nu_{0}y_{n}(1-y_{n})$, therefore they cannot see the extinction phenomenon in the bifurcation diagram directly as in our study.

\begin{landscape}
    \begin{figure}[!htbp]
	\centering
	\includegraphics[width=\textheight,height=\linewidth,keepaspectratio,angle=-90]{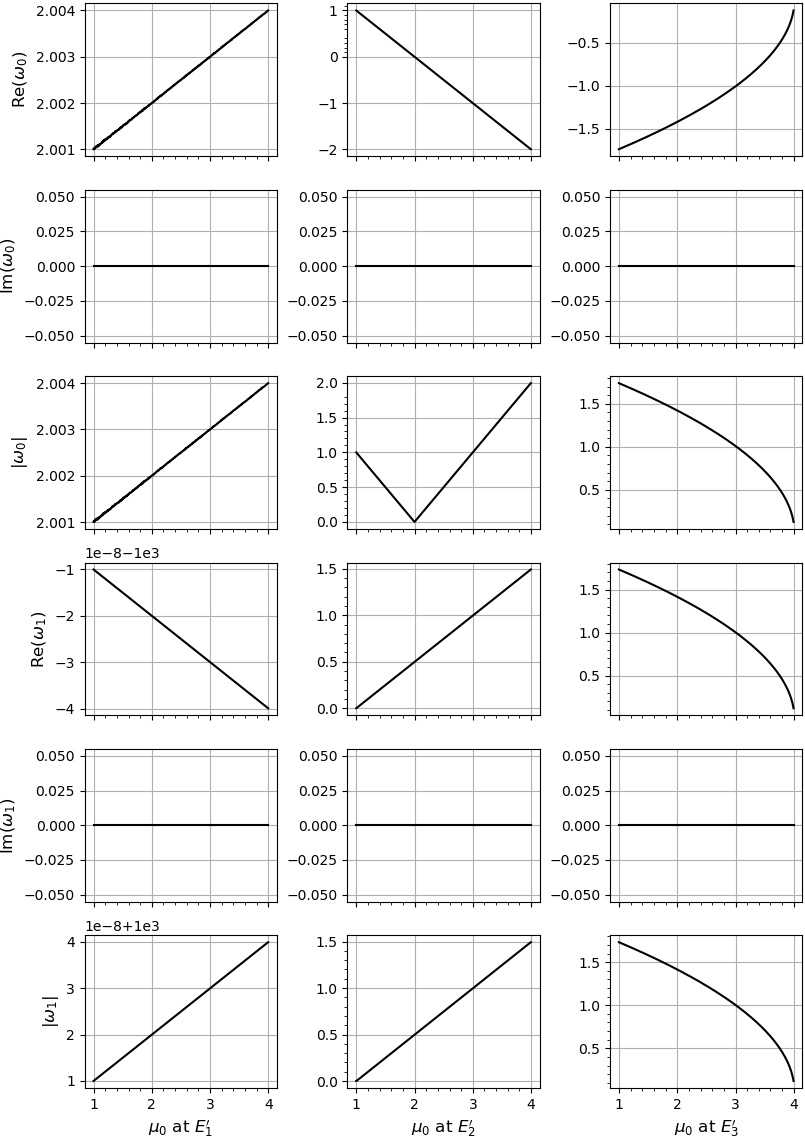}
	\caption{Absolute values of eigenvalues vs. growth rate at fixed points for Extinction. From the figures in the first column, it is clearly shown that $E^{\prime}_{1}$ is a source. Also, at $\mu_{0}=3$ where flip bifurcation occurs, $E^{\prime}_{2}$ changes its stability from a sink to a source, while $E^{\prime}_{3}$ from source to sink.} 
	\label{fig:Extinction_absEval_vs_mu0}
    \end{figure}
\end{landscape}

The stability of fixed points may also be examined in Figure $\ref{fig:Extinction_absEval_vs_mu0}$. The figures in the first column show that $E^{\prime}_{1}$ is a source, since both eigenvalues have absolute values greater than $1$. In the second column, we see that $E^{\prime}_{2}$ changes its stability from a sink to a source when $\mu_{0}$ varies at $3$, at which a flip bifurcation occurs; while, $E^{\prime}_{3}$ changes from source to sink at $\mu_{0}=3$.

\begin{figure}[!htbp]
	\centering
	\includegraphics[width=1.0\textwidth]{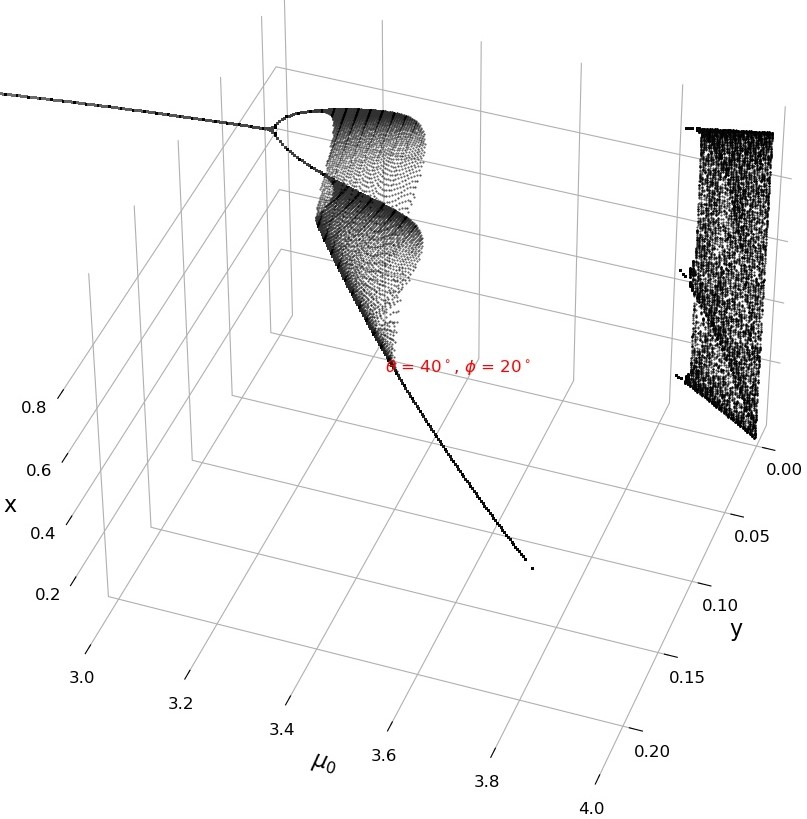}
	\caption{3D phase portrait of Extinction viewed at elevation $40^{\circ}$ and azimuth $20^{\circ}$.}
	\label{fig:Extinction_phasePortrait3D_040_020}
\end{figure}

Figure $\ref{fig:Extinction_phasePortrait3D_040_020}$ shows a three-dimensional phase portrait with varying $\mu_{0}$ of \textbf{Extinction}, viewed along the elevation angle $40^{\circ}$ and the azimuthal angle $20^{\circ}$. Now a clearer idea of the shaded areas is manifested within $3.165<\mu_{0}<3.434$ for $x$ and $3.10<\mu_{0}<3.343$ for $y$ in Figure $\ref{fig:Extinction_bifurcation}$, which is quasi-periodic, forming a shape resembling a butterfly wing, subtending to the $(x,\mu_{0})$ plane where $y$ is extinct.

\begin{figure}[!htbp]
        \centering \includegraphics[width=\linewidth,height=\textheight,keepaspectratio,angle=0]{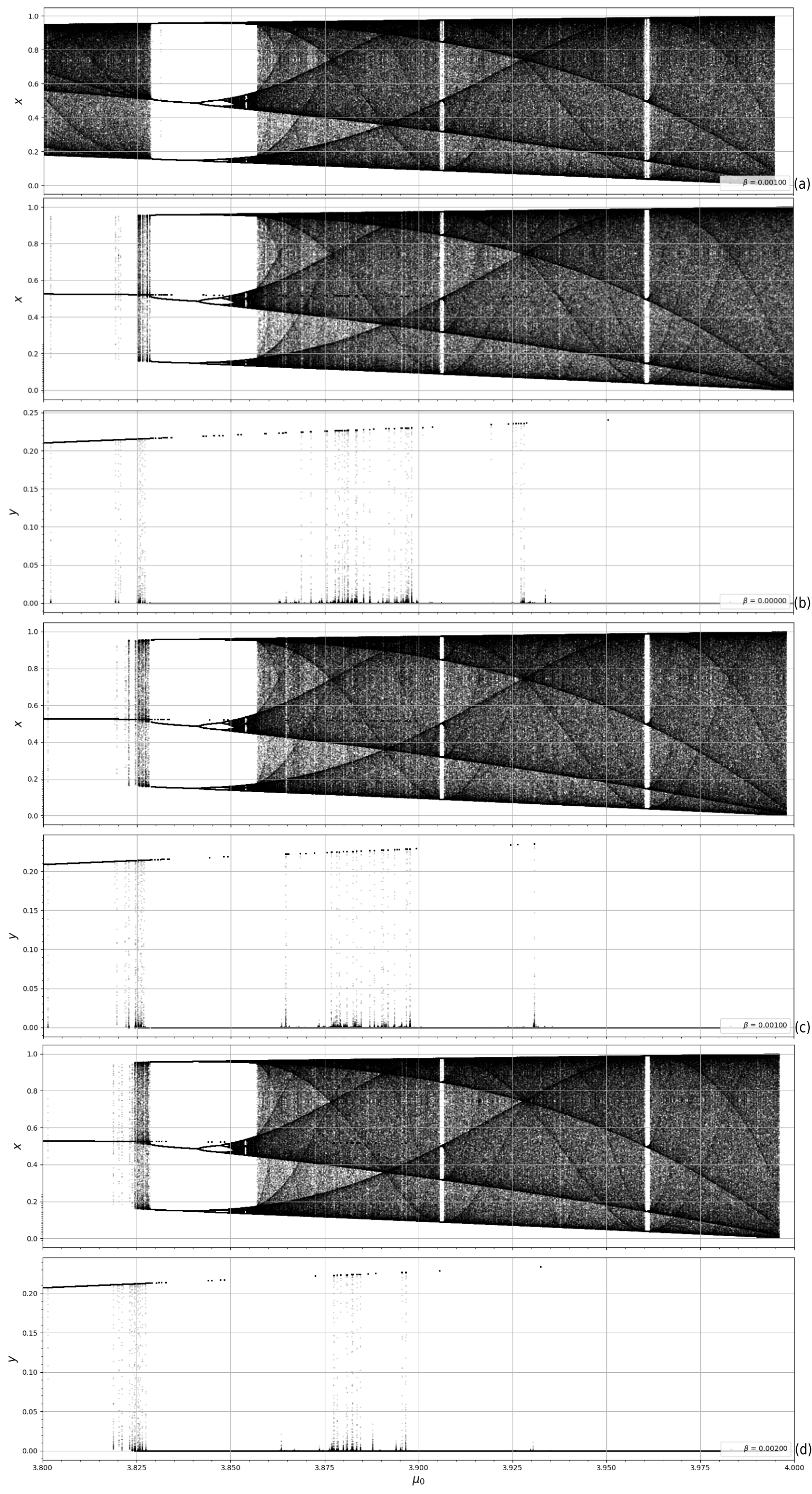}
	\caption{Comparison between Standard and Extinction. Figure $\ref{fig:Standard_Extinction}a$, Standard. Figure $\ref{fig:Standard_Extinction}b$, Extinction with $\beta=0.000$, Figure $\ref{fig:Standard_Extinction}c$ with $\beta=0.001$, and Figure $\ref{fig:Standard_Extinction}d$ with $\beta=0.002$.}
        \label{fig:Standard_Extinction}
    \end{figure}

Figure $\ref{fig:Standard_Extinction}$ represents the figures zoomed in within $3.8\le\mu_{0}\le4$ of the comparisons between \textbf{Standard} (in Figure $\ref{fig:Standard_Extinction}a$) and \textbf{Extinction} (in Figure $\ref{fig:Standard_Extinction}b-d$) with different values $\gamma$ as indicated in each figure. A three-period window of \textbf{Standard} within $3.83<\mu_{0}<3.86$ in $x$ is clearly shown in Figure $\ref{fig:Standard_Extinction}a$, with $y$ relatively small (also compared to Figure $\ref{fig:Standard_bifurcation}$). As shown in Figures $\ref{fig:Standard_Extinction}b-d$, \textbf{Extinction} breaks the chaos region before the three-period window (see Figure $\ref{fig:Standard_Extinction}a$), making a predictable $x$ value. On the other hand, the bifurcation diagrams for $x$ in all subfigures are relatively very similar relative to fine structures, such as supertracks, after the three-period windows. This means that \textbf{Extinction} of $y$ does not influence $x$ after the three-period window. Finally, a very interesting phenomenon is observed within the three-period window in Figure $\ref{fig:Standard_Extinction}b-d$ . As we can see, within $3.825<\mu_{0}<3.850$, near the middle of the three-period window, there are some isolated orbitals for $x$, and we can find exactly the same orbitals at the same $\mu_{0}$ in the corresponding $y$. For example, in Figure $\ref{fig:Standard_Extinction}(b)$ , there are five isolated orbitals near the center of the three-period window in $x$, and we can find exactly the same number of isolated orbitals at the corresponding $\mu_{0}$ for $y$. Furthermore, closer inspection reveals that for the residual orbitals of $y$ in \textbf{Extinction}, there are always counterparts in $x$. Despite the orbitals before the three-period window in $x$, the rest are cloaked by the chaotic region and cannot be seen obviously.

\begin{figure}[!htbp]
	\centering
	\includegraphics[width=1.0\textwidth]{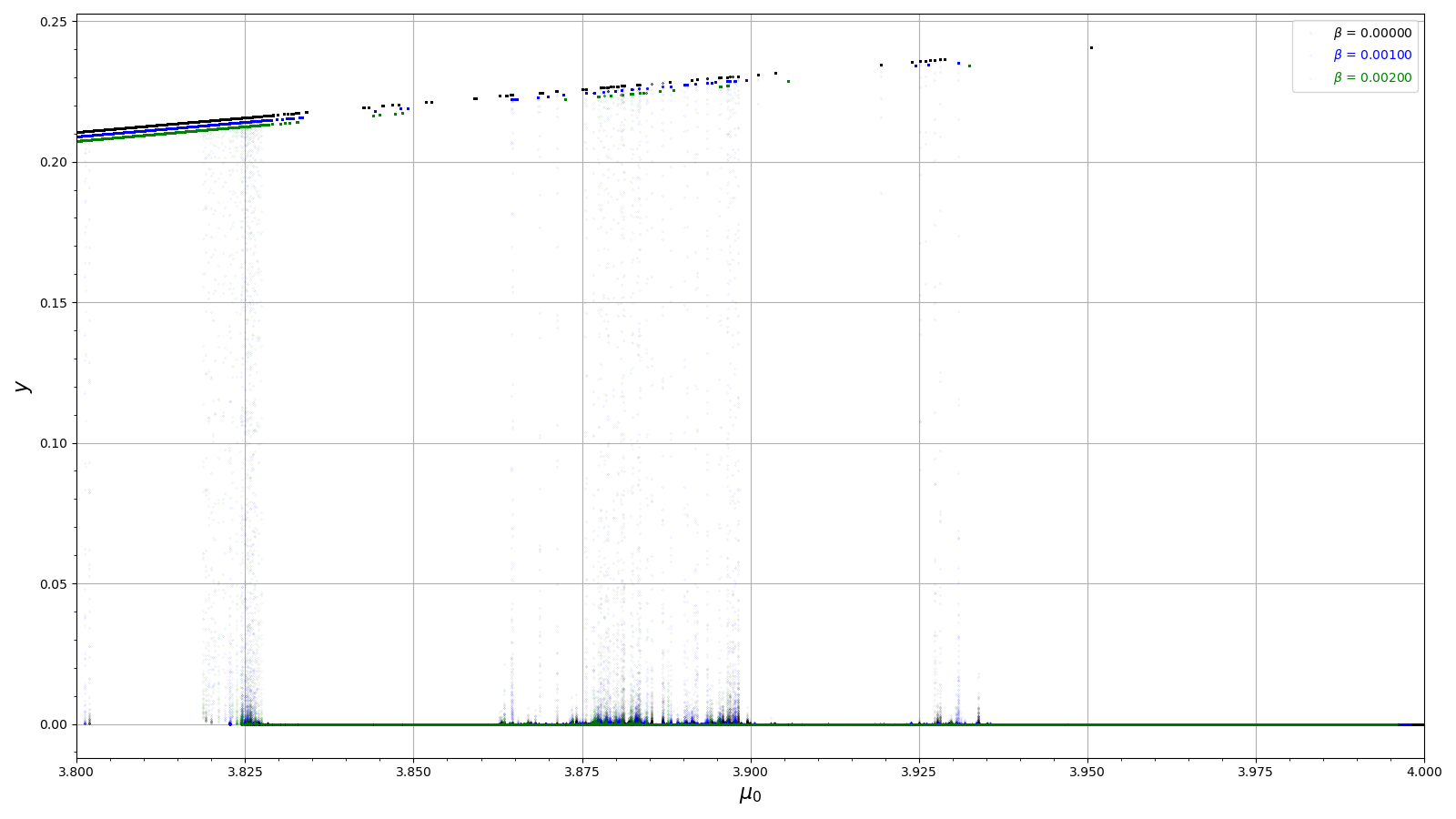}
	\caption{Comparison of $\beta$ values for the predator population $y$ of Extinction within $3.8<\mu_{0}<4.0$.}
	\label{fig:threeCurvesExtinction}
\end{figure}

Figure \ref{fig:threeCurvesExtinction} shows comparisons of the predator population ($y$) of \textbf{Extinction} for the three values $\beta=0.001$, $0.002$ and $0.003$ with the range $3.8<\mu_{0}<4.0$. In general, the survival population of the predator is very close for different values of $\beta$. But there is still some minor difference. Careful analysis shows that increasing $\beta$ causes a reduction in the number of survival orbitals of the predator, which, not surprisingly, means that increasing the death rate of the predator would diminish the possibility of survival of the predator in the era of \textbf{Extinction}. However, if we increase $\beta$ further, for example $\beta\geq0.003$, \textbf{Extinction} phenomenon cannot hold. This counterintuitive result could infer that chaotic extinction cannot occur with a slightly higher death rate of the predator.

\newpage
\subsection{Vorticella}\label{subs:vorticella}
\begin{center}
	\begin{figure}[!htbp]
		\begin{subfigure}[b]{1.0\textwidth}
			\centering
			\includegraphics[width=1.0\linewidth]{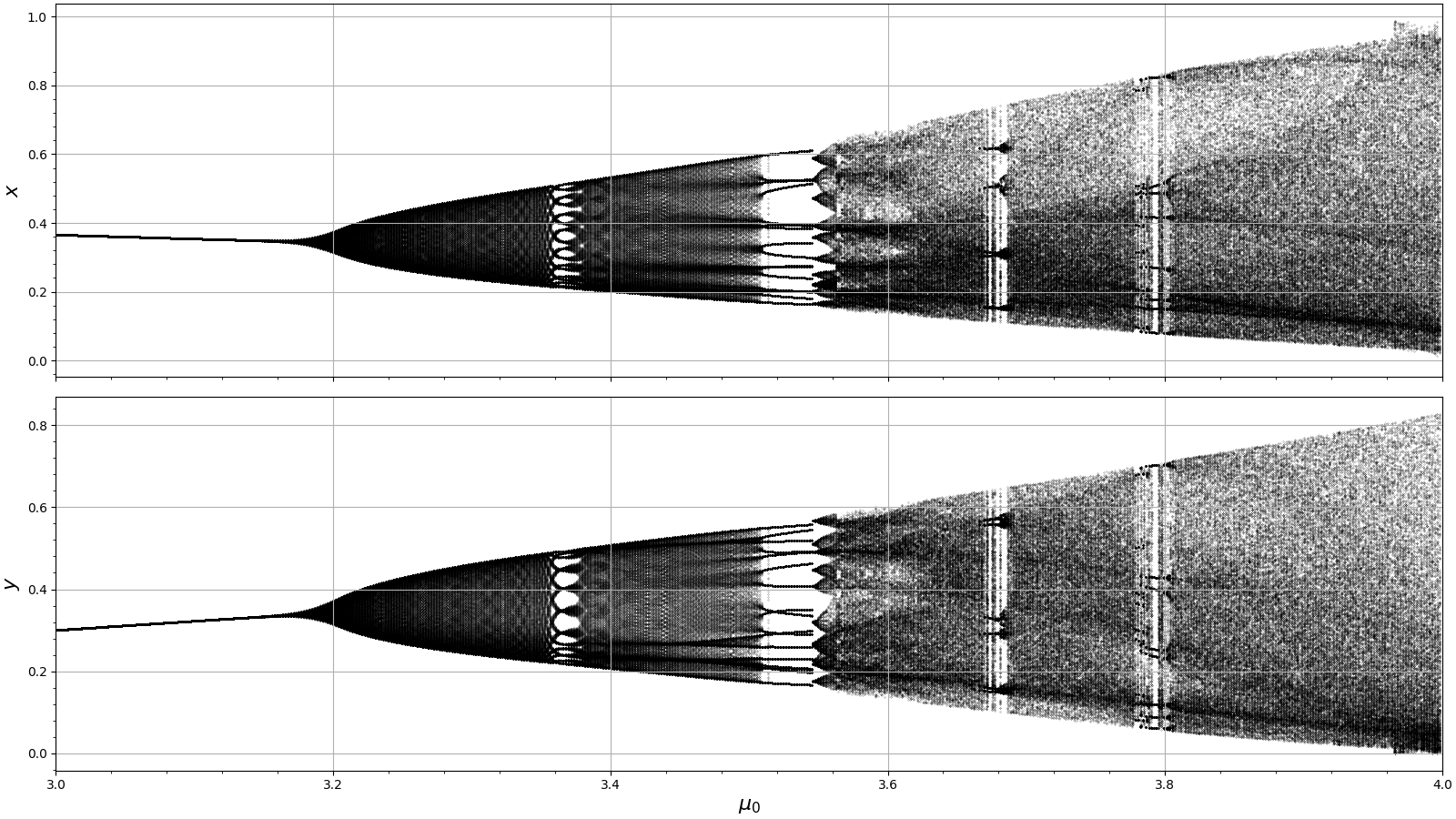}
			\caption{}
			\label{fig:Vorticella_bifurcation}
		\end{subfigure}
		\begin{subfigure}[b]{1.0\textwidth}
			\centering
			\includegraphics[width=1.0\linewidth]{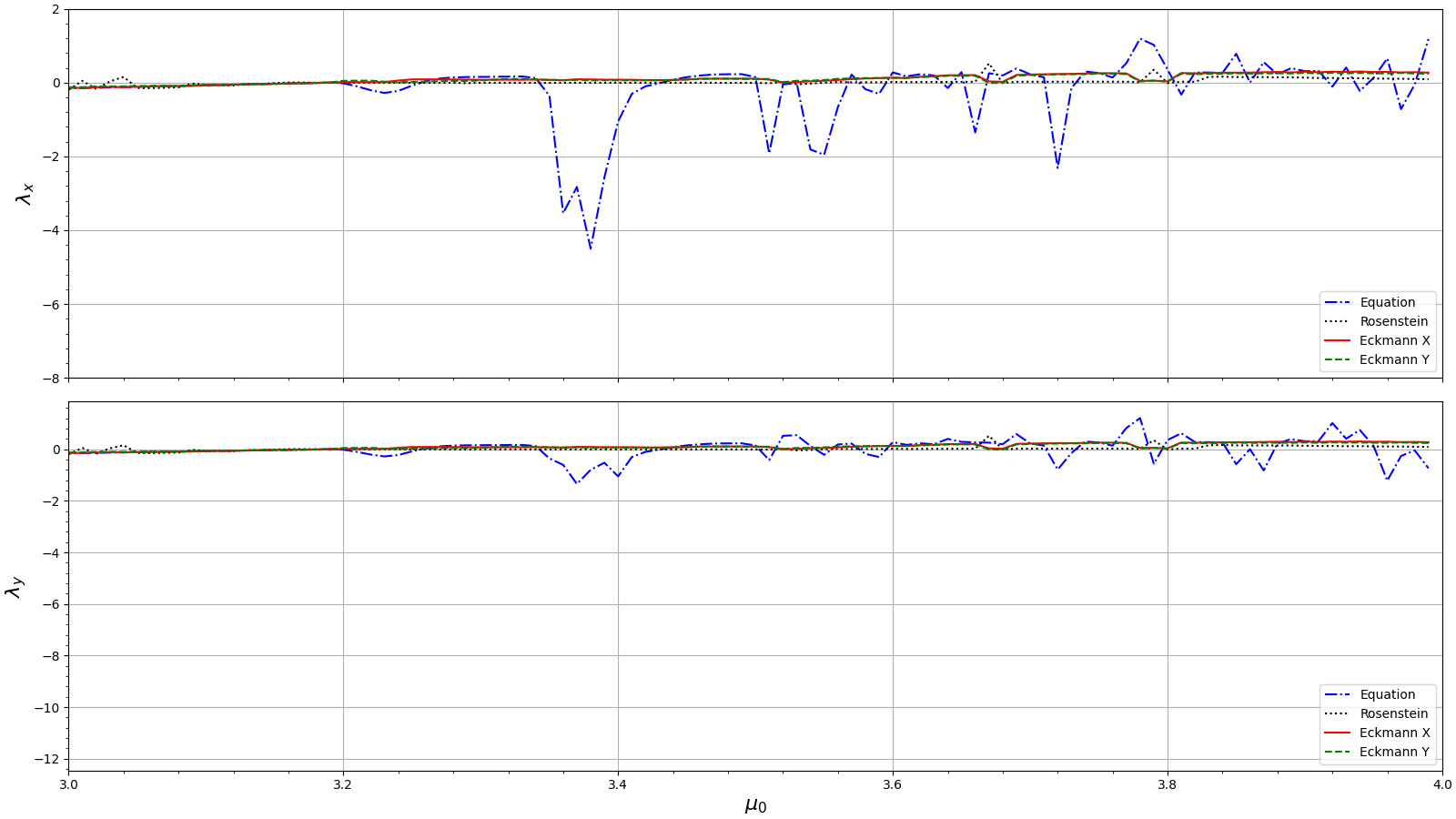}
			\caption{}
			\label{fig:Vorticella_lyapunovExponents}
		\end{subfigure}
		\caption{Bifurcation diagram and Lyapunov exponents of Vorticella.}
		\label{fig:VS}
	\end{figure}
\end{center}

After $\gamma>0.694$, the bifurcation type is Vorticella, named because of its appearance. When $\mu_{0}<2.0$, $x$ grows steadily without $y$. In the presence of the predator, the population of the prey starts to decrease. Both species show a normal population before $\mu_{0}=3.195\approx3.2$, at which point we classify a Neimark-Sacker bifurcation, as shown in Figure $\ref{fig:Vorticella_bifurcation}$. Figure $\ref{fig:Vorticella_lyapunovExponents}$ shows its Lyapunov spectrum. All four algorithms barely show positive spectra for $x$ and $y$ before $\mu_{0}=3.195\approx3.2$. On the contrary, after $\mu_{0}=3.195\approx3.2$, four algorithms show positive and negative values in the Lyapunov exponents, where the system goes into chaos.

\begin{figure}[!htbp]
	\centering
	\includegraphics[width=\linewidth,height=0.95\textheight,keepaspectratio,angle=0]{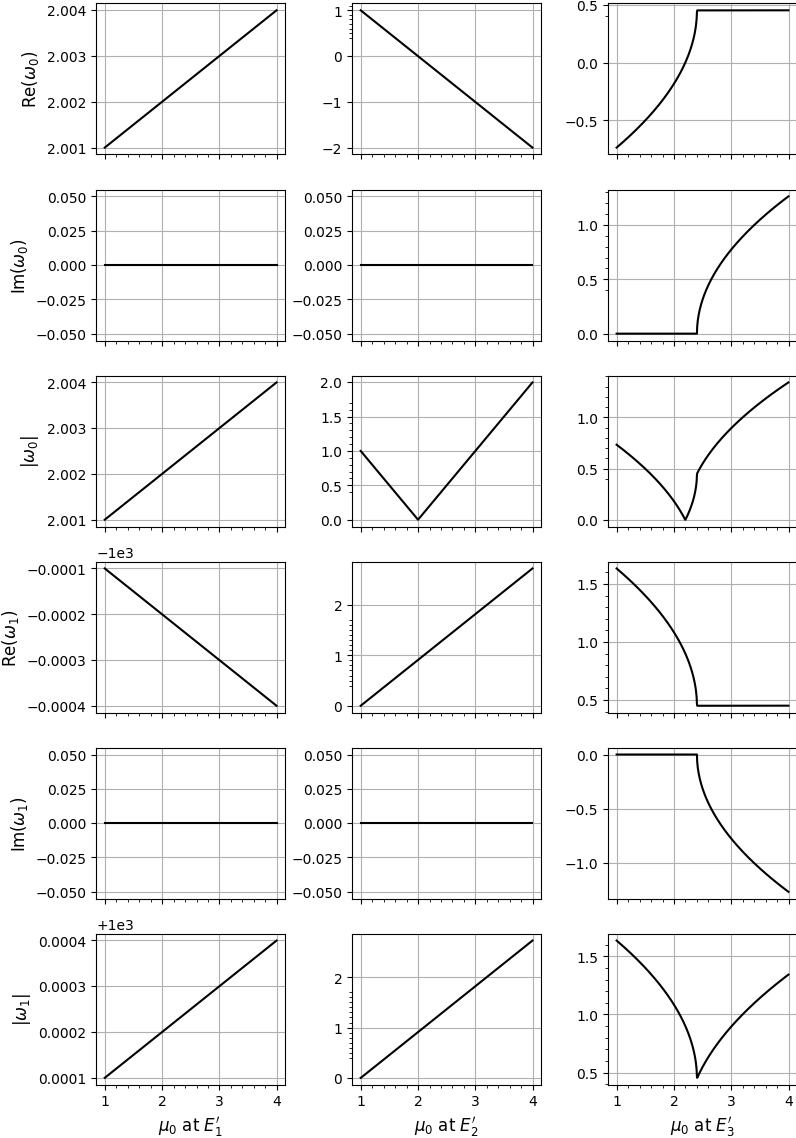}
	\caption{Absolute values of eigenvalues vs. growth rate at fixed points for Vorticella. We observe that $E^{\prime}_{1}$ is always a source. $E^{\prime}_{2}$ is a sink when $1<\mu_{0}<2.1$, a saddle when $2.3<\mu_{0}<3$, and a source when $3<\mu_{0}<4$. In addition, $E^{\prime}_{2}$ is a non-hyperbolic at $\mu_{0}=2.1$ and $3$. For $E^{\prime}_{3}$, the Neimark-Sackar bifurcation occurs at $\mu_{0}=3.195\approx3.2$.} 
	\label{fig:Vorticella_absEval_vs_mu0}
    \end{figure}

\begin{figure}[!htbp]
	\begin{subfigure}[b]{0.5\textwidth}
		\centering
		\includegraphics[width=1.0\linewidth]{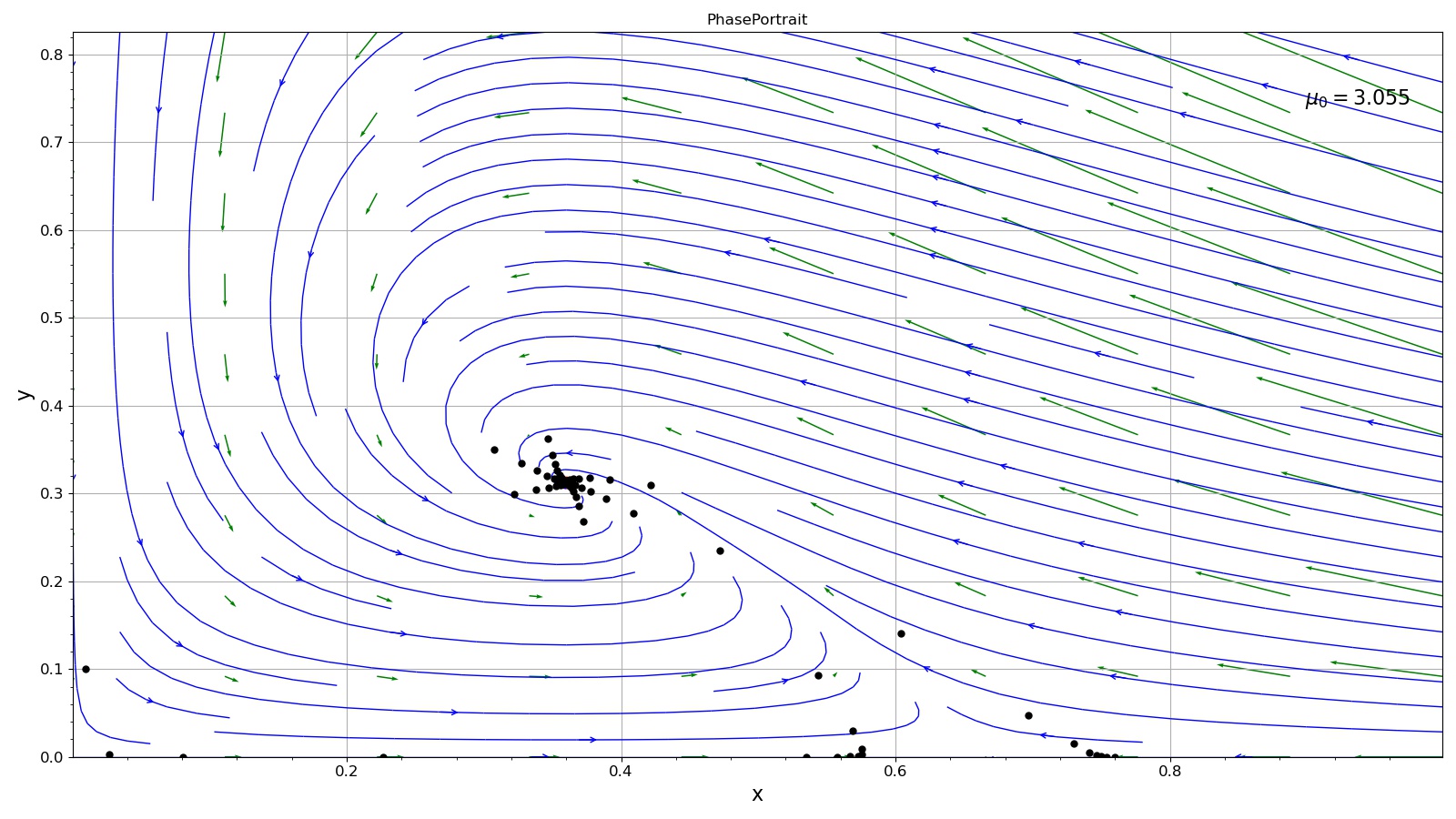}
		\caption{}
		\label{fig:Vorticella_Trajectory_3.05500}
	\end{subfigure}
	\begin{subfigure}[b]{0.5\textwidth}
		\centering
		\includegraphics[width=1.0\linewidth]{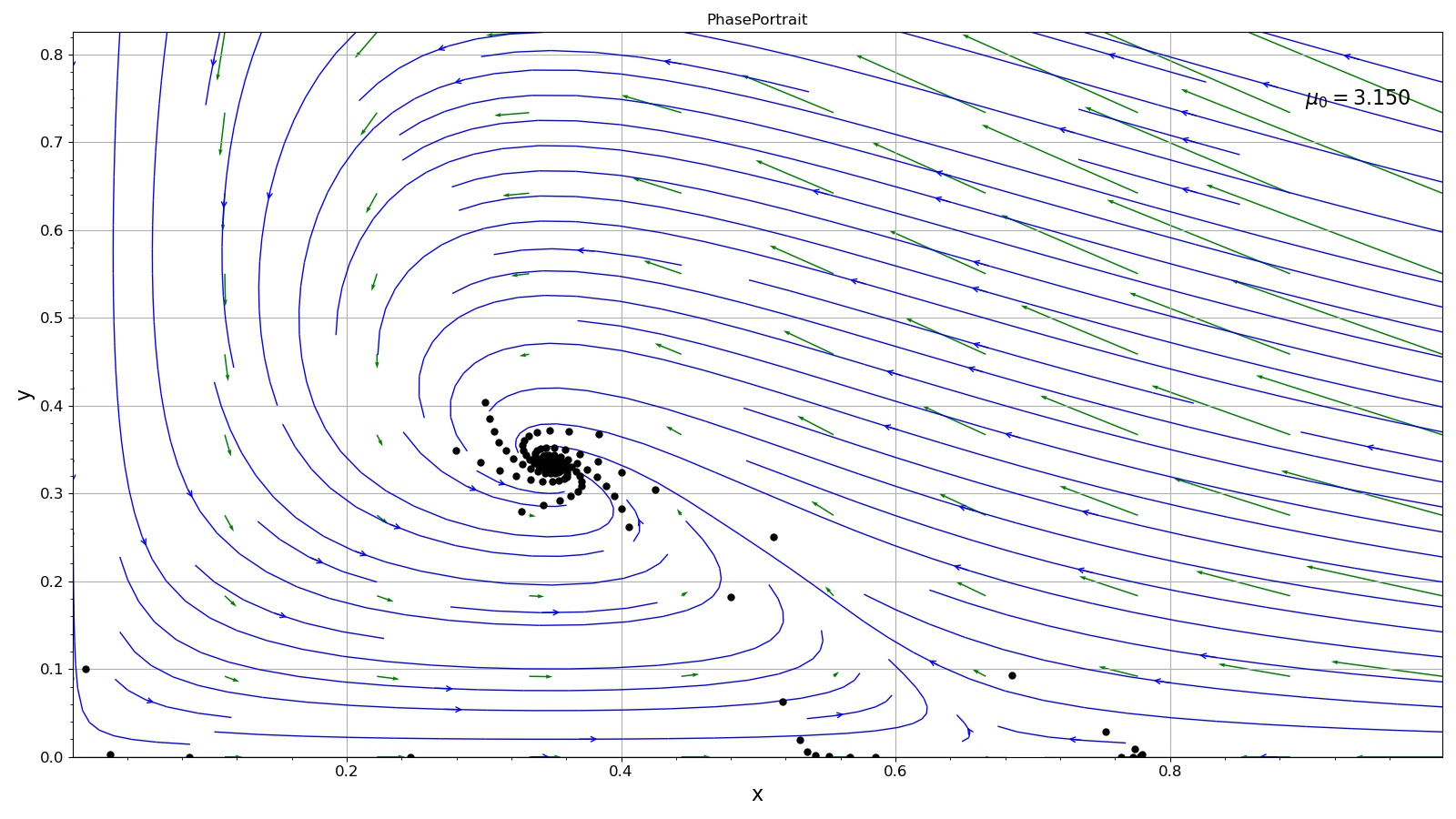}
		\caption{}
		\label{fig:Vorticella_Trajectory_3.15000}
	\end{subfigure}
	\begin{subfigure}[b]{0.5\textwidth}
		\centering
		\includegraphics[width=1.0\linewidth]{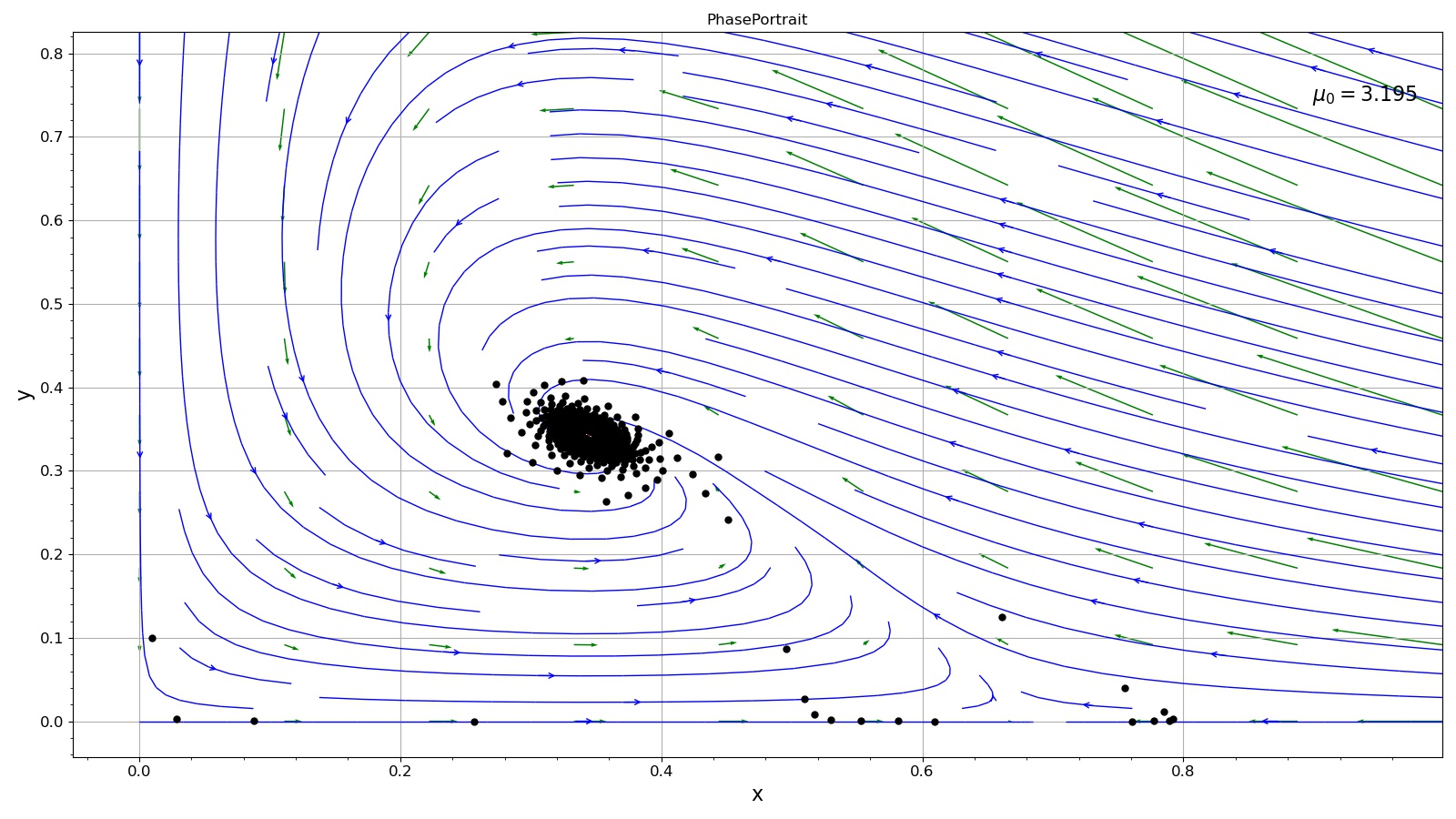}
		\caption{}
		\label{fig:Vorticella_Trajectory_3.20000}
	\end{subfigure}
	\begin{subfigure}[b]{0.5\textwidth}
		\centering
		\includegraphics[width=1.0\linewidth]{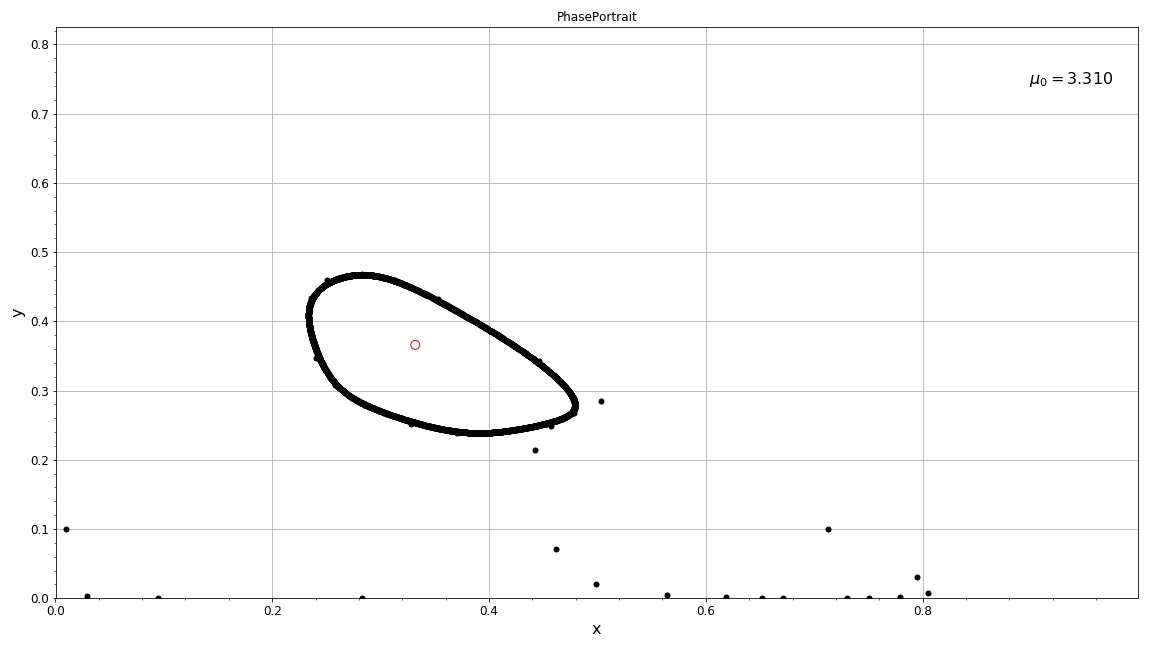}
		\caption{}
		\label{fig:Vorticella_Trajectory_3.31000}
	\end{subfigure}
	\begin{subfigure}[b]{0.5\textwidth}
		\centering
		\includegraphics[width=1.0\linewidth]{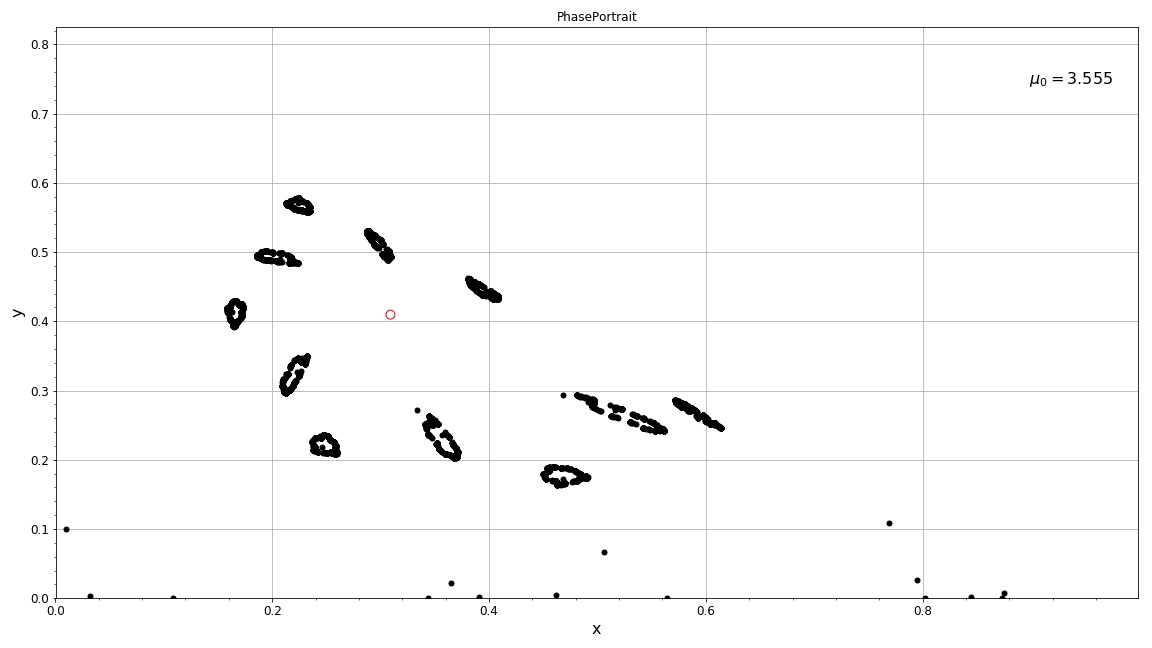}
		\caption{}
		\label{fig:Vorticella_Trajectory_3.55500}
	\end{subfigure}
		\begin{subfigure}[b]{0.5\textwidth}
		\centering
		\includegraphics[width=1.0\linewidth]{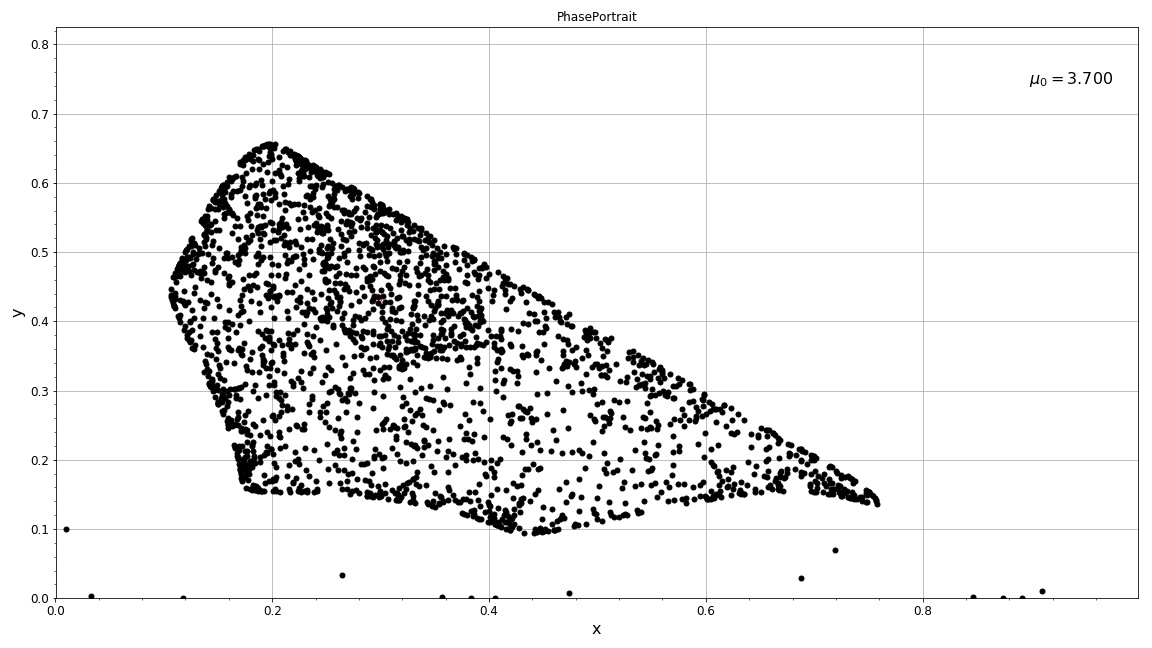}
		\caption{}
		\label{fig:Vorticella_Trajectory_3.70000}
	\end{subfigure}
	\caption{Some outstanding phase portraits of Vorticella.}
	\label{fig:Vorticella_Trajectory}
\end{figure}

The first row of Figure $\ref{fig:Vorticella_absEval_vs_mu0}$ shows the eigenvalues of $E^{\prime}_{1}$, both of which have zero imaginary parts, with absolute real parts greater than $1$, showing that $E^{\prime}_{1}$ is a source. A similar analysis may also be performed on $E^{\prime}_{2}$, as shown in the second column of Figure $\ref{fig:Vorticella_absEval_vs_mu0}$. For $1<\mu_{0}<2$, $E^{\prime}_{2}$ is a sink and turns from a sink to a saddle at $\mu_{0}=2.1$, where it is a non-hyperbole ($\omega_{0}\approx0.166$ while $\omega_{1}=1$). Within $2.1<\mu_{0}<3$, $E^{\prime}_{2}$ is a saddle, and it turns from a saddle to a source at $\mu_{0}=3$, where it is a non-hyperbole ($\omega_{0}=1$ while $\omega_{1}\approx1.867$). After $\mu_{0}>3$, $E^{\prime}_{2}$ is a source. Finally, the third column in Figure $\ref{fig:Vorticella_absEval_vs_mu0}$ shows that the Neimark-Sacker bifurcation occurs at $\mu_{0}=3.195\approx3.2$, with coordinates ($0.346,0.339$) in the phase portrait, as shown in Figure $\ref{fig:Vorticella_Trajectory_3.20000}$ due to the following facts: first, $\omega_{0}$ and $\omega_{1}$ are complex conjugates with modulus $1$, and second, as $\mu_{0}$ varies between $3.2$ from smaller to larger value, the topological type of $E^{\prime}_{3}$ changes from a sink (stable) to a source (unstable)$\cite{Mareno and English}$.  

Hu et al. $\cite{Hu Teng and Zhang}$ found a similar bifurcation diagram that was identified as the Hopf bifurcation. However, the criteria for Hopf bifurcation in the two-dimensional system include that the two eigenvalues are purely conjugate imaginary pairs with zero real part$\cite{Hale and Kocak}$. Since Figure $\ref{fig:Vorticella_absEval_vs_mu0}$ shows that $\omega_{1}$ has a nonzero real part, our Vorticella cannot be a Hopf bifurcation.

Figure $\ref{fig:Vorticella_Trajectory}$ shows some interesting phase portraits in this case. Figure $\ref{fig:Vorticella_Trajectory_3.05500}$ shows a star with $\mu_{0}=3.055$, and Figure $\ref{fig:Vorticella_Trajectory_3.15000}$ shows a spiral with $\mu_{0}=3.150$, both of which have counterclockwise vector flows spiraling into the fixed point (sink). However, when the Neimark-Sacker bifurcation occurs at $\mu_{0}=3.195\approx3.2$ as shown in Figure $\ref{fig:Vorticella_Trajectory_3.20000}$, whose stability cannot be determined by linearization and we have to exploit center manifold of mappings, a limit cycle appears and divides the phase portrait into two regions. At $\mu_{0}=3.310$ (Figure $\ref{fig:Vorticella_Trajectory_3.31000}$), inside the limit cycle, the vector flow is unstable because the fixed point, represented by the read cycle, is an unstable source, while outside the limit cycle, the vector flow is still stable spiraling toward the edge of the limit cycle. Before chaos, there are attractor islands that spread around the fixed point (source), as we can see in Figure $\ref{fig:Vorticella_Trajectory_3.55500}$. After that, the phase portrait shows chaos, as in Figure $\ref{fig:Vorticella_Trajectory_3.70000}$. Similar results were found in Danca et al.$\cite{Danca et al}$ and Song $\cite{Ning Song}$. 

\begin{figure}[!htbp]
	\centering
	\includegraphics[width=1.0\textwidth]{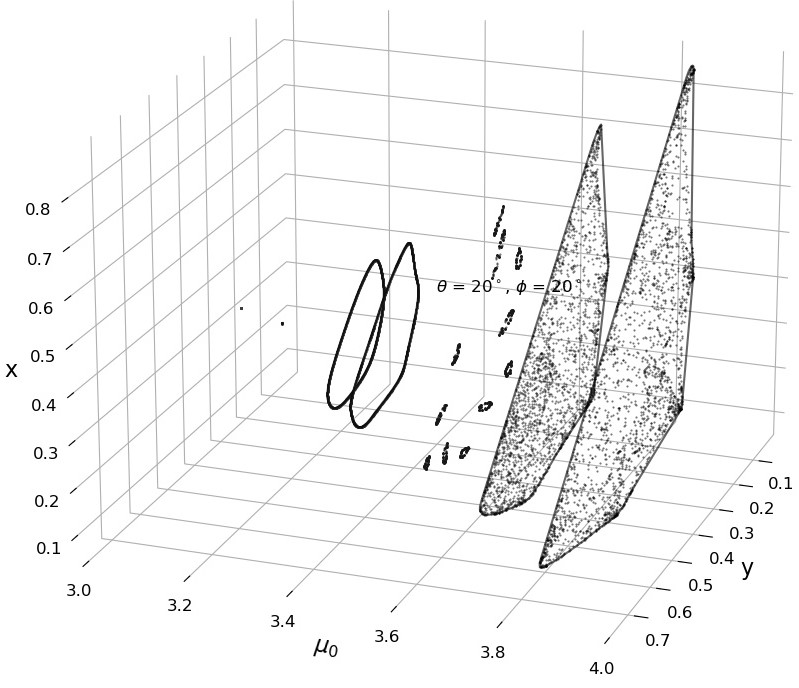}
	\caption{Neimark-Sackar}.
	\label{fig:Vorticella_NSLoops_020_020}
\end{figure}

It is much more interesting to show the phase portrait in three dimensions, with $\mu_{0}$ varying as another axis. In Figure $\ref{fig:Vorticella_NSLoops_020_020}$, we clearly see that within $3.2<\mu_{0}<3.4$, the phase portraits are loops, although they appear blacked in the corresponding region in the bifurcation diagram (Figure $\ref{fig:Vorticella_bifurcation}$.) Within the range $3.4<\mu_{0}<3.6$, there is a phase portrait consisting of attraction islands. After $\mu_{0}>3.6$, the phase portraits demonstrate chaos regions with boundaries delimited by black curves. The figure shows that it is important to keep in mind that even if there is a black region in the bifurcation diagram, the system may not exhibit chaos because the black region just exhibits a loop in a higher-dimensional space.

\newpage
\section{Conclusions}
We successfully built up a discrete-time model of two-dimensional mapping that resembles features of differential Lotka-Volterra Equations. We studied the topological types of fixed points and found that fascinating feather-like structures were formed around limit circles pierced through by the axis of fixed points in the phase portraits and phase-space diagrams under various growth rates with Neimark-Sacker bifurcation. We further divided our dynamical systems into four categories depending on different shapes of bifurcation diagrams. \textbf{Normal}, \textbf{Standard}, \textbf{Vorticella},  \textbf{Extinction},  and \textbf{Vorticella}. In each case, we study the stability and topological types of the fixed points using criteria discussed in \textbf{Lemma \ref{lemma}}. In addition to plotting the population vs. iteration, we also calculated Lyapunov exponents both by Jacobian eigenvalues of the mapping functions and by the Rosenstein algorithm and Eckmann et al. algorithm. Discrepancies clearly showed that Lyapunov exponents calculated by time-series algorithms may be unreliable within the range of chaos and at low growth rates, as well as when the Lyapunov exponents change dramatically. Furthermore, our model not only regained the logistic mapping $1$ D of the prey under zero predator yet with nonzero initial predator population and nonzero interspecies constants, but also showed the normal competitiveness of the prey and the predator without chaos. The quintessence of the current research was that, in addition to the possibility that the prey and predator become chaotic altogether, it is also probable that the predator will go extinct in the chaotic state of the prey. In other words, human overpopulation would cause chaos in natural resources and, ultimately, in return erase the entire human race. Fortunately, even under this difficult circumstance, a slim chance is still left upon us to continue our race under some specific growth rate, as we may see some isolated fixed points remain in the predator bifurcation diagram. 

Last but not least, our model may inspire conjecture on other relationships between two quantities coupled with the form of discrete-time difference equations as in our study because mathematically, what we demonstrate is that one of the two quantities may dramatically reduce to zero at the onset state of chaos of the other. Therefore, it could be highly possible that, for example, the superconducting state, which refers to the zero resistance, may be achieved with the chaos of the applied magnetic field. Another possible application may be that we may suppress viruses or pests by triggering the chaotic states of the prey on which the viruses or pests feed for survival.  
\newpage
      
\section{Acknowledgment}
The authors thank Pui Ching Middle School in Macau PRC for its kindness in supporting this research project.

\end{document}